\documentclass[12pt]{article}
\usepackage[utf8]{inputenc}
\usepackage{amssymb}
\usepackage{amsmath}
\usepackage{amsthm}
\usepackage{float}
\usepackage{setspace}
\usepackage{subcaption}
\usepackage{siunitx}
\usepackage{bbm}
\usepackage[margin=1in]{geometry}
\usepackage{hyperref}
\usepackage{multicol}
\usepackage{multirow}
\usepackage{caption}
\usepackage{float}
\usepackage{placeins}
\usepackage{afterpage}
\usepackage{pdflscape}
\usepackage{longtable}
\usepackage{array}
\usepackage{booktabs}
\usepackage{makecell}

\newcolumntype{P}{>{\raggedright\arraybackslash}p{3cm}}
\newcolumntype{T}{>{\raggedright\arraybackslash}p{1.5cm}}
\newcolumntype{R}{>{\raggedright\arraybackslash}p{5cm}}

\usepackage[authordate, 
            backend=biber,
            ibidtracker=false,
]{biblatex-chicago}
\AtEveryBibitem{
    \clearfield{urlyear}
    \clearfield{urlmonth}
}
\DeclareSourcemap{
  \maps[datatype=bibtex]{
    \map{
      \step[typesource=dataset, typetarget=article, final]
    }
  }
}

\DeclareSourcemap{
  \maps[datatype=bibtex]{
    \map[overwrite]{
      \step[fieldsource=doi, final]
      \step[fieldset=url, null]
      \step[fieldset=eprint, null]
    }  
  }
}
    
\usepackage{xcolor}
\definecolor{Pink}{RGB}{245, 169, 184}
\definecolor{Blue}{RGB}{91, 206, 250}

\hypersetup{
    colorlinks=true,
    linkcolor=magenta,
    filecolor=magenta,  
    citecolor=magenta,
    urlcolor=magenta,
    pdftitle={The Short- and Long-Term Impacts of Expanding Public Education for Disabled Students}
    }

\theoremstyle{definition}

\usepackage{accents}

\usepackage{graphicx}
\author{Laura Caron \footnote{lkc2142@columbia.edu. The author acknowledges the support of the National Science Foundation Graduate Research Fellowship Program. I am  grateful to Alessandra Casella, Simon Lee, and Ebonya Washington, as well as Bitsy Perlman and Erwin Tiongson, for their excellent feedback and support. I also thank Elizabeth Ananat, Cassandra Benson, Sandra Black, Claudia Goldin, Rajeev Dehejia, Bilge Erten, Mary Eschelbach Hansen, Donn Feir, Maxwell Grozovsky, Emily Guiterrez, Rey Hernández-Julián, Tatiana Homonoff, Michelle Jiang, Jo King, Elira Kuka, Nicole Maestas, Katie Malhotra, Tatiana Mocanu, Victoria Mooers, Suresh Naidu, Amanda Pallais, Douglas Ready, Bernard Salanié, Evelyn Skoy, and Leanna Stiefel for many helpful discussions and suggestions for improvement. I acknowledge attendees of the Applied Micro/Labor and Econometrics colloquia and seminars at Columbia University, attendees of the Graduate Workshop in Economic History at Harvard University, and seminar participants at American University for their comments and questions.}}
\title{{\vspace{-2.0cm} \singlespacing \Large The Short- and Long-Term Impacts of Expanding Public Education for Disabled Students} \\  
}
\date{October 2025}
\begin{document}

\maketitle
\null\vspace{-1cm}

\begin{abstract}
\singlespacing
\noindent Between 1949 and 1980, every U.S. state mandated public schools to provide educational services for disabled students. This is one of the largest education reforms in U.S. history, but little is known about its impacts. Given scarce data in this period, I compile survey and administrative datasets and set up a difference-in-difference design using variation in the mandates' timing. I show that the mandates increased both services for disabled students and preschool enrollments. In adulthood, disabled individuals below school age at a mandate's implementation became about 20\% less likely to have no education, attained up to 0.23 more years of education, and were more likely to have worked. Although this policy could have taken away resources from non-disabled students, in fact, education and employment also increased for non-disabled individuals. These effects align with evidence that the mandates increased spending per student by up to 15\%. Families were also impacted: the mandates increased employment among mothers of disabled children and the probability that disabled individuals became household heads. Over the long term, the mandates paid for themselves by generating government revenues in excess of their cost. These results provide new evidence on the large, broad impacts of expanding access to education for disabled students.

\end{abstract}

\vfill

\noindent \textbf{Note:} This paper contains quotes from historical sources including terms now commonly considered slurs or derogatory among disabled people.
\vspace{1cm}

\newpage

\setcounter{section}{0}

\section{Introduction}

Historically, many public schools did not provide educational services or support for students with disabilities (also known as special education). Instead, disabled students struggled in their education or were excluded from public schools altogether. But, between 1949 and 1980, as activists pushed to improve the education of disabled students, every state in the US enacted legislation mandating public schools to identify and provide educational services for children with all types of disabilities. This is arguably one of the largest expansions of public education in recent US history, potentially enabling access to education for as many students as other major programs like Head Start. Today, about 15\% of students in the US receive these services \parencite{ncesStudentsDisabilities2024}. 

Given that they affected so many students, these services have potentially large and wide-reaching economic impacts, and understanding these is a key question for policy. In this paper, I study the causal impacts of these state mandates on disabled students and their non-disabled peers from preschool to adulthood and on a wide range of outcomes, including education, labor market outcomes, and family life. 

Before the mandates required states to provide services for disabled students, little data was collected on their education, in part because it was considered outside the scope of or only a marginal part of the public school system. This lack of data is one possible reason for the scarcity of prior work on these mandates despite their importance. I begin by assembling a database of the relevant legislation in each state. To study relevant outcomes, I identify and compile a number of datasets from survey, administrative, and Census sources. Together, these data sources allow me to conduct, to my knowledge, the first quantitative causal study of these mandates and to offer detailed new evidence on how access to education for disabled students impacts their economic and family lives. 

Using variation in the timing of the mandates between states and a staggered difference-in-difference design, I estimate causal impacts separately for both disabled and non-disabled students. 
I validate this empirical design using a novel state-level dataset on the number of students receiving services for a disability from 1952 forward.

Using this design and several additional datasets on individuals' outcomes, I find that mandating the provision of services for disabled students improved adulthood education and labor market outcomes for disabled individuals and had positive spillovers for their parents and peers. I build on previous work that has shown the benefits of providing services for disabled students for short-term outcomes like test scores, but which has rarely been able to observe outcomes beyond a student's exit from education (eg, \textcite{ballisLongRunImpactsSpecial2021, hurwitzSpecialEducationIndividualized2019}). I  add evidence on the positive impacts of these services on a broader set of outcomes extending even later into adulthood. I also use data on school finances to understand how school spending may act as a mechanism for these impacts. Throughout the paper, I present evidence ruling out factors other than the mandates that could explain these gains.

I begin by showing that the mandates resulted in immediate improvements in the services disabled students received at school. To analyze the services received on the individual level, I use a little-known health survey, the National Health Examination Survey (NHES), which gives one of the richest pictures of children's disabilities and the services they received at school over this period.
I find that the mandates made disabled students much more likely to receive a school recommendation to receive services and to actually receive these services, a large increase of about 20 percentage points. Disabled students also became more likely to be transferred into ``special education'' classes and less likely to be frequently absent from school. 

Along with improving the services received by disabled students, I show that the mandates increased overall school enrollments. I quantify this increase using individual-level data on enrollment from the Current Population Survey (CPS) October Supplement. In line with their provisions, the mandates caused large increases in preschool enrollments. They also increased enrollments among students over the age of compulsory schooling, indicating that students stayed in school longer. As the services offered by public education improved, I find that the mandates caused shifts from private to public education.  

Having shown that the mandates expanded the public school services available to disabled individuals and school enrollments, I next present evidence that they improved educational attainment in adulthood. To study adulthood outcomes, I use data from the Census and American Community Survey (ACS) from 1970-2007. I use a similar difference-in-differences design, exploiting variation in an individual's birth year relative to the timing of the legislation in their state. I find that, by age 25-35, the mandates caused large improvements in educational attainment among disabled individuals. Disabled individuals under school age at the time of the mandates' implementation had an average increase in educational attainment of approximately 0.23 years through grade 12. They also became much less likely to have very poor education outcomes, such as no schooling at all. 

Turning to impacts on their non-disabled peers, the theoretical sign of any spillover effect could be positive or negative: improvements in the quality of public education could benefit non-disabled students, but redirecting resources away from them could instead hurt them. My evidence suggests that spillovers are positive. Using a difference-in-difference design analogous to that used for disabled individuals, I find that non-disabled individuals who were young when a mandate was enacted in their state also experienced an increase in educational attainment, with an average increase of 0.24 years of education for those under school age. However, this increase occurred at higher levels of education: while disabled students mostly experienced increases in their primary and secondary education, non-disabled students experienced increases in higher education.   

Why did the mandates have such positive impacts, especially for non-disabled individuals? I unpack the mechanism for these improvements using a database of state and school district finances. As expected given that the mandates did not include state funding provisions, state governments did not increase spending in response to these mandates. Instead, school districts funded increases in expenditures via local property taxes. In the long term, school districts experienced large increases of about 15\% in spending per student, driven by increased employment in public education. Using estimates from prior work of the impact of increased funding on educational attainment, I highlight that this increase in spending can plausibly explain a sizeable share of the positive spillover experienced by non-disabled students.

I find that these improvements in education also came with improvements in labor market outcomes for both disabled and non-disabled individuals. At age 25-35, the mandates increased the probability of disabled individuals having some work experience by 2.9 percentage points and reduced the receipt of Social Security disability benefits for those unable to work by the same amount. For non-disabled individuals, the mandates increased employment by 2.8 percentage points, with correspondingly large increases in income.

The mandates also had positive spillovers for another group: the parents of disabled children. Before the mandates, mothers of disabled children may have had greater care responsibilities for children who were out of school or struggling in school. Following the mandates, I show that employment of mothers of disabled children increased.

Beyond education and employment, the mandates' impacts also extended into the families and social lives of disabled people. As would be expected if improved education led them to be more independent, I find evidence that, in adulthood, disabled people affected by the mandates became more likely to head their own households and to become parents. 

Finally, I quantify the monetary costs and benefits of the mandates. I find that the mandates have large monetary benefits for affected individuals and, from a public funds perspective, pay for themselves in the form of increased tax revenue resulting from higher incomes. On top of these monetary benefits, it is likely that the largest benefits of these mandates for disabled individuals, such as overall improved wellbeing, are not quantifiable. 

\textbf{Contributions.} In this paper, I document and provide the first causal quantitative estimates of the impacts of these state-level mandates to educate disabled students, which constituted a major expansion of public education services. Previous work studying mandates to educate disabled children has been limited to qualitative or descriptive studies (eg, \textcite{leafstedtWasItWorth2007, wrightLocalImplementationPL1980, hobbsProgressFreeAppropriate1979}). The introduction of these mandates expanded the availability of public education services for all disabled children, that is, for up to 15\% of the child population. This can be compared to the impact of the introduction of compulsory schooling laws in the nineteenth century US, which are estimated to have increased school attendance by up to 10 percentage points \parencite{margoCompulsorySchoolingLegislation1996}. 
The mandates also constituted an expansion of public preschool provision for disabled children, in a vein similar to Head Start, which is estimated to have served about 10\% of children born between 1964 and 1977 \parencite{garcesLongerTermEffectsHead2002} and to have increased educational attainment, earnings, and health \parencite{garcesLongerTermEffectsHead2002, ludwigDoesHeadStart2007a, baileyPrepSchoolPoor2021,johnsonReducingInequalityDynamic2019}. I also contribute to a literature which has debated the impacts of school funding on educational attainment (eg, \textcite{jacksonEffectsSchoolSpending2016}) by adding evidence from a new source of variation. 

Relative to the existing literature on providing educational services to disabled children, the results from this paper provide new insights on the short- and long-term impacts of these services on a rich set of outcomes. Prior work has used smaller sources of variation -- in the US, limited to a single state or school district -- to show that the provision of services for disabled students improves test scores in the short- and medium-term \parencite{schwartzEffectsSpecialEducation2021, hanushekInferringProgramEffects2002, hurwitzSpecialEducationIndividualized2019, nielsenHowCopeDyslexia2021}. On the other hand, some papers have indicated that services can generate negative impacts for disabled students, particularly when they are stigmatized, there are tracking effects, or services are low-quality \parencite{bensonThreeEssaysLocal2019, morganPropensityScoreMatching2010}. Given that I study a reform that occurred across states, I am able to identify average effects for a much broader group of students. Meanwhile, although a small but growing strand of the literature suggests that the benefits of these services can persist in the long-term, it focuses only on academic achievement and educational attainment (eg, \textcite{hurwitzSpecialEducationIndividualized2019}). For example, \textcite{ballisLongRunImpactsSpecial2021} show that reductions in services for disabled students cause negative and lasting impacts on educational trajectories, including reductions in high school graduation and college enrollment. I study outcomes that also extend beyond school years and add a number of outcomes, including benefit receipt, employment, and household structure, to this literature. Further, I study school funding and employment of teachers as factors that shape the impacts of these services. 

Beyond impacts on the disabled students themselves, I provide evidence to answer an open question in the literature on the spillovers of services for disabled students. Positive spillovers for non-disabled students may arise from improved peer effects, while negative spillovers may arise if resources are redirected away from non-disabled students. Prior work has generally not found evidence of negative spillovers, with \textcite{hanushekInferringProgramEffects2002} finding no spillovers on non-disabled students and \textcite{ballisLongRunImpactsSpecial2021} finding that reductions in services for disabled students worsen outcomes for non-disabled students, implying a positive spillover. I contribute to this literature by offering estimates of the spillover impact of a large increase in the availability of educational services for disabled students on non-disabled students and funding as a possible mechanism for these spillovers. 

Finally, this work also sheds light on a potential mechanism driving gender gaps in labor supply during a period of rapidly expanding maternal employment \parencite{goldinQuietRevolutionThat2006}. A broad literature has shown maternal labor supply reductions in response to having a child, e.g., \textcite{angristChildrenTheirParents1998, lundborgCanWomenHave2017, angelovParenthoodGenderGap2016}. These effects may be particularly strong when the child has a disability \parencite{wasiHeterogeneousEffectsChild2012a, zhuMaternalEmploymentTrajectories2016a, burtonChildHealthParental2017, powersNewEstimatesImpact2001, porterfieldWorkChoicesMothers2002}. I connect this work with a literature indicating that public school provision, specifically for young children, can increase maternal labor supply \parencite{fitzpatrickRevisingOurThinking2012, gelbachPublicSchoolingYoung2002, cascioMaternalLaborSupply2009}. I offer some of the first estimates of how services for disabled children, in particular, affect parental labor supply and how these effects differ by gender.

The paper proceeds as follows. In section 2, I outline the historical background of education for disabled students before the mandates and describe the mandates' origins and provisions. Section 3 describes the main difference-in-difference approach and provides evidence supporting its underlying assumptions. Results showing the mandates' impacts on services received by students, adulthood educational attainment and labor market outcomes, school finance, and families and household formation appear in sections 4, 5, 6, and 7, respectively. In section 8, I discuss the costs and benefits of the mandates. Section 9 concludes. 

\section{History of educational services for disabled students in the United States}
Between 1949 and 1980, every state in the United States implemented a mandate requiring public schools to provide educational services for disabled students. Despite a long history of special education systems in the US, the majority of disabled students did not receive adequate support for their education before these mandates. The mandates, enacted mostly in the 1970s, were motivated by a movement for more equal education and activism by parents of disabled children. 

\subsection{Educational services before the mandates}

Some forms of educational services for disabled children existed long before the mandates, but early forms offered limited services. In 1817, the American Asylum, At Hartford, For The Education And Instruction Of The Deaf was chartered as one of the first such public programs \parencite{gallaudetHistoryEducationDeaf1886}. Although schools for the d/Deaf, along with schools for the blind, were supported with state funds in the 19th century, they served only very small populations of severely disabled children and were residential programs. \textcite{lazersonOriginsSpecialEducation1983} argues that education for the disabled only attracted broader policy attention when states began to more strongly enforce compulsory schooling laws in the 1890s and early 1900s. As early as 1911, as more disabled children entered public schools, states began to establish some forms of ``special education'', but services were sporadically-provided and low-quality for many decades \parencite{lazersonOriginsSpecialEducation1983, winzerHistorySpecialEducation1993}.

By 1930, gaps in educational services for disabled students gained more attention. During this period, the rapid growth in education in the US \parencite{goldinAmericasGraduationHigh1998}, along with the financial struggles of the Great Depression \parencite{winzerHistorySpecialEducation1993}, made these gaps more obvious. In that year, a committee convened by President Herbert Hoover recommended that the federal government take action to ensure education for ``all types of handicapped children''. The committee found that at least 10 million children (22\% of the child population) could be considered to have some kind of “deficiency” (including 6 million undernourished), and that 80\% of them lacked the services they needed \parencite{longWhiteHouseConference1931}.

However, the situation did not improve much over the next decades. Schools continued to struggle with how to educate disabled students and place them in classrooms, and even denied enrollment to students with disabilities. For example, a 1951 letter from a school official in the public school system in Carlstadt, NJ highlights:

\begin{quote}
``We have in our community, a family in which two of the three children are definitely mentally retarded. ...They are unable to benefit by formal education. ... It has been recommended that these children be committed to a state institution, but this the mother has refused to do, and apparently has been unable to accept the fact that her children cannot be tought [sic] in the Public or Trade schools'' \parencite{novellaLetterBarringSchool1951}
\end{quote}

As a result, many disabled children received poor quality education. In 1967-1968, only 36\% of disabled children were receiving the services they needed at school \parencite{usofficeofeducationBetterEducationHandicapped1969}. Many disabled children found themselves out of the public school system, suffering in classes without support, or confined to institutions. For example, a 1970 report from the Task Force on Children Out of School in Boston, Massachusetts found that ``In general, crippled children in Boston are not allowed to attend school. ... A number of services are available to other children in the school system, but are denied to children in ‘special classes’'' \parencite{taskforceonchildrenoutofschoolWayWeGo1970}. Others, including many children with learning disabilities like dyslexia, were able to attend school but did not receive proper services to support their education \parencite{gaoDisparitiesStillExist1981}. For example, describing the struggle of some disabled students deemed ``slow learners'', \textcite{madisonNCTEERICREPORT1971} points out, ``In the past he was ‘held back’ once or twice in elementary school, ‘quit’ school soon after it was legally possible, or was promoted year after year until he finally ‘graduated’.'' Even when they nominally received services, these did not constitute a proper education: as one disabled person in school in the 1960s recalled, ``At that time, `special ed' was gluing peas on cardboard, and cleaning the windows of the high school'' \parencite{pelkaWhatWeHave2012}. 

Those who were considered too severely disabled to participate in public education may have been institutionalized for long periods of their lives and received little education. In 1967, over 107,000 people in the United States were confined to ``institutions for the mentally retarded'' \parencite{frohlichWhoAreDisabled1971}. These institutions received local, state, and federal funding, and often confined people for long periods with few opportunities for education or employment. For example, in the same year, the majority of those over 18 in institutions for the ``mentally retarded'' had been at the same institution at least 10 years. Further, 66\% had received no education at all. Another 353,000 people were institutionalized in psychiatric hospitals or chronic disease facilities, where more than 25\% reported having received no education. 

Conditions in these institutions and residential state schools were often poor and offered few opportunities for education. Visiting Rome State School and Willowbrook State School in New York City in 1965, then-Senator Robert F. Kennedy described the dismal conditions of residents: 

\begin{quote}
``And what do they do during the day? Many just rock back and forth. They grunt and gibber and soil themselves. They take off their clothes. They struggle and quarrel -- though great doses of tranquilizers usually keep them quiet and passive. ....     we observed no on-going programs with any purpose or direction. The classrooms at Rome were empty, as were the shops. The playrooms at Willowbrook were also empty.'' \parencite{kennedyExcerptsStatementKennedy1965}
\end{quote}
    
\subsection{The expansion of educational services for disabled students}
In this section, I discuss the process by which access to educational services began to expand for disabled students. Motivated largely by activism based on the belief that education is essential to improve the lives of disabled individuals, between 1949 and 1980, every state implemented a mandate requiring public schools to provide educational services for all disabled students. These mandates, which had similar provisions across all states, marked a major reform and expansion of the special education system in the United States.

\begin{figure}[hbt]
    \centering
    \caption{Google Ngrams data showing interest in disability and education}
    \includegraphics[width=.7\textwidth]{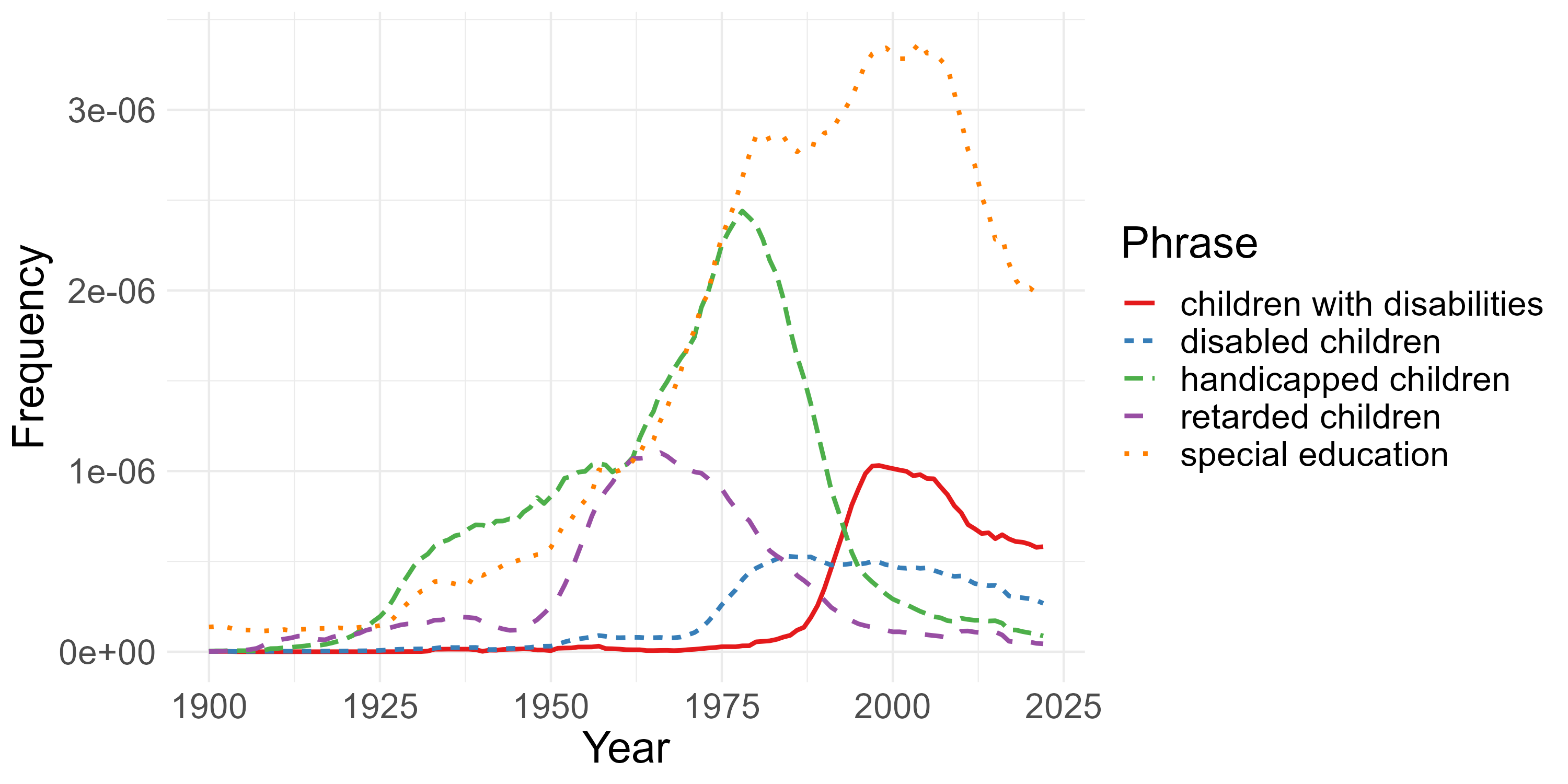}
    \label{fig:google_ngrams}
\end{figure}

As part of a landscape of civil rights and education reform in the 1960s and 70s, this period saw a major increase in national attention around the education of disabled children. Between the 1967-68 school year and the 1977-78 school year, the number of disabled children receiving services at school increased by  70\%, from 2.1 million to 3.6 million \parencite{usofficeofeducationBetterEducationHandicapped1969, usofficeofeducationProgressFreeAppropriate1979}, even as the total child population fell. Data from Google Ngrams, which measures the relative frequency of words and phrases published in books, shows the sharp increase during this time period in mentions of ``handicapped children'', ``special education'', and ``retarded children'' (Figure \ref{fig:google_ngrams}).\footnote{This figure also highlights the changing language of disability, as by the 1990s, ``disabled children'' and ``children with disabilities'' dominated ``handicapped children'' and ``retarded children''.}

\begin{figure}[hbt]
\caption{Parents' advocacy for educational services for disabled children}
\centering
\begin{subfigure}{.25\textwidth}
     \centering
    \caption{Parents' handbook}
    \includegraphics[width=\textwidth]{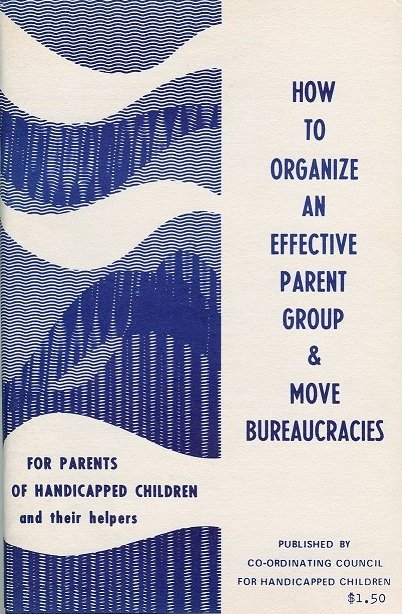}
    \label{fig:how_to_organize}   
    \caption*{\raggedright Source: Smithsonian Museum of American History }
\end{subfigure}
\begin{subfigure}{.55\textwidth}
    \centering
    \caption{Parent protest, NYC c. 1960}
    \vfill
    \includegraphics[width=\textwidth]{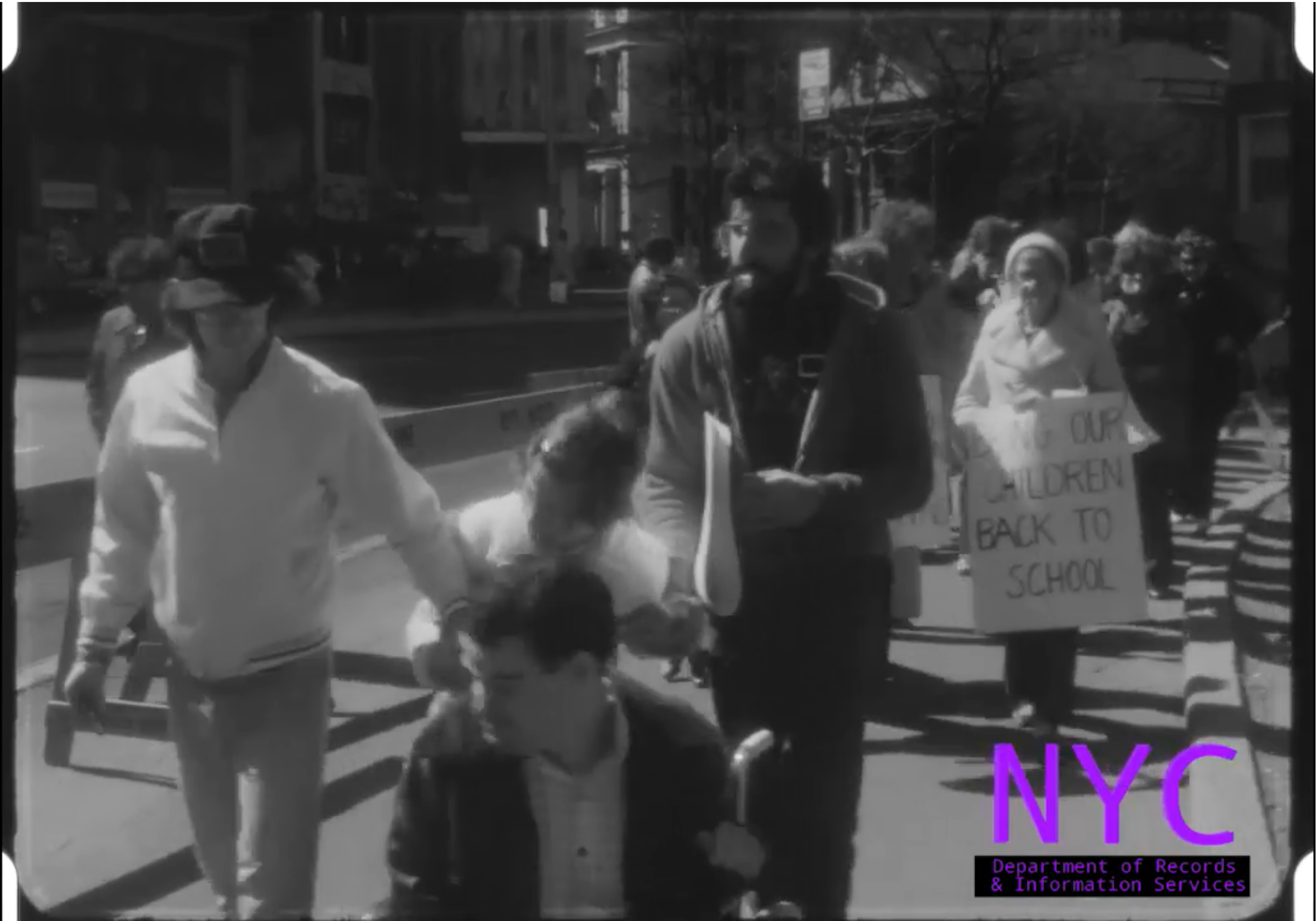}
    \vfill
    \caption*{Source: NYC Municipal Archives}
    \label{fig:nyc_archives} 
\end{subfigure}
\end{figure}

During this time, parental advocacy began to set the stage for improved access to education and disability rights in general. In the late 1960s and early 1970s, the movement for deinstitutionalization of disabled individuals built support, along with a number of high-profile exposés highlighting poor conditions in institutions (eg, \textcite{riveraWillowbrookLastGreat1972,baldiniSufferLittleChildren1968}). Illustrative of the efforts to organize parents of disabled children, a 1971 manual called \textit{How to Organize an Effective Parent Group and Move Bureaucracies: For Parents of Handicapped Children and their Helpers} offered an instruction manual for parents hoping to advocate for better education for their children (Figure \ref{fig:how_to_organize}). This volume provided a step-by-step guide to organizing an advocacy group and interfacing with policymakers and school officials and suggested, ``If you [sic] state does not have mandatory legislation, you should try to get such a law passed (see Chapter VIII, ‘How to Lobby and Get Results’).'' Figure \ref{fig:nyc_archives} shows a protest held in New York City in the 1960s with parents and their disabled children in attendance, carrying signs with messages such as ``Bring our children back to school''. One school administrator in Virginia recalled, ``Parents would shout and plead at school board meetings. It was a cry that never changed: `Why won't you teach our children?' '' \parencite{specterEducationHandicappedUndergoes1985}. Oral histories of what became known more broadly as the parents' movement are documented in \textcite{pelkaWhatWeHave2012}.

Meanwhile, several federal laws aimed to provide some funding for, but not require, educational services to be provided to disabled students \parencite{dragooIndividualsDisabilitiesEducation2019}. In 1971 and 1972, two key court cases, \textit{PARC v. Commonwealth of Pennsylvania} and \textit{Mills v. Board of Education of the District of Columbia}, reaffirmed the right to services in public education for all disabled students. Following these court cases, both states and the federal government began to seriously consider the need for mandates that would require public schools to provide services for disabled students. For example, in 1972, Senator Harrison Williams (NJ-D) called federal funding for these services up to that point only ``token expenditures'' and noted the urgency of a federal mandate after these decisions \parencite{dragooIndividualsDisabilitiesEducation2019}. Meanwhile, in 1973, Section 504 of the Rehabilitation Act forbade discrimination on the basis of disability in federally funded programs (although enforcement of this legislation was delayed until mass protests in 1977 pushed regulators to implement it) \parencite{carmelADAThereWas2020,pelkaWhatWeHave2012}.

\begin{figure}[hbt]
    \centering
    \caption{Year of first mandate to provide educational services to disabled students}    \label{fig:year_of_state_law}
    \includegraphics[width=\textwidth]{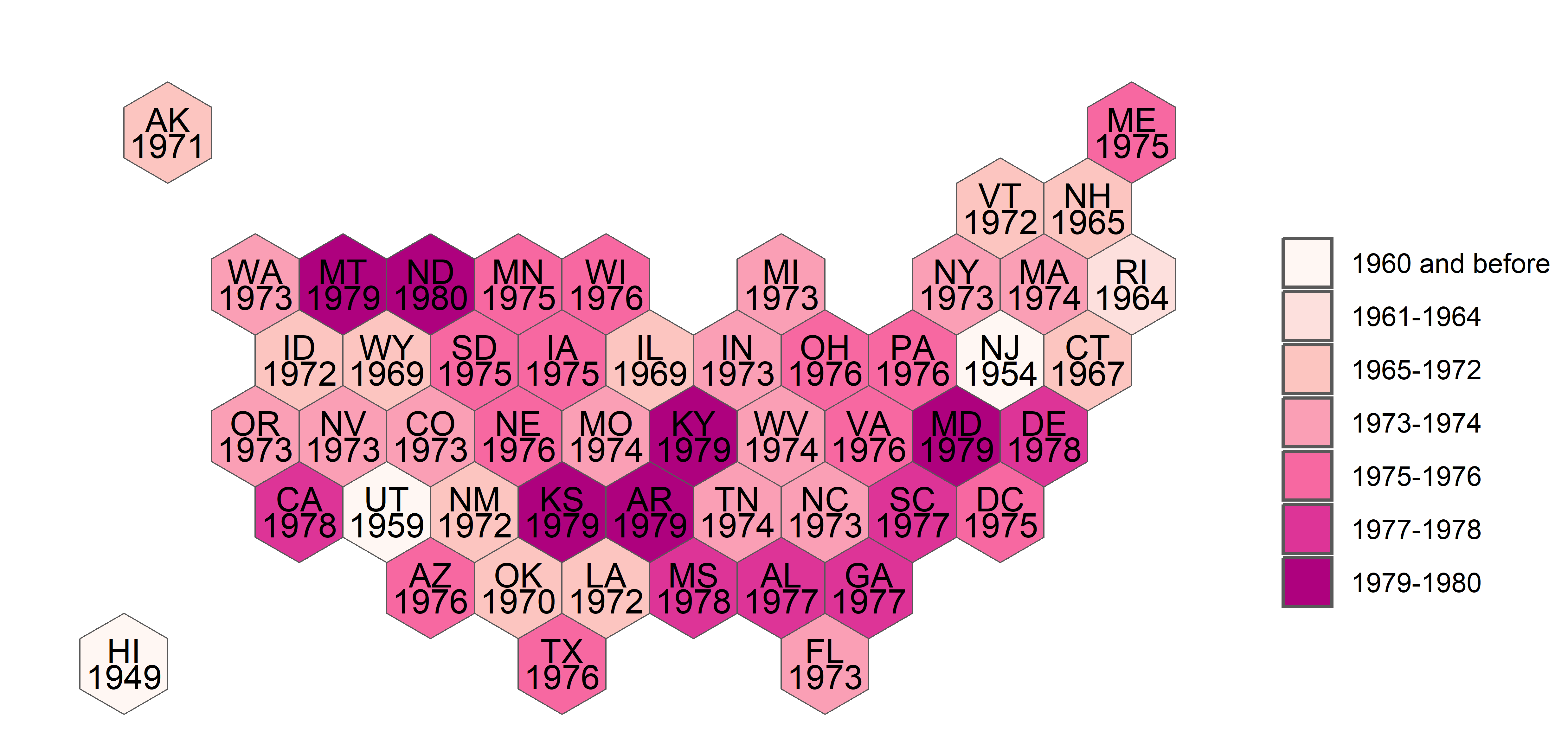}
    \caption*{\footnotesize Sources: \textcite{nationalassociationofstatedirectorsofspecialeducationStateProfilesSpecial1977, hensleyQuestionsAnswersEducation1975, uscongressCongressionalRecord1975}, and author's  review of the legislation in each state}
\end{figure}

Figure \ref{fig:year_of_state_law} shows the year in which each state implemented its first mandate, each with similar provisions. I compiled information on state mandates by cross-referencing a 1977 book of state profiles in special education policy \parencite{nationalassociationofstatedirectorsofspecialeducationStateProfilesSpecial1977}, a 1975 Q\&A report for legislators discussing special education legislation \parencite{hensleyQuestionsAnswersEducation1975}, and a state-by-state summary appearing in the US Congressional Record in 1975 \parencite{uscongressCongressionalRecord1975}, completed and verified by a manual review of the legislation in each state. 

These mandates all required school districts to identify children with disabilities and provide them services and accessible public education, although eligibility criteria and other supporting services varied between states. For example, in Arizona, by school year 1976-1977, school districts were required to ``provide special education and required supportive services for all handicapped, except emotionally handicapped, children'', hire staff to provide these services, educate students in mainstream environments whenever possible, and provide transportation for them. The program also offered vouchers for disabled children to attain other forms of education until the programs were fully established. Appendix \ref{appendix_laws} contains information on each state law.

Many of these mandates also included provisions requiring schools to provide services to disabled children as young as 3 years old. This was motivated by a number of studies showing that early education for disabled individuals could help them adapt or catch up to their peers before beginning school \parencite{kirkEvolutionPresentStatus1982}. Small-scale experiments from this period suggested that, for example, for intellectually disabled orphans in state institutions, preschool education could increase adulthood educational attainment by several years \parencite{skeelsAdultStatusChildren1966}.

These state-level mandates were soon superseded by federal law. In 1975, Congress passed the law known as P.L. 94-142, which required that a ``free appropriate public education'' be available for all children 3-18 by September 1, 1978 and through age 21 by September 1, 1980, as well as requiring schools to evaluate students' educational needs, create individualized education programs (IEPs) for them in consultation with parents, and include them in mainstream classrooms if possible. By the 1980-1981 school year, 4 million children received services under this program \parencite{gaoDisparitiesStillExist1981}.

\section{Data and empirical approach}\label{section:empirical_approach}
I identify causal effects of the mandates using a difference-in-difference approach that exploits the variation in timing of the mandates between states. Data sources are limited over this time period because detailed information on disabled children and their education was not regularly collected in a way that was comparable between states and over time before 1976. Instead, I compile several sources both on the state and individual level to analyze the impacts of the mandates in the short and long term. These sources include a little-known health survey, the Current Population survey, Census and American Community Survey data, and a dataset on state and school district finances. Each dataset is presented in more detail in its respective section below. 

\subsection{Difference-in-difference approach}
I exploit differences in the timing of the mandates to provide educational services to disabled students across states in order to identify causal impacts of the mandates (relative to no mandate) for both disabled and non-disabled students. To do this, I use a difference-in-differences setup, which requires an assumption of parallel trends in counterfactual outcomes between early- and late-adopter states.

My design follows the approach suggested by \textcite{callawayDifferenceinDifferencesMultipleTime2021} for staggered treatment designs. I estimate average treatment effects on the treated (ATT) for each treatment cohort (year of mandate enactment, $g$) and each time period (year, $t$). These estimates can then be aggregated into a simple overall average treatment effect on the treated or event-study estimates.

For a given period $t$, estimation of the $ATT(g,t)$ using the approach suggested by \textcite{callawayDifferenceinDifferencesMultipleTime2021} requires the existence of a clean control group which is not treated at any time before $t$. This can take the form of a never-treated unit, which does not receive treatment at any point, or a not-yet-treated unit, which receives treatment after $t$. In this setting, given that all states eventually implemented a mandate, I take never-treated units to be the states which did not implement a mandate before the federal mandate took effect in 1978. These states are never-treated in the sense that their state mandates were superseded by the federal mandate, which took effect across all states. 

This control group is appropriate under two conditions: (1) to the extent that the parallel trends assumption is satisfied, that is, to the extent that changes in outcomes over this period would have been the same between these control states and the treated states had the mandates not been implemented; and (2) that the mandates in these states did not have any effect and the federal mandate did not have a differential effect in these states relative to states with their own mandates, so that any effect of the federal mandate is removed by making comparisons within a year. To support assumption (1), throughout the analysis, I test for evidence of parallel trends between these states before the implementation of the mandates. I address support for assumption (2) in Section \ref{section:num_served}. 

To estimate each $ATT(g,t)$, I use the repeated cross-section estimator proposed by \textcite{callawayDifferenceinDifferencesMultipleTime2021} and implemented in Stata by \textcite{rios-avilaCSDIDStataModule2025,rios-avilaCsdid22025}. Let $Y_{it}$ represent the outcome of unit $i$ in year $t$. Let $G_i$ represent the treatment cohort to which an individual belongs, where $G_i = \infty$ indicates the never-treated group, and $\mathbb{E}_n[X_{it}] = \frac{1}{n}\sum_n X_{it}$. Then,
\begin{align*}
    \widehat{ATT}(g,t) & = \mathbb{E}_n [Y_{it} | G_i = g] -  \mathbb{E}_n[Y_{it} | G_i = \infty] \\
    & - \Big(\mathbb{E}_n[Y_{ig-1} | G_i = g] - \mathbb{E}_n[Y_{ig-1} | G_i= \infty] \Big)
\end{align*}

As suggested by \textcite{callawayDifferenceinDifferencesMultipleTime2021}, I create event-study estimates by aggregating $ATT(g,t)$ according to the shares in the population. That is, as in their paper, given $t$ and $g$, defining time relative to the treatment $e = t-g$ and the maximum time period to be $\tau$, 
\begin{align*}
    \widehat{ATT}(e) & = \sum_{g} \mathbbm{1}[g + e \leq \tau] P(G_i=g | G_i+e\leq \tau) \widehat{ATT}(g,g+e)
\end{align*}

I also create summaries by taking simple averages of these event-study estimates over the event-time periods of interest. Finally, I also report ``simple'' overall aggregations as suggested by \textcite{callawayDifferenceinDifferencesMultipleTime2021}.

All standard errors are clustered at the state level, that is, at the level of treatment assignment, as recommended by \textcite{abadieWhenShouldYou2023}. 

\subsection{Validating the empirical approach}\label{section:num_served}
In this section, I address two potential concerns with the empirical approach. First, one potential concern is that control states, whose state mandates were implemented after the federal mandate, may still have been affected by state mandates, which could bias estimated impacts downward. I show that this is not likely the case by showing the different patterns in the number of children receiving services in treated and control states around the time of a mandate's implementation. Second, I test for correlation between the timing of a state's mandate and other state characteristics, which could violate the parallel trends assumption, and find no such correlation.

To do this, I digitize state-level data from several reports and surveys to construct a series on the number of students receiving services for a disability at school from 1952 forward. The sources of this data are documented in Appendix Table \ref{tab:data_handicapped}. This dataset fills a gap in national statistics on educational services for a disability before the 1980s.   

\begin{figure}[hbt]
    \caption{Share of children receiving educational services for a disability by state over time}
    \centering
    \includegraphics[width=0.5\textwidth]{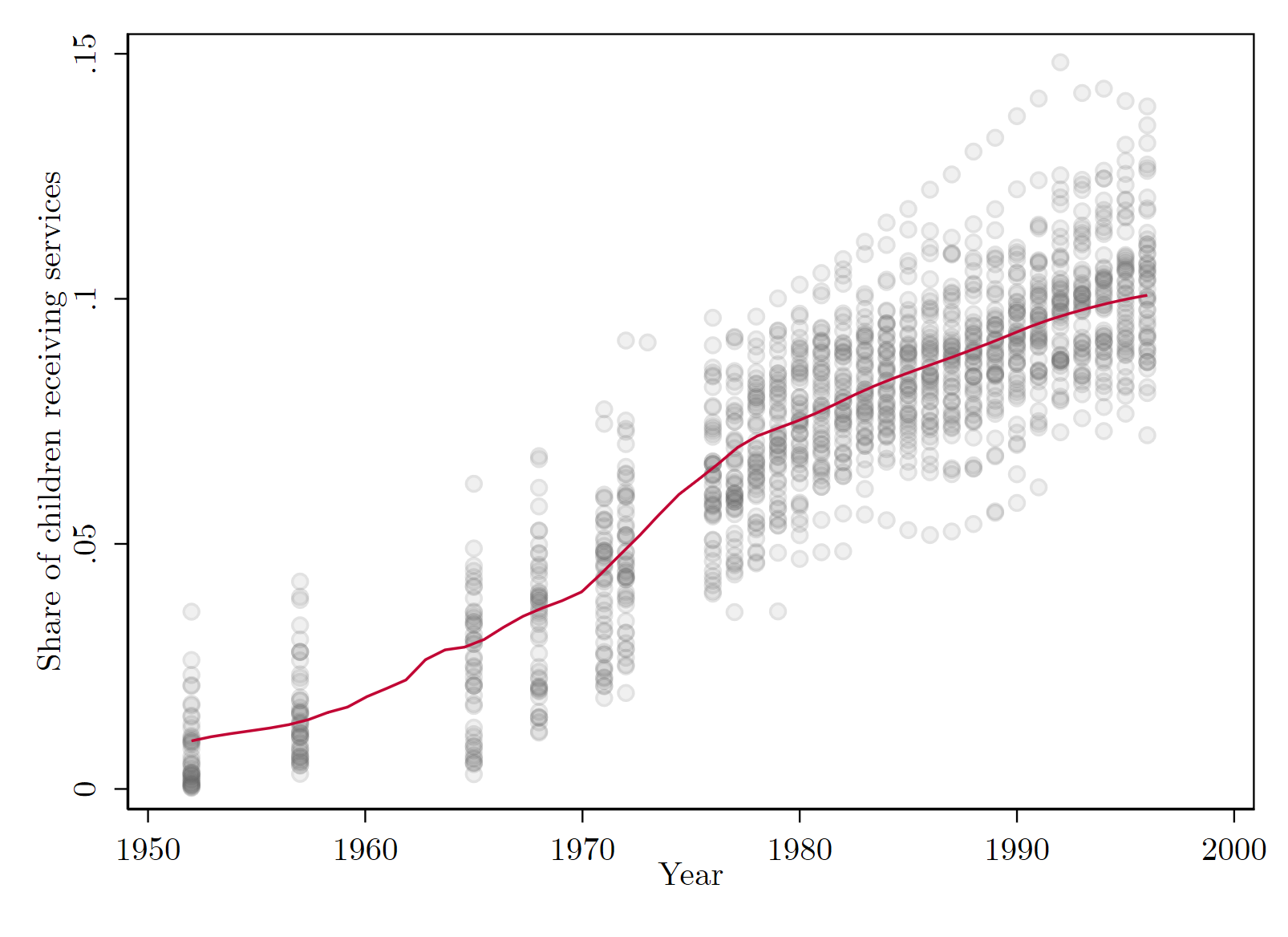}
    \label{fig:numserved_by_year}
    \caption*{\footnotesize Note: This figure shows the share of children receiving services for a disability at school in each state and year based on a compilation of state-level data on children receiving services for a disability and Census estimates of the child population in each state. The line plots a local constant regression of the share of children receiving services on the year with a bandwidth of 2.75 years.}
\end{figure}

Using the state-level dataset I have compiled, Figure \ref{fig:numserved_by_year} shows the rapid growth in services after 1952. The figure plots the share of children in each state receiving educational services for a disability in each year for which data is available. Although no state provided services to more than 4\% of the child population in 1952, multiple states provided services to over 10\% of children by 1990. The increase continues over the entire period, with the most rapid period occurring in the 1970s, coinciding with the rapid expansion of the mandates.

\begin{figure}[htb]
    \caption{Share of children receiving educational services for a disability by state, event time}
    \centering
    \includegraphics[width=0.8\textwidth]{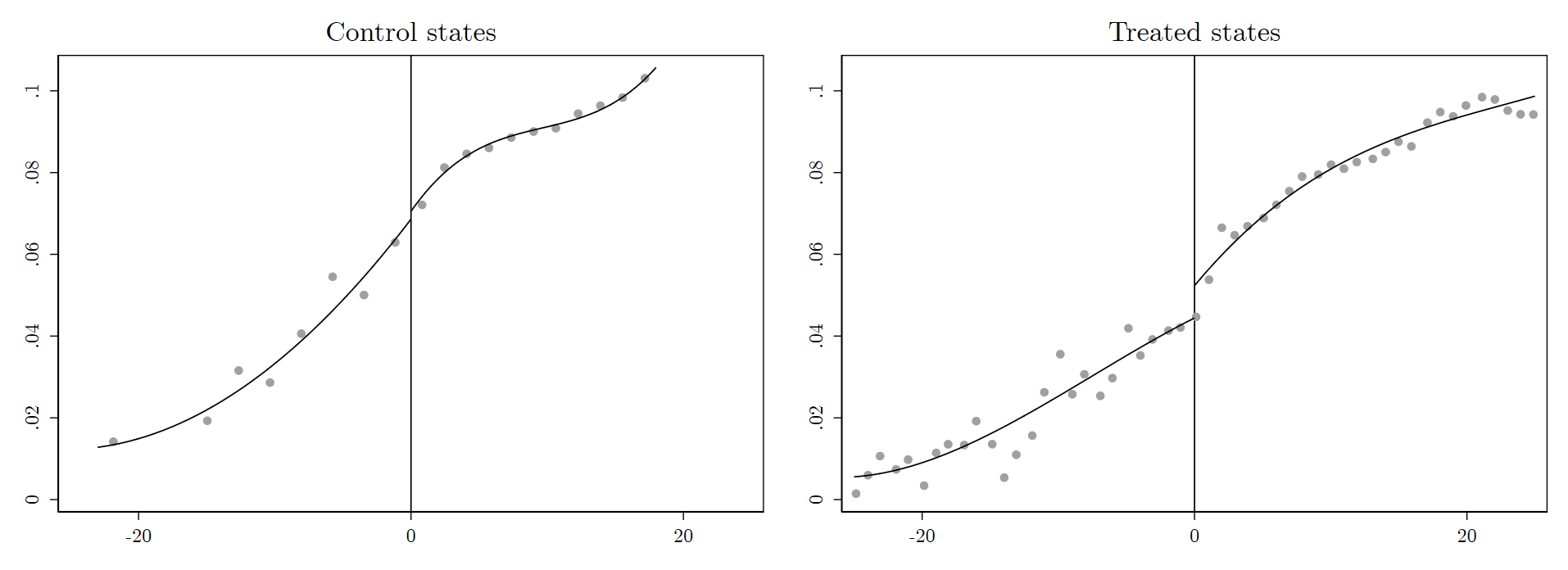}
    \label{fig:numserved_per_Cpop2_rdplot_combined_3}
    \caption*{\footnotesize Note: The figure plots a binned scatter plot of the share of children receiving services for a disability at school in each state and year, plotted according to event time relative to the implementation of the mandate in each state. Treated states are those implementing mandates before 1978, while control states are those implementing mandates in 1978 and after. A polynomial of order 3 is fitted to the pre-period and post-period data, showing any discontinuity following the mandate.}    
\end{figure}

Studying these patterns relative to the year of a mandate's implementation in each state, I find evidence of a jump in the number of children receiving services, which happened immediately after the mandates and in treated states only. 
Figure \ref{fig:numserved_per_Cpop2_rdplot_combined_3} presents a binned scatter plot showing the share of children receiving services in each state according to time since the mandate's implementation. Both treated and control states experienced an increase in the number of children receiving services over this period. However, the implementation of a mandate is only associated with a jump in the number of children receiving services in treated states, with a magnitude of about 2 percentage points within the first two years following the mandate's implementation. Meanwhile, in the control states, there is no evidence of such a jump, validating their suitability as a control group. Although this setting is not a traditional regression discontinuity design due to the long-term nature of the mandates' likely impacts, Appendix Table \ref{tab:rdrobust} uses these methods to document evidence of a statistically significant discontinuity around the first year of the mandate's implementation in the treated states, with an estimated 1.1 percentage point increase in the number of children receiving services at this discontinuity.

Another potential concern about the empirical approach is that the timing of the mandate in each state might be correlated with other characteristics or policies in the state that would affect trends in education or other outcomes. The discussion in Appendix \ref{appendix_rdrobust} and Appendix Figure \ref{fig:balance_coefplot} present results of bivariate regressions between the year of state's mandate and its pre-period per capita income, state-level education spending per capita, and Democrat vote share of the governor, as well as any court-ordered school finance reform or equalization and property tax limits. The results show that there is little evidence of a significant correlation between  these variables and the year of a state's mandate or its probability of being a control state. The results support the idiosyncratic nature of the timing of the state mandates with respect to these characteristics. 

\FloatBarrier

\section{Impacts on educational services received by disabled students}
\label{section:school}
Did the mandates actually improve the services provided to disabled students? In this section, I first give a descriptive picture on the most common types of services received by students over this period. I then show that the mandates led to large increases in the probability that disabled students received services. They also increased public school enrollments, driven by three major shifts: (1) increases in early childhood education, (2) increases for students beyond the age of compulsory schooling, and (3) shifts from private to public education. 

\subsection{A descriptive picture of services for disability}\label{section:descriptives}
Before the state or federal mandates to educate disabled children, many students did not have their needs met in educational settings. Although the scope of this deficit is difficult to quantify due to a lack of data  before 1976, I use a little-known health survey to offer one of the richest available pictures of the services available to disabled students before and after the mandates. These descriptive statistics give a sense of the increase in services over this period and its most common forms. 

Individual-level data on children's experiences is drawn from the National Health Examination Survey (NHES) II and III, which is one of the only data sources from this period containing information about children's disabilities and school experiences \parencite{dhhsNationalCenterHealth1965, dhhsNationalCenterHealth1970}. The NHES II was fielded in 1963-1965 and covered over 7000 children ages 6-11 representative of the continental US. The NHES III was fielded in 1966-1970 and covered over 7000 children ages 12-17 using the same sampling approach. This dataset includes information about children's disabilities, medical history, current health and wellbeing, and school experiences. For each child, this dataset contains information from surveys of their parents and (for older children) the child themselves, the child's birth certificate, surveys of school personnel at the child's school, and a psychological and physical examination in the field. Importantly, this includes details provided by school officials on whether a child has been recommended to receive any additional services at school and whether they are currently receiving them. 

A challenge for studying disability over this time period is a lack of a harmonized definition of disability across surveys. Purely medical or psychiatric definitions of disabilities, which rely on professional diagnosis, may exclude many disabled children, particularly those who have not been able to access services like diagnostic evaluations in school systems. Medical and psychiatric diagnoses may also change over time, for example, with revisions of the Diagnostic and Statistical Manual of Mental Disorders and other diagnostic tools.\footnote{For example, although researchers diagnosed cases of autism as early as 1943, autism did not appear in the Diagnostic and Statistical Manual of Mental Disorders (DSM) until the DSM-III was released in 1980. The DSM did not apply the term autism to children diagnosed after the age of 30 months until the DSM-III-R was released in 1987 \parencite{volkmarDSMIIIDSMIIIRDiagnoses1988}. In the DSM-IV (1994), Asperger's syndrome appeared as a separate diagnosis but was recategorized under the label autism spectrum disorder in the DSM-V (2013) \parencite{berneyAutism2019}. The International Classification of Diseases (ICD) also had substantial changes in autism diagnosis criteria over this period. As such, any attempt to identify autistic children over this period would be affected by the diagnostic categories available at the time. } Because of this, and in line with modern international standards for measuring disability (eg, \cite{whoMeasuringHealthDisability2010}), I rely on functional definitions of disability, that is, definitions that reflect the things an individual is able or unable to do. In analyses of children using the NHES, I will take disability to be defined according to parents' reports of whether the child has difficulty walking, talking, hearing, or moving limbs, or is limited in playing or exercising.\footnote{Although a more continuous and nuanced definition of disability would likely provide important insights, as would being able to analyze heterogeneity by type of disability, this analysis is limited to a binary definition given the information available in this dataset.}

Descriptively, data from the 1965 and 1966 waves of the NHES show that students' needs were likely in line with estimates of needs for special education today, even if they were not being met. These descriptive statistics also highlight the broad variation in and most common types of services offered.

\begin{figure}[htb]
\caption{Descriptive statistics on recommendations for services for a disability, 1965-66}    \label{fig:desc_resources_needed_type}   

\centering
\begin{subfigure}{.45\textwidth}
     \centering
    \includegraphics[width=\textwidth]{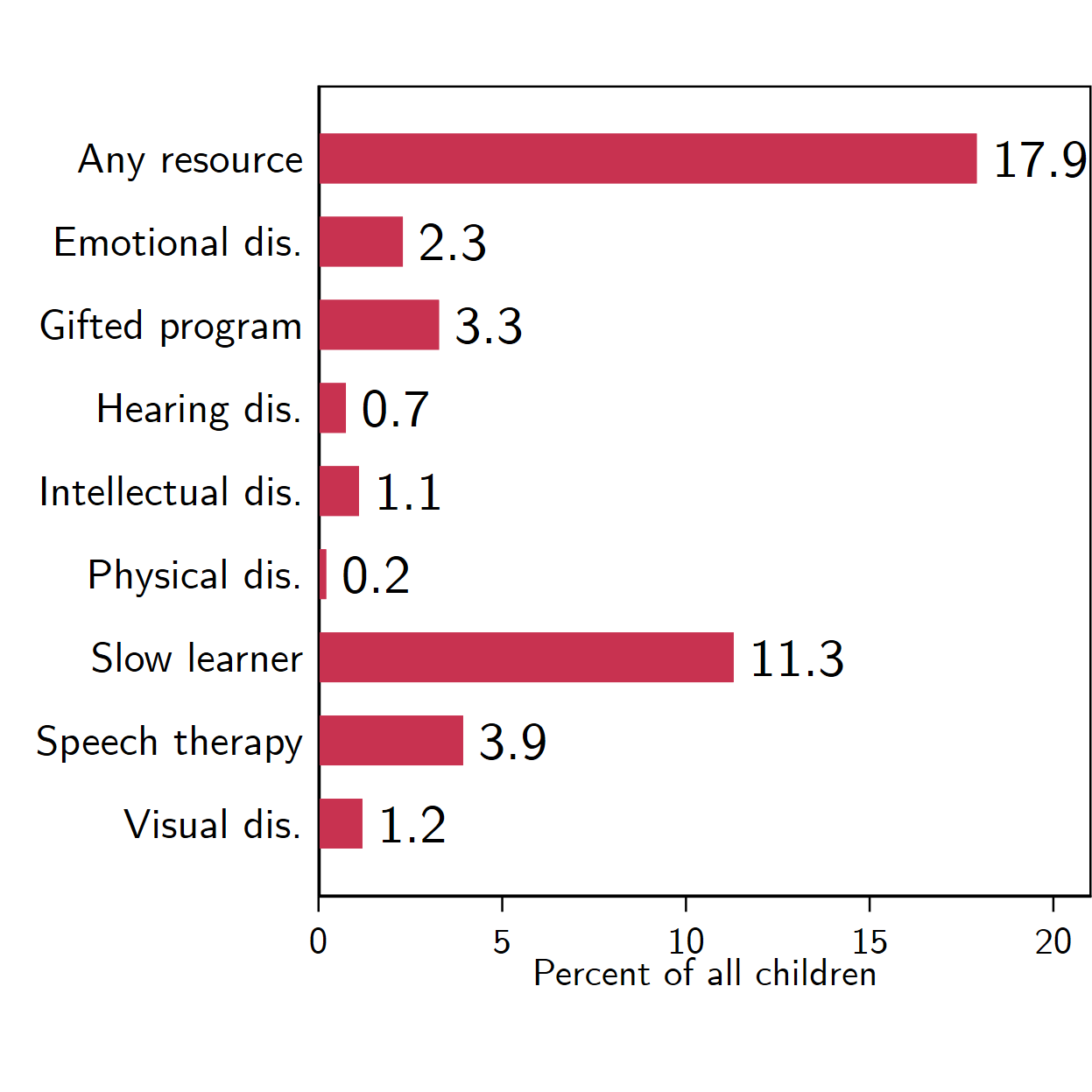}
\end{subfigure}
    \caption*{\footnotesize Note: This figures shows the share of children in the NHES sample in 1965-66 recommended by their schools to receive certain support services at school.}    
\end{figure}

Figure \ref{fig:desc_resources_needed_type} shows the share of students recommended to receive each type of service for a disability. Overall, 17.9\% of this sample of children was recommended to receive some additional support services at school. This is in line with modern estimates which suggest that, as of 2018, 17.3\% of children have at least one disability or developmental delay \parencite{cogswellHealthNeedsUse2022} and that, as of 2023, about 15\% of students receive services at school \parencite{ncesStudentsDisabilities2024}. The most common services needed in 1965-66 were classes or resources for ``slow learners'' (distinguished from resources for those with intellectual disabilities, known in the original survey as ``mentally retarded''). In the language of this period, the group of slow learners may have included those who had a learning disorder or another kind of disability as well as those who had difficulty learning for another reason, such as unstable housing \parencite{madisonNCTEERICREPORT1971}. In this sample, 11.3\% of students are recommended to receive services for slow learners. After excluding services for slow learners, 8.6\% of children are recommended to receive other kinds of services. 

Rates of recommendations for more specific services, such as speech therapy, are lower, around 3.9\% of the sample in 1965-66. Rates of recommendation for the most specific services, such as those for intellectually disabled students, hard of hearing/Deaf/deaf, or blind or visually impaired students are much lower, each around 1\% of the sample.\footnote{These are in line with what would be expected based on modern medical data: in 2018, about 0.6\% of US children are estimated to have significant hearing loss or d/Deafness and 1.1\% to have an intellectual disability \parencite{cogswellHealthNeedsUse2022}.} To put these numbers in context, the figure also shows that the number of students recommended for gifted programs in 1965-66 is 3.3\%.  

\begin{figure}[htb]
    \centering
    \caption{Resources unavailable, by whether state has a mandate to provide services for disabled students}
    \includegraphics[width=.5\textwidth]{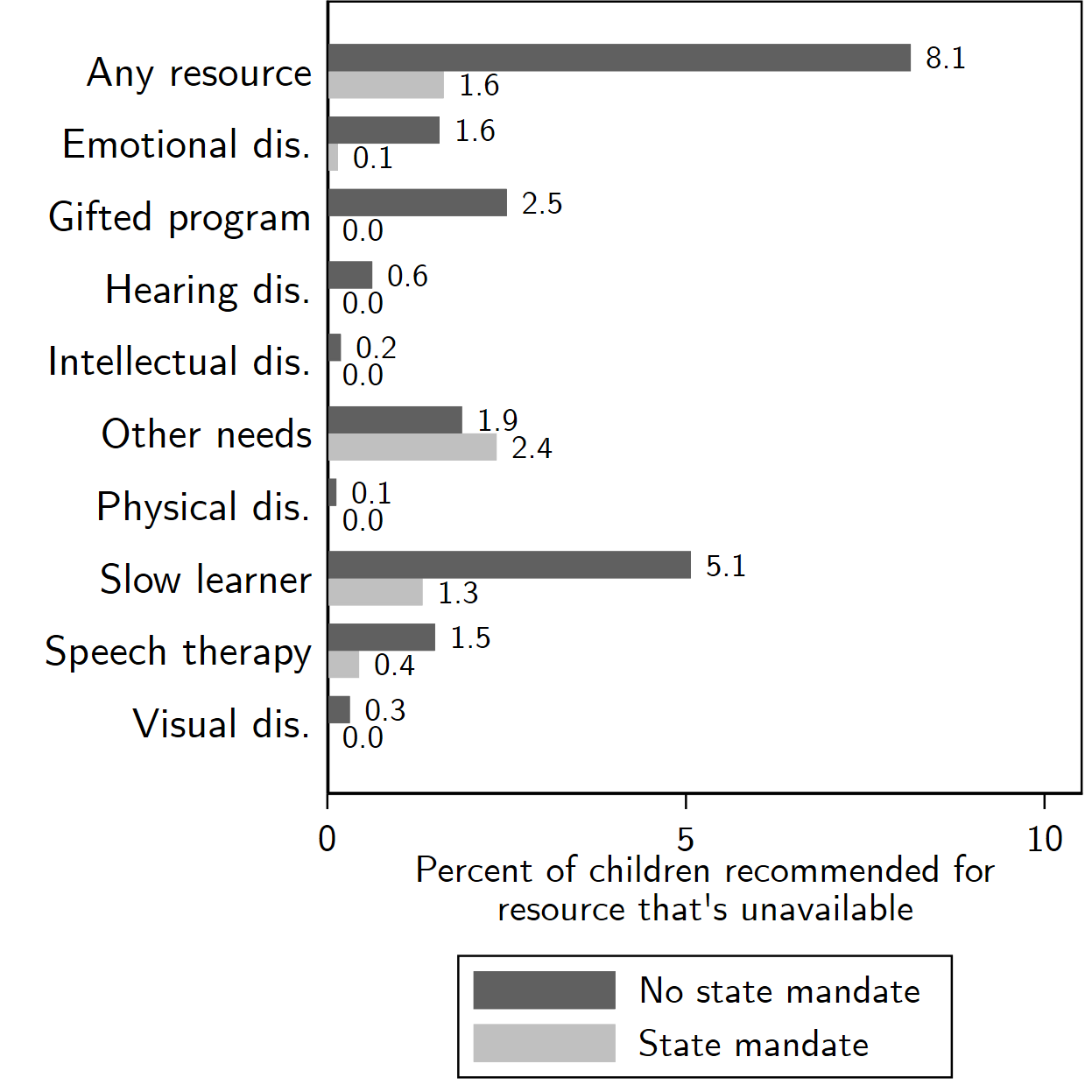}
    \label{fig:desc_resources_na_by_mandate}
    \caption*{\footnotesize Note: This figure shows the share of children in the NHES being recommended for a resource that is unavailable at their school, before and after the implementation of a mandate in their state. The sample is limited to states that introduced a mandate between 1963-1970.}
\end{figure}

Descriptive statistics also show that students were much more likely to find the services they needed to be available following the passage of a state mandate. Contingent on a student having been identified as needing services, the NHES dataset indicates whether the student is receiving that service and, if not, why not, including whether the resource was not available. Looking at states that adopted a mandate to provide these services in the time period covered by the NHES dataset (1963-1970) and comparing before and after these mandates were adopted, Figure \ref{fig:desc_resources_na_by_mandate} shows that states with active mandates provided more services to their students. Before these mandates came into effect, about 8.1\% of the sample was recommended to receive a service that was not available to them. Afterwards, this fell to only 1.6\%. Remarkably, after the state mandates took effect, unavailability of many types of services -- such as those for emotional, hearing, intellectual, physical, and visual disabilities -- fell to nearly zero. For comparison, the number of students recommended for gifted programs who found them not available also fell to 0\%, while the those finding resources for other needs (like remedial reading classes) unavailable increased slightly, from 1.9\% to 2.4\%.

However, causal effects are not identified by this descriptive analysis, as there are no controls for other factors that could affect both mandates and the provision of services. For example, a general trend towards more services becoming available over time, with or without a mandate, would confound this analysis. In the next section, I turn to results from the difference-in-difference design to explore causal impacts. 
\FloatBarrier

\subsection{Impacts on services received in school}\label{section:nhes_education}

Using the difference-in-difference design described in Section \ref{section:empirical_approach}, I show that, as a result of these mandates, disabled students became much more likely to be recommended to receive services at school and to actually receive them, an increase on the order of 20 percentage points. Disabled students also became more likely to be transferred to special education classes (that is, to be provided services in the form of a different teacher or classroom environment) and less likely to be frequently absent from school. 

\begin{figure}[h]
\caption{Effects on receiving services for a disability}
\label{nhes_resource}
    \begin{subfigure}{.45\textwidth}
        \caption{Being recommended to receive services}
        \label{fig:cs_event_resource_needed}
        \centering
        \includegraphics[width=\textwidth]{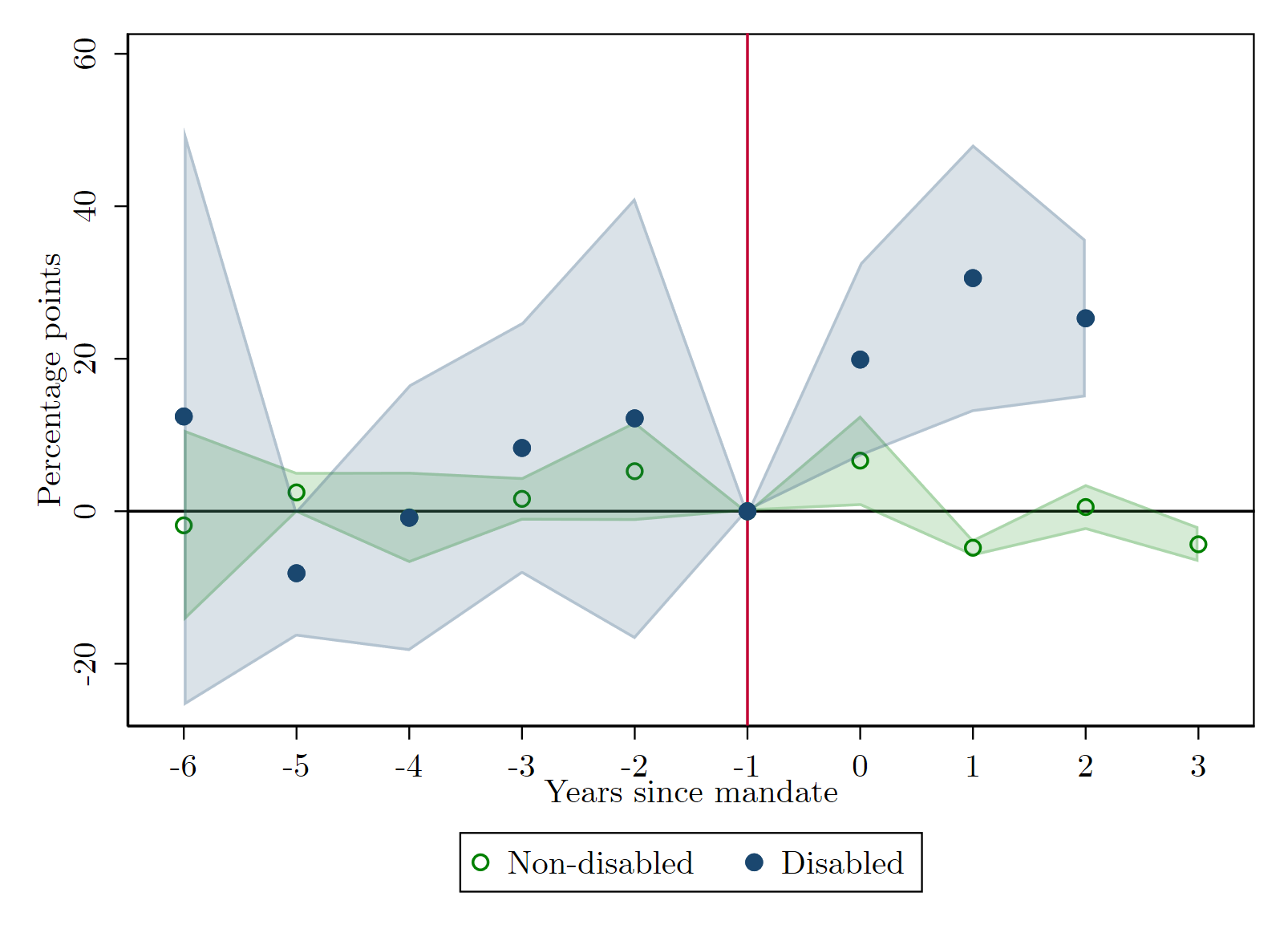}
    \end{subfigure}
    \begin{subfigure}{.45\textwidth}
        \caption{Using services}
        \label{fig:cs_event_resource_used}
        \centering
        \includegraphics[width=\textwidth]{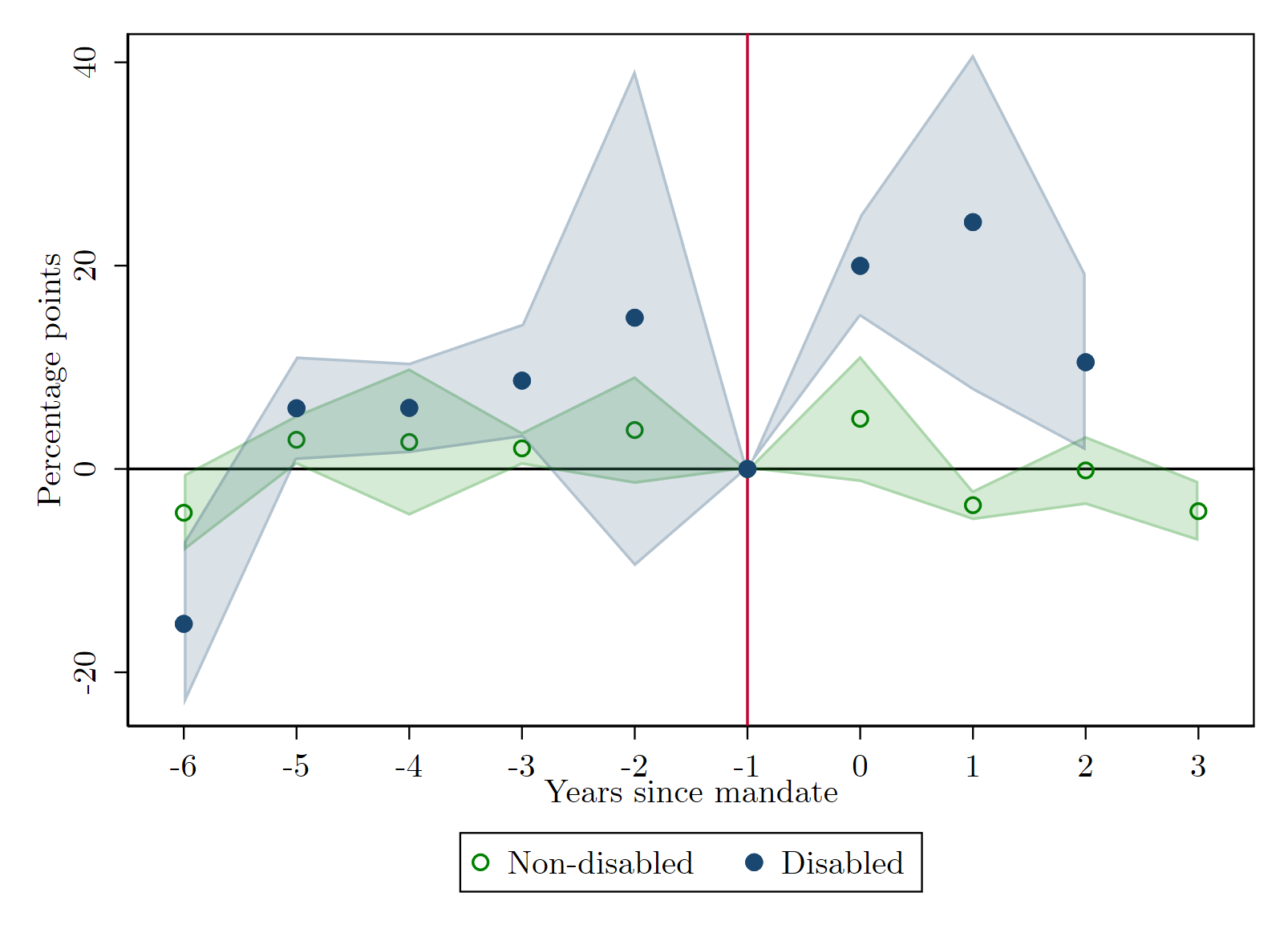}
    \end{subfigure}
    \begin{subfigure}{.45\textwidth}
        \caption{Transferring to special education}
        \label{fig:cs_event_transfered_special}  
        \centering
        \includegraphics[width=\textwidth]{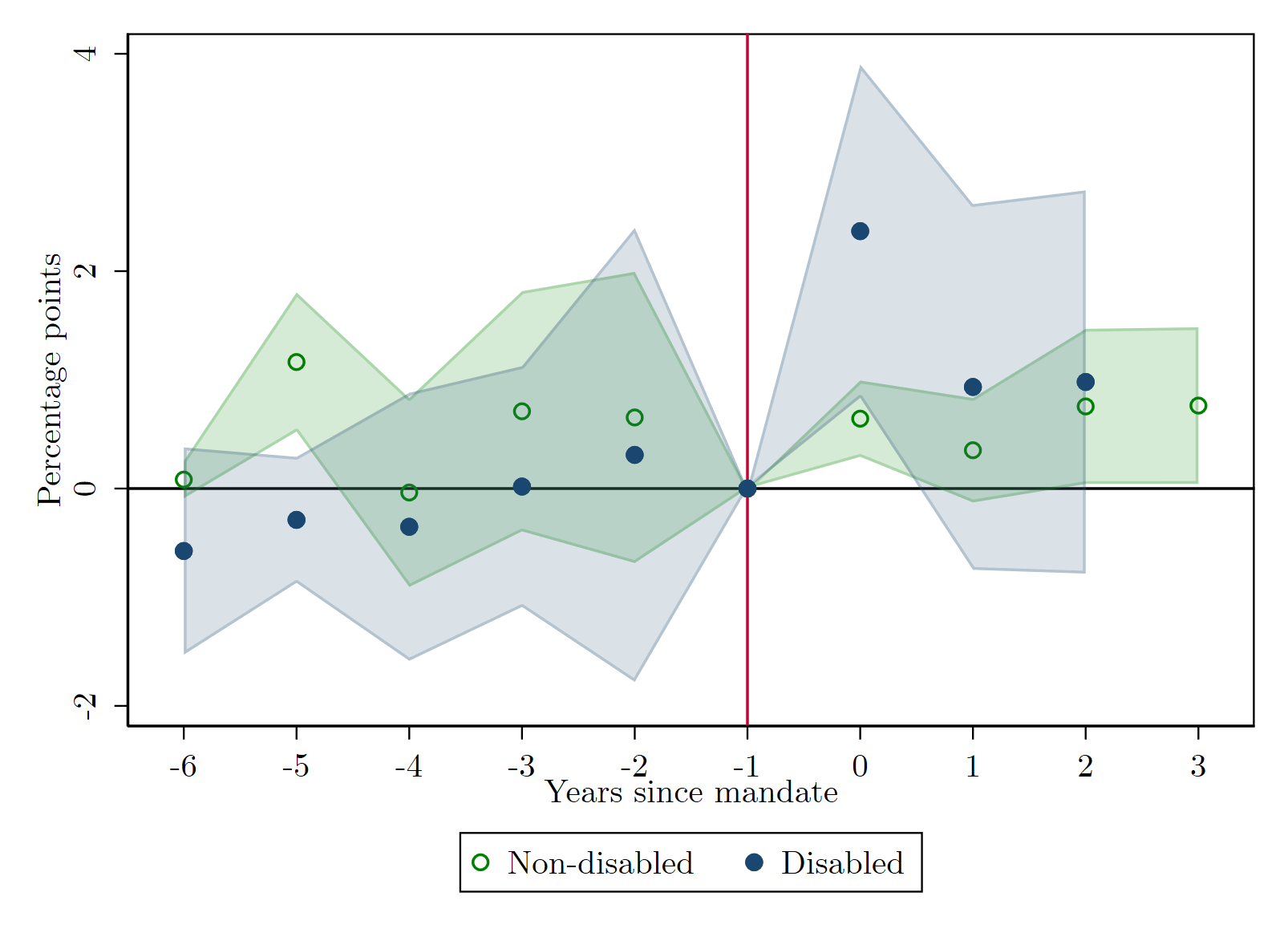}
    \end{subfigure}
        \caption*{\footnotesize Note: 95\% confidence intervals shown, standard errors clustered at the state level. The figures plot difference-in-difference event-study estimates of the impact of the mandates on educational outcomes for disabled and non-disabled students in the NHES. Panel A shows effects on the probability of a student being recommended to receive services for a disability by the school/teacher. Panel B shows effects on the probability of a student actually using these services. Panel C shows effects on being transferred into ``special education''. }
\end{figure}

Since the purpose of these laws was to ensure that children were receiving services, I first show evidence that the laws did increase both children's probability of being recommended for and actually receiving services at school. The magnitude of these impacts is quantified in Table \ref{tab:nhes_rsce}. The table presents aggregations of the $ATT(g,t)$ parameters estimated in the difference-in-difference setup. Pre-period average refers to an average of estimated effects for years prior to the implementation of the mandates and serves as a test for parallel trends in the pre-period. Post-period average refers to a simple average of effects following the mandates' implementation. Callaway and Sant'Anna average provides an alternative average of the post-period effects, using weights proposed in \textcite{callawayDifferenceinDifferencesMultipleTime2021}. The sample is split between disabled and non-disabled children. The corresponding event study coefficients are plotted in Figure \ref{nhes_resource}.

\begin{table}[hbt]
    \centering
        \caption{Results on recommendations for and use of resources}
    \label{tab:nhes_rsce}
    \resizebox{\textwidth}{!}{%
    \begin{tabular}{lcccccccc}
    \hline \hline
                  &\multicolumn{1}{c}{(1)}   &\multicolumn{1}{c}{(2)}   &\multicolumn{1}{c}{(3)}   &\multicolumn{1}{c}{(4)}   &\multicolumn{1}{c}{(5)}   \\
            &Resource needed   &Resource used   &Transferred to special ed.   &      Absent   &Repeated grade   \\
 \hline  \textbf{Disabled} & & & & & \\
Pre-period average&       0.048   &       0.041   &      -0.002   &       0.116   &       0.057   \\
            &     (0.081)   &     (0.032)   &     (0.003)   &     (0.077)   &     (0.048)   \\
Post-period average&       0.253***&       0.183***&       0.014*  &      -0.154***&       0.150** \\
            &     (0.055)   &     (0.043)   &     (0.007)   &     (0.038)   &     (0.065)   \\
Callaway \& Sant'Anna average&       0.223***&       0.185***&       0.019** &      -0.074** &       0.108*  \\
            &     (0.062)   &     (0.032)   &     (0.008)   &     (0.037)   &     (0.066)   \\
\hline
Observations&        1636   &        1636   &        1505   &        1555   &        1505   \\
Pre-mandate mean&        0.32   &        0.16   &        0.00   &        0.15   &        0.23   \\
\hline\hline
\textbf{Non-disabled} & & & & & \\
Pre-period average&      -0.008   &       0.008   &       0.005*  &       0.032*  &      -0.058** \\
            &     (0.016)   &     (0.010)   &     (0.003)   &     (0.017)   &     (0.027)   \\
Post-period average&      -0.005   &      -0.007   &       0.006** &      -0.031*  &      -0.085***\\
            &     (0.011)   &     (0.012)   &     (0.003)   &     (0.016)   &     (0.027)   \\
Callaway \& Sant'Anna average&       0.024   &       0.016   &       0.006***&      -0.014   &       0.002   \\
            &     (0.023)   &     (0.023)   &     (0.002)   &     (0.019)   &     (0.065)   \\
\hline
Observations&       10270   &       10270   &        9574   &        9815   &        9574   \\
Pre-mandate mean&        0.14   &        0.06   &        0.00   &        0.09   &        0.15   \\

    \\ \hline
    \end{tabular}
    }
    \caption*{\footnotesize Note: The table shows difference-in-difference estimates of the impacts of the mandates on students' educational experiences. The top panel shows results for disabled students and the bottom panel shows results for non-disabled students. Column (1) shows effects on the probability of being identified as needing services for a disability, column (2) of using that service or resource, column (3) of being transferred to ``special education'', column (4) of being frequently absent from school, and column (5) of repeating a grade. Pre-period average and post-period average refer to a simple average of event-study coefficients before and after the implementation of a mandate, respectively. Callaway \& Sant'Anna average refers to a weighted average of estimated impacts, with weights given by the share belonging to each treated cohort in the sample. Standard errors clustered at the state level shown in parentheses. \\
    * p \textless 0.1, ** p \textless 0.05, *** p \textless 0.01 }
\end{table}

Very quickly after the mandates became effective, disabled students became much more likely to be identified as needing services. Column (1) of Table \ref{tab:nhes_rsce} studies whether a student was recommended to receive support services by their school or teacher. Following the mandate's implementation, a simple average of the post-period event study estimates suggests that this increase amounted to 25.3 percentage points. The Callaway and Sant'Anna average provides a similar estimate of 22.3 percentage points. This increase is quite large relative to a pre-mandate share of 32\% of disabled students being identified as needing services. Figure \ref{fig:cs_event_resource_needed} plots the event study estimates for this outcome and highlights that this increase was immediate and persistent in the years after the mandate. Further, the results in column (1) of Table \ref{tab:nhes_rsce} and Figure \ref{fig:cs_event_resource_needed} both show no evidence of a significant pre-trend that would raise concerns about endogeneity of the mandates. There are also no significant impacts on non-disabled students. 

Along with being more likely identified as needing services, disabled students also became much more likely to use these services. Column (2) of Table \ref{tab:nhes_rsce} studies impacts on the probability of actually receiving recommended services. Both the simple post-period average and the Callaway and Sant'Anna average effects indicate that the mandates caused an increase in services received of about 18 percentage points. Again, this effect is remarkably large relative to a pre-mandate mean of 16\% of disabled students receiving these services. There are also no impacts on non-disabled students' usage of services. The dynamics of these impacts, as shown in Figure \ref{fig:cs_event_resource_used}, are again similar to the above. 

Appendix Table \ref{tab:nhes_rsce_by_type} shows the increase in use of resources for the four most popular types of resources used: those for students with intellectual disabilities, those for ``slow learners'', speech therapy, and those for students with emotional disabilities. The results highlight that the largest increases were driven by those for intellectual disability and ``slow learners''. 

Although the NHES data provide detail on only a sample of children, the magnitudes of these estimates are in line with the estimated discontinuity on the state level using the state-level series on children receiving support for a disability discussed in Section \ref{section:num_served}. That analysis found an increase of 1-2 percentage points in the number of children receiving services in the first few years following the mandate's enactment. An increase of about 18 percentage points concentrated among the 14\% of children with disabilities in this sample would give an estimate of $18 \times .14 = 2.5$ percentage points, a similar magnitude. 

Along with receiving additional services part-time, like speech therapy, many children in this period participated in full-time ``special education'' classrooms, and I find that the mandates also led to an increase in students reporting transferring into special education. Column (3) of Table \ref{tab:nhes_rsce} shows the mandates' impact on the probability of disabled students reporting that they were transferred into ``special education'' (likely meaning a separate classroom with specialized instruction). The probability of being transferred to special education increased by 1.4-1.9 percentage points for disabled students, as well as by 0.6 percentage points for non-disabled students.\footnote{However, it should be noted that this outcome is extremely rare and identified on few cases in this data, with fewer than 1\% of children reporting this before the mandates.} Figure \ref{fig:cs_event_transfered_special} plots the corresponding event study coefficients and shows this increase in the years following the mandate's implementation.

Consistent with improved school services, disabled children also became less likely to be frequently absent from school. Before the mandates, 15\% of disabled children in this sample were frequently absent from school, compared with only 9\% of non-disabled children. Column (4) of Table \ref{tab:nhes_rsce} shows that the mandates decreased the probability of having frequent absences by 7.4-15.4 percentage points, with no effect for non-disabled students.

Considering the quality of students' education, in theory, effects on the probability of repeating a grade might go either way. Students might be less likely to repeat grades if they are more successful in learning. But, they might be more likely to repeat grades given increased standards and attention to the educational performance of disabled students, rather than nominal promotions from year to year, as noted by some contemporary writers \parencite{madisonNCTEERICREPORT1971}. Column (5) of Table \ref{tab:nhes_rsce} shows that disabled students became more likely to repeat a grade following the mandates. This increase amounts to 10.8-15.0 percentage points, with unclear evidence of an effect for non-disabled students due to a significant pre-trend estimate.

\FloatBarrier

\subsection{Impacts on school enrollments }

Given that the mandates expanded the scope of the public schools and the services they were required to offer, we should expect to find impacts on enrollments in the public school system. For example, the mandates may have meant that families no longer had to rely on the private education system to meet the needs of their disabled children who were denied enrollment or services in the public schools \parencite{mandelPrivateSchoolsFill1962}. In this section, using data from the CPS, I find that the mandates caused large increases in public school enrollments, driven by three important channels: (1) increases in preschool, the availability of which was expanded for disabled children by the mandates; (2) increases in high school, indicating longer stays in school beyond the age of compulsory schooling; and (3) a shift from private to public schools. 

For this analysis, I use data from the Current Population Survey (CPS) October Education Supplement from 1968-1990 \parencite{hauserCurrentPopulationSurveys2006}. This dataset offers some of the best information on school enrollment over this period and was used by the Census Bureau to produce official enrollment statistics. For my sample of individuals age 0-20, it includes individual-level information on school enrollment and type of school. However, it does not contain a disability measure, so effects can only be studied overall. 

\begin{figure}[htb]
    \centering
    \caption{Probability of being enrolled in school, age 0-20}
    \label{fig:cps_enroll}
    \includegraphics[width=.6\textwidth]{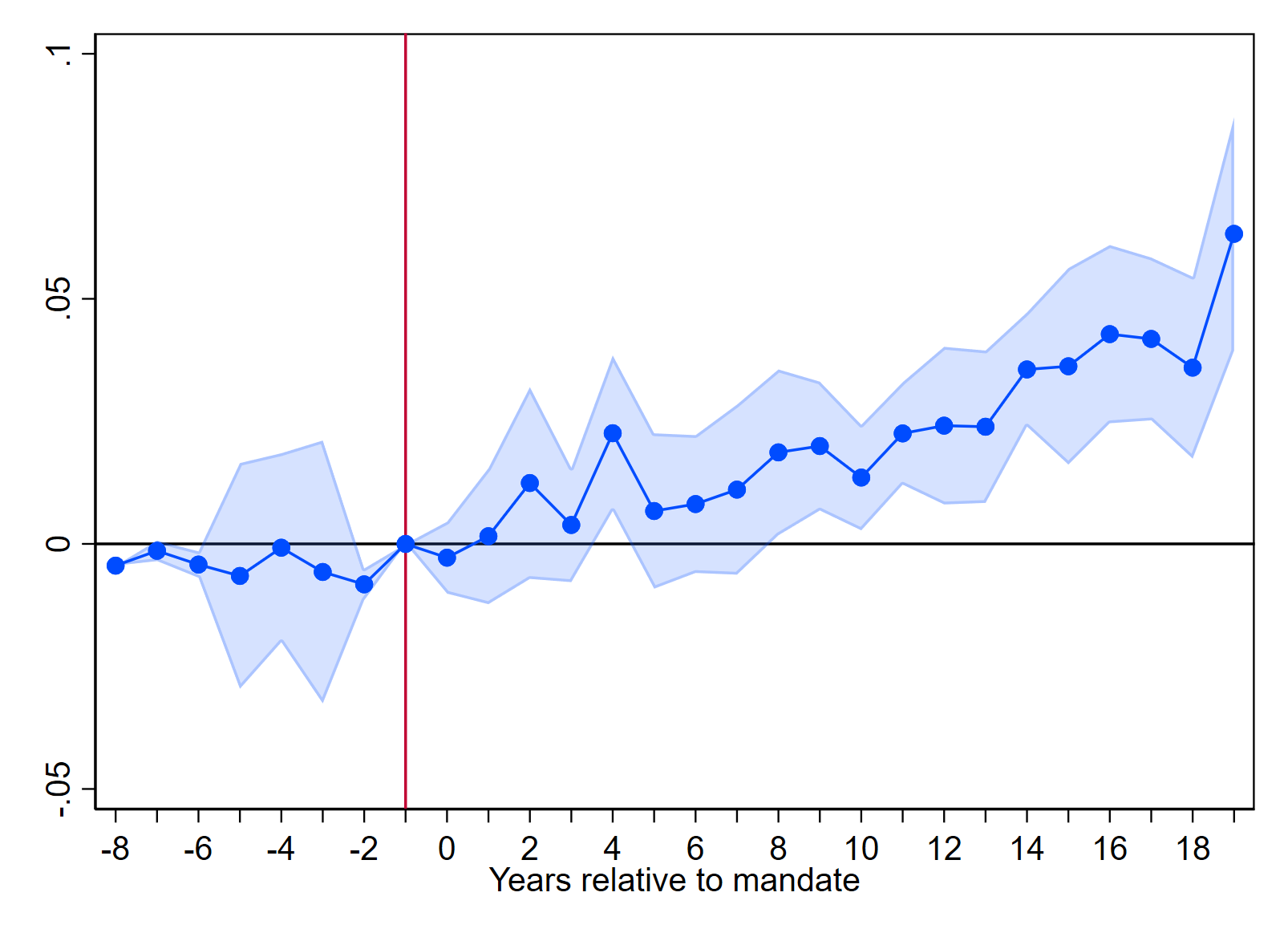}
        \caption*{\footnotesize Note: 95\% confidence intervals shown, standard errors clustered at the state level. The figures plot difference-in-difference event-study estimates of the impact of the mandates on the probability of being enrolled in school among individuals under age 20 in the CPS. }
\end{figure}

Consistent with the intention of the mandates in increasing access to education, I first show that the mandates led to overall large increases in school enrollments. Treatment effects estimated using the same difference-in-difference design appear in Table \ref{tab:cps_enroll}. Again, pre-period average and post-period average represent simple averages of the event-study coefficients in the respective periods, and Callaway and Sant'Anna average offers another aggregation of post-period effects. To highlight the dynamics of the effects, the tables also include average effects for years 0-9 following the mandates and years 10+ following the mandates. Column (1) of Table \ref{tab:cps_enroll} shows the overall impacts on enrollments among young people under age 20. First, there is little evidence of concerning pre-trends, with a small and insignificant estimated impact in the pre-period. In contrast, average effects over the post-period are large, positive, and significant, indicating that the mandates led to a 2.2-2.6 percentage point increase in enrollments. The effects took time to ramp up, with impacts on enrollment insignificant over years 0-9 and significant and larger for years 10 and beyond. After year 10, the mandates are estimated to have led to a 3.9 percentage point increase in enrollments. This increasing impact over time can also be seen in Figure \ref{fig:cps_enroll}, which plots the event-study coefficients for this outcome.

\begin{table}[htb]
    \centering
    \caption{Enrollment effects of mandates}
    \label{tab:cps_enroll}
        \resizebox{\textwidth}{!}{%
    \begin{tabular}{lcccccc} 
        \hline \hline 
                    &\multicolumn{1}{c}{(1)}   &\multicolumn{1}{c}{(2)}   &\multicolumn{1}{c}{(3)}   &\multicolumn{1}{c}{(4)}   &\multicolumn{1}{c}{(5)}   &\multicolumn{1}{c}{(6)}   \\
            &    Enrolled   & Under age 6   &    Age 6-15   &   Age 16-20   &Public school   &Private school   \\
\hline
Pre-period average&      -0.004   &       0.001   &      -0.002** &      -0.013   &      -0.008   &       0.004   \\
            &     (0.005)   &     (0.008)   &     (0.001)   &     (0.009)   &     (0.007)   &     (0.004)   \\
Post-period average&       0.026***&       0.103***&       0.002   &       0.019***&       0.049***&      -0.023*  \\
            &     (0.006)   &     (0.032)   &     (0.002)   &     (0.007)   &     (0.011)   &     (0.012)   \\
Callaway \& Sant'Anna average&       0.022***&       0.094***&       0.002   &       0.014   &       0.043***&      -0.021*  \\
            &     (0.006)   &     (0.029)   &     (0.001)   &     (0.009)   &     (0.012)   &     (0.012)   \\
Years 0-9 average&       0.010   &       0.059** &       0.000   &       0.001   &       0.027***&      -0.017** \\
            &     (0.006)   &     (0.024)   &     (0.001)   &     (0.010)   &     (0.009)   &     (0.009)   \\
Years 10+ average&       0.039***&       0.140***&       0.003   &       0.034***&       0.068***&      -0.028*  \\
            &     (0.006)   &     (0.040)   &     (0.003)   &     (0.006)   &     (0.012)   &     (0.016)   \\
\hline
Observations&      298984   &       47417   &      168322   &       83245   &      298984   &      298984   \\
Pre-mandate mean&        0.82   &        0.49   &        0.99   &        0.65   &        0.71   &        0.09   \\
 \\
        \hline
    \end{tabular}
        }
        \caption*{\footnotesize Note: The table shows difference-in-difference estimates of the impacts of the mandates on school enrollments using the CPS data. Column (1) shows effects on the probability of being enrolled among individuals below age 20, column (2) among those under age 6, column (3) among those age 6-15 (compulsory schooling years), and column (4) among those age 16-20. Column (5) shows impacts on being enrolled in a public school and column (6) of being enrolled in a private school. Pre-period average and post-period average refer to a simple average of event-study coefficients before and after the implementation of a mandate, respectively. Callaway \& Sant'Anna average refers to a weighted average of estimated impacts, with weights given by the share belonging to each treated cohort in the sample. Years 0-9 average and Years 10+ average refer to simple averages of event-study coefficients for those years. Standard errors clustered at the state level shown in parentheses. \\
    * p \textless 0.1, ** p \textless 0.05, *** p \textless 0.01
        }
\end{table}

This increase in school enrollments included a large increase in preschool enrollments. Column (2) of Table \ref{tab:cps_enroll} shows impacts on school enrollment for children under age 6. Among this group, the mandates led to an increase in enrollments of 9.4-10.3 percentage points, with long-term impacts of up to 14.0 percentage points. This estimate is quite large, representing an increase of nearly 20\% relative to a pre-mandate mean enrollment of 49\% among these children. The large magnitude of these estimates would be consistent with near universal preschool attendance of disabled children,  possibly with spillovers in preschool availability for non-disabled children. State reports from the time confirm the plausibility of this result; for example, Massachusetts reported in 1975, one year after its mandate, that 170 ``public and private early education programs were established to serve special needs students. Planning efforts during this first year indicate that next year the number will double'' \parencite{massdeptofeducationAnnualReport19741975}.

On the other hand, for children ages 6-15, who would be of compulsory schooling age, there is little impact of the mandates, as school enrollment was already near-universal. Since nearly all children of this age group were at least nominally enrolled in school before the mandates, column (3) of Table \ref{tab:cps_enroll} shows little impact on this group.

Above compulsory schooling age, that is, for those age 16-20, the mandates again increased enrollments. Column (4) of Table \ref{tab:cps_enroll} shows that, beyond 10 years after the mandates, these individuals had a 3.4 percentage point increase in the probability of being enrolled in school. Appendix Figure \ref{fig:enrolled615_1620} shows the event-study plot for this outcome.

As the mandates improved the services offered by public education, families may have chosen to switch from private to public schools for their children, and I find evidence that this is the case.\footnote{Another question is whether families switched from home schooling to public education. Neither the NHES nor the CPS collected data on homeschooling, and homeschooling in the 1970s and 1980s was quite rare. Given the lack of official data collection on the number of homeschooled children, \textcite{linesEstimatingHomeSchooled1991} uses data from sellers of homeschool curricular materials to construct estimates from the late 1970s and early 1980s which suggest that only about 10,000-15,000 students were homeschooled over this period. This is very small compared to a school-age population in 1976 of over 51 million \parencite{usofficeofeducationProgressFreeAppropriate1979}.} Columns (5) and (6) of Table \ref{tab:cps_enroll} study the probability of an individual being enrolled in public school and private school, respectively. The results highlight that increases in enrollments were concentrated in public schools, with the mandates causing long-term increases of up to 6.8 percentage points in the probability of being enrolled in a public school. Part of this increase can be explained by the mandates causing a shift away from private schools. Column (6) shows that private school enrollments declined by 2.8 percentage points in the long term. These results are consistent with a story in which parents of disabled children no longer had to rely on the private school system to meet the needs of their children. 
\FloatBarrier

\section{Impacts on adulthood education and labor market outcomes}
\label{section:census_education}

Having shown that the mandates increased school enrollments, I next study their impacts on adulthood education and labor market outcomes. The mandates led to large increases in educational attainment for disabled individuals. Despite concerns that resources may have been redirected away from non-disabled individuals, I find evidence that their educational attainment improved as well. Along with improved education, the mandates also improved labor market outcomes for both groups.

To study impacts in adulthood, I use individual-level data from the Census 1970-2000 samples, as well as the 2001-2007 ACS samples, from IPUMS \parencite{rugglesIPUMSUSAVersion2024}.\footnote{Specifically, the Census samples are the 1970 1\% neighborhood form 1, 1\% state form 1, 1\% metro form 1, 1980 5\%, 1990 5\%, and 2000 5\% samples.} This dataset contains information on an individual's demographic characteristics, disability status, education, and employment status. Since this dataset contains a measure of disability that is consistent over time, it can be used to study long-term outcomes for disabled and non-disabled individuals. 

In the Census data, disability is only assessed for those over age 15, and an individual is identified as disabled if they report a health or physical condition which limits the kind or amount of work they can do at a job. Although this definition is only available for adults and differs from how disability might be assessed in childhood, it is similar to the definition used in Section \ref{section:school} to identify disabled children in that they both relate to major functional limitations an individual may have. 
I address further concerns about this definition, including evidence that the effects I document cannot be explained by shifts in disability identification due to the mandates, in Section \ref{acs_disability}.

In order to study impacts in adulthood, I make a slight modification to the difference-in-difference event-study design. The Census data only contain disability information for adults and beginning in 1970, resulting in a lack of a pre-period for a traditional difference-in-difference analysis. Instead, rather than comparing the year of the mandate's implementation to the year of the Census data, I compare an individual's year of birth to the year of the mandate's implementation in their state of birth. That is, rather than event-time, I study an individual's age at the time of the mandate's implementation. This allows me to compare individuals who were older than school age (older than 25, belonging to the ``pre-period'' birth years) at the time of the mandate with those who were younger (belonging to the ``post-period'' birth years). The age 25 is used because no state law required services to be provided to individuals older than age 25.

\subsection{Impacts on educational attainment}

Using this design, I show that the mandates resulted in large increases in educational attainment for both disabled and non-disabled individuals who were young when a mandate was implemented in their state. Table \ref{tab:census_edu} summarizes these results by presenting several aggregates of the event-study coefficients. The first row presents the average event-study coefficient for those over age 25 at the time of the mandate. This represents a test for parallel pre-trends, as there should be no estimated effect for this group,  which was older than schooling age at the time of the mandate. Next, the average for those under age 25 at the time of the mandate and the Callaway and Sant'Anna average both represent the average effect in the post-period, with different weights aggregating group-time estimates as described in the empirical strategy. Aggregates are also presented for those age 6 to 25 at the time of the mandate (that is, for those of schooling age at the time) and for those under age 6 at the time (that is, younger than schooling age). Finally, results for those age -10 to 5 are presented to provide more precise estimates for the group younger than schooling age at the time, omitting event-study coefficients identified on relatively few individuals at the extremes of the window studied.

The mandates generated large increases in the years of education attained by disabled individuals, especially for those who were below school age at the time of their implementation. Column (1) of Table \ref{tab:census_edu} shows that, for all the disabled individuals under age 25 at the time of the mandates, education increased by 0.08 years, although this estimate  is not significant. Effects are larger for individuals who were younger at the time of the mandates, up to 0.245 years for those age -10 to 5.  
Further, the insignificant estimate for those over age 25 indicates that there is no evidence of a violation of the parallel trends assumption. That is, this result supports the validity of the design by showing that there is no estimated effect for individuals older than schooling age at the time of the mandate's implementation and helps rule out any underlying trends that could confound the analysis. 

\begin{figure}[H]
\caption{Effects on educational attainment}
\label{census_school}
    \begin{subfigure}{.5\textwidth}
        \caption{Years of education attained}
        \centering
        \includegraphics[width=\textwidth]{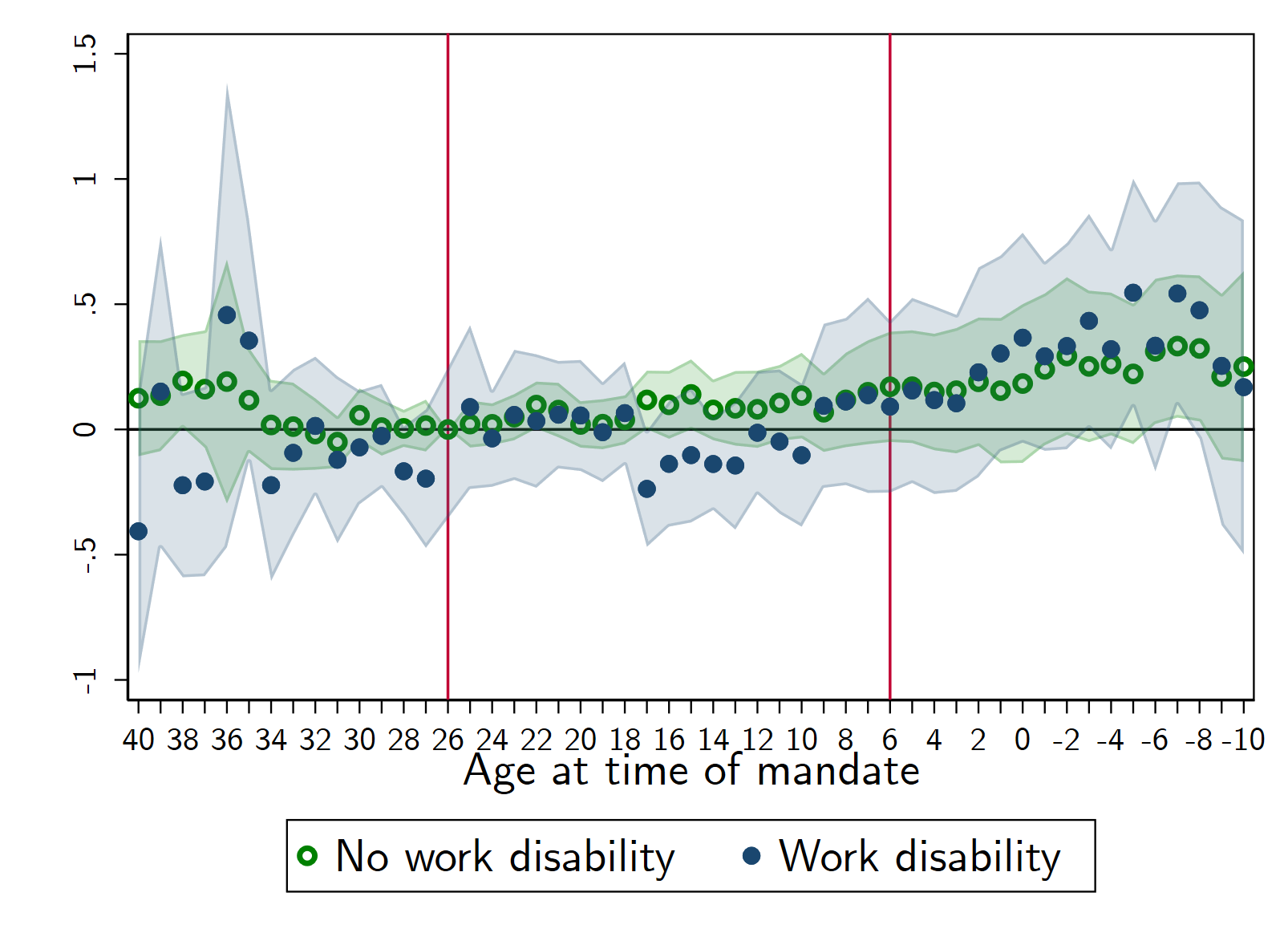}
        \label{fig:census_yr_edu}  
    \end{subfigure}
    \begin{subfigure}{.5\textwidth}
        \caption{Years of education up to 12th grade}
        \centering
        \includegraphics[width=\textwidth]{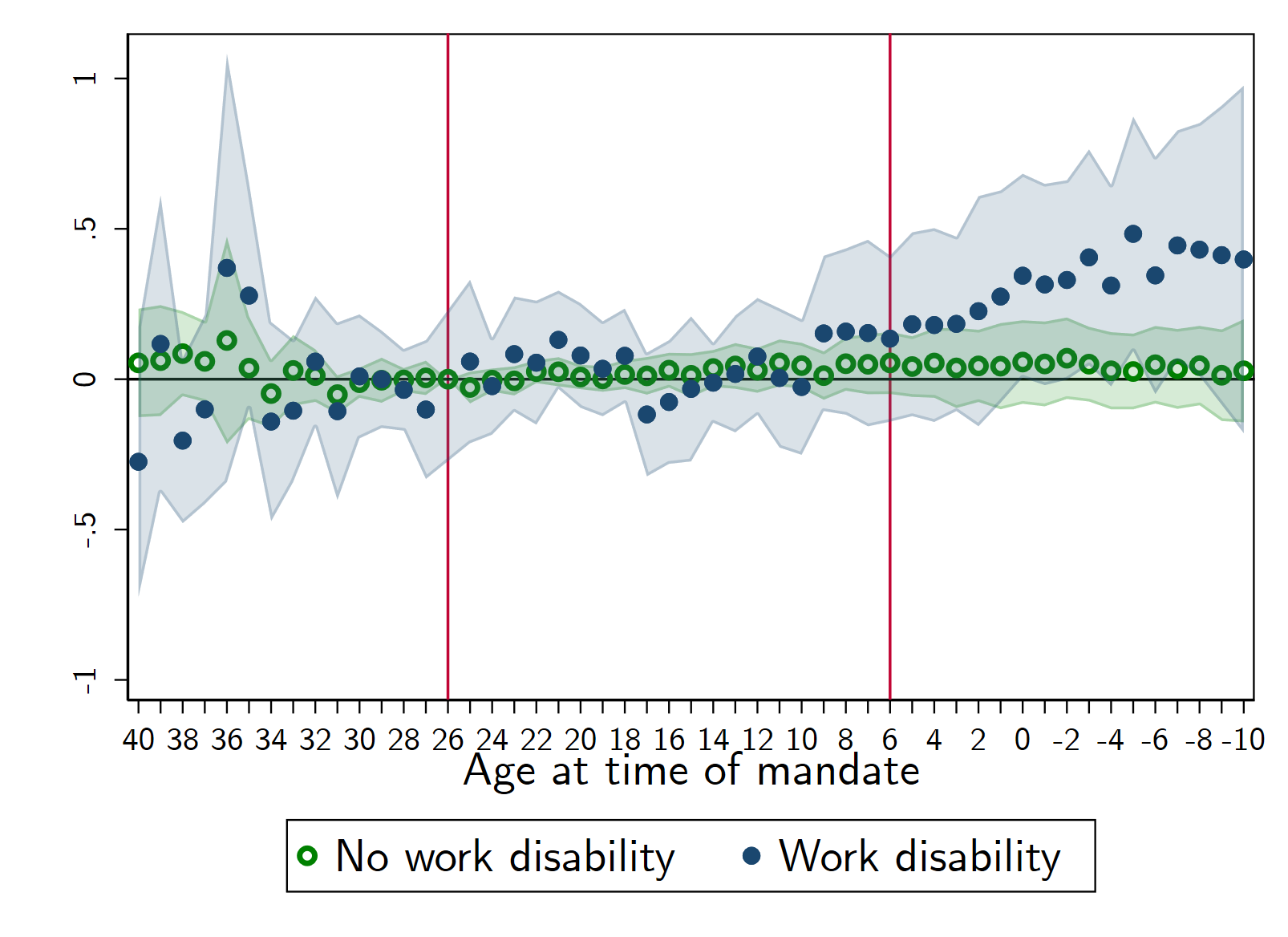}
        \label{fig:census_yr_edu_bHS}  
    \end{subfigure}
    \begin{subfigure}{.5\textwidth}
        \caption{Probability of having no schooling}
        \centering
        \includegraphics[width=\textwidth]{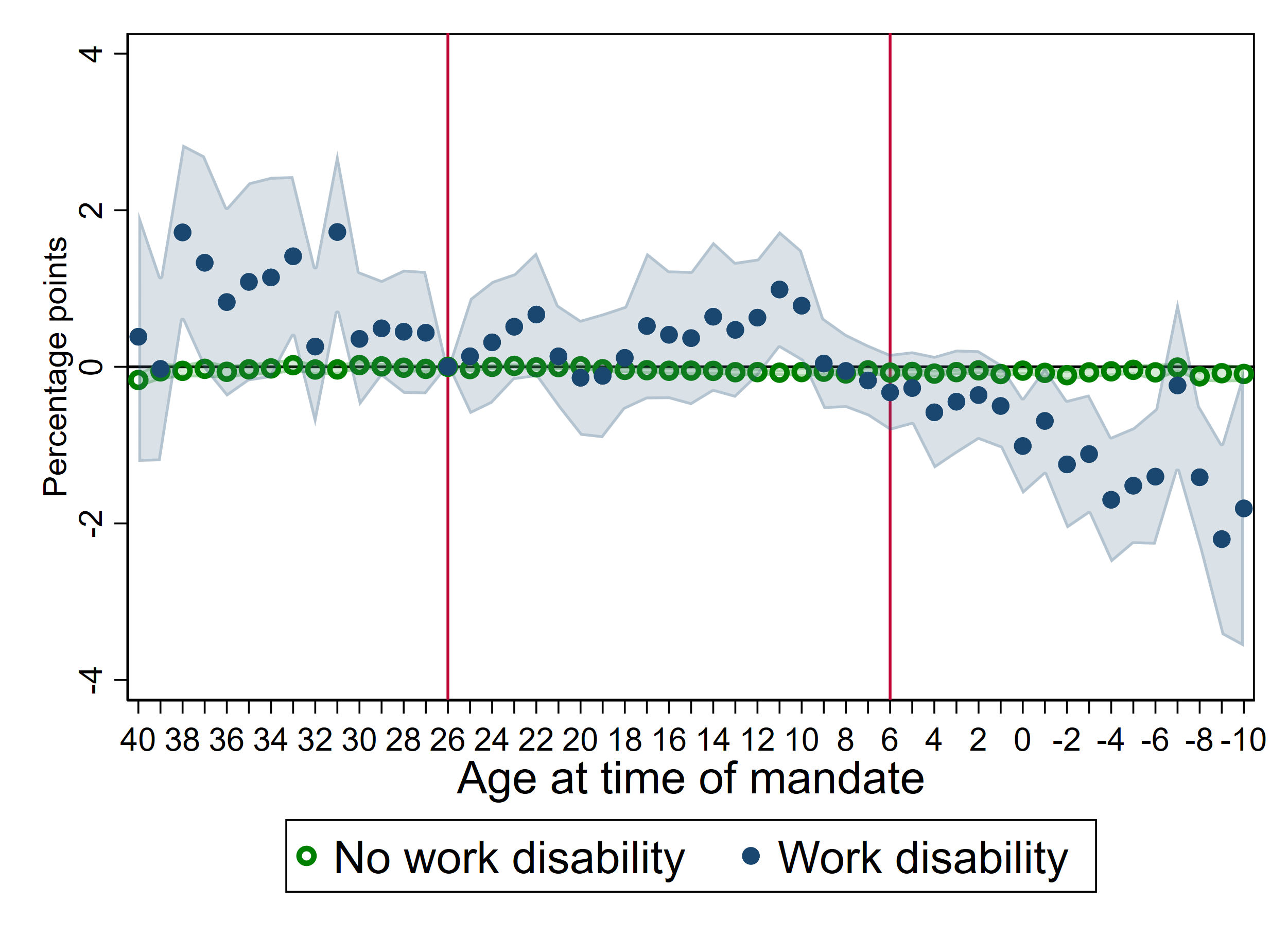}
        \label{fig:census_no_school}  
    \end{subfigure}
    \begin{subfigure}{.5\textwidth}
        \caption{Probability of finishing 9th grade}
        \centering
        \includegraphics[width=\textwidth]{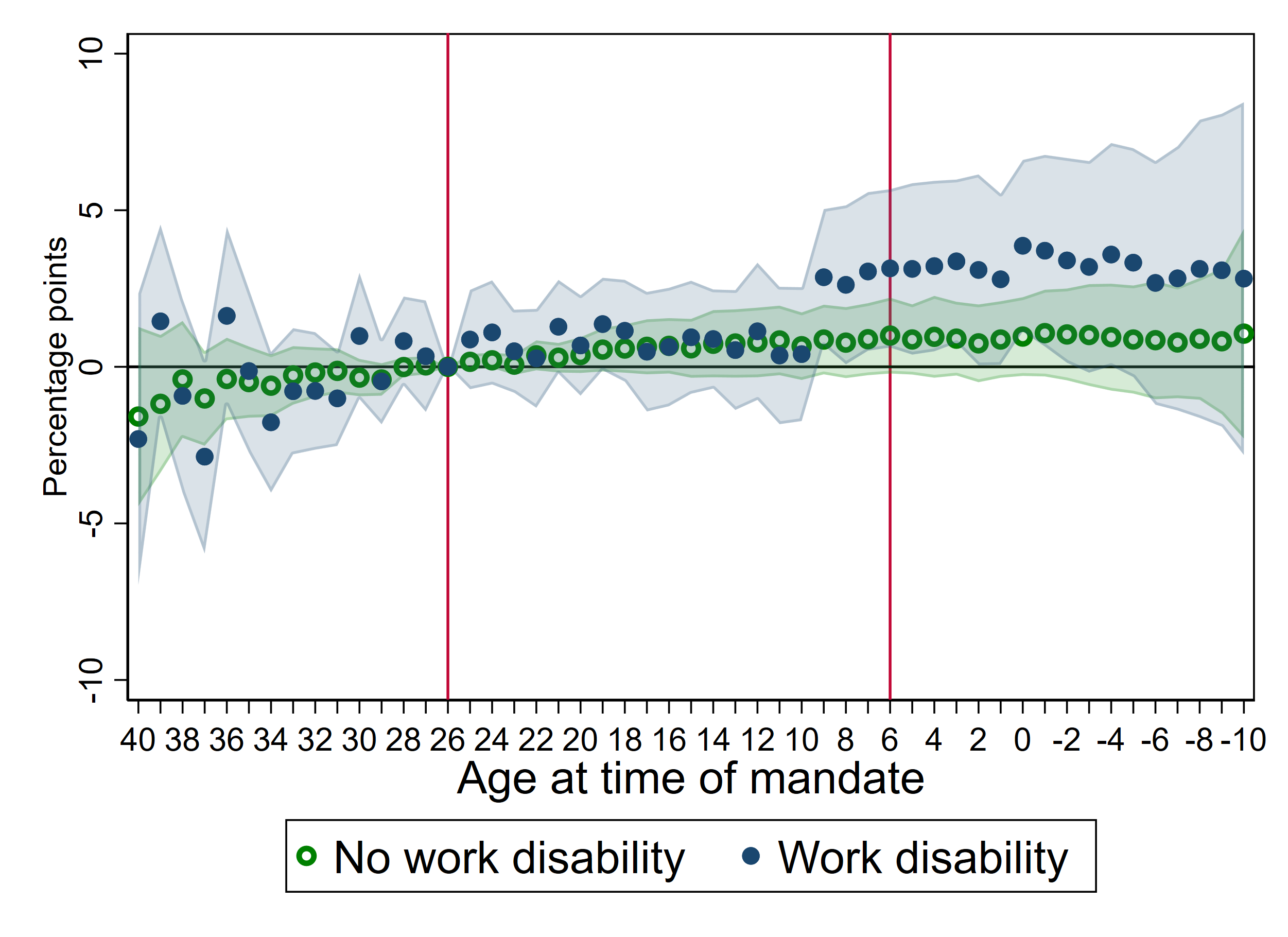}
        \label{fig:census_9}  
    \end{subfigure}
        \caption*{\footnotesize Note: 95\% confidence intervals shown, standard errors clustered at the state level. The figures plot difference-in-difference event-study estimates of the impact of the mandates on the probability of having no schooling, finishing 9th grade, years of education attained, and years of education attained up to 12th grade (imputed from the education codes in the data in line with \textcite{jaegerEstimatingReturnsEducation2003}), with the sample split between disabled and non-disabled individuals age 25-35 in the Census and ACS.}
\end{figure}

As suggested by Table \ref{tab:census_edu}, Figure \ref{fig:census_yr_edu} highlights that effects are largest for those below school age at the time of the mandate's implementation. The figure plots event-study coefficients estimating the impact on years of education. Rather than years since the mandate's implementation, these are plotted in terms of an individual's age at the time of the mandate. Those who were younger at the time of the mandate, that is, with later birth years, are to the right of the figure (later in time). Effects are increasingly large and significant for those who were under age 6 at the time of the mandate's implementation.

\begin{table}[H]
    \centering
        \caption{Effects on educational attainment}
    \label{tab:census_edu}
    \resizebox{.9\textwidth}{!}{%
    \begin{tabular}{lcccccccc}
    \hline \hline \\
                 &\multicolumn{1}{c}{(1)}   &\multicolumn{1}{c}{(2)}   &\multicolumn{1}{c}{(3)}   &\multicolumn{1}{c}{(4)}   &\multicolumn{1}{c}{(5)}   &\multicolumn{1}{c}{(6)}   \\
            &  Years edu.   &\makecell{Years edu. \\ up to 12}   &   No school   &\makecell{Less than \\ 4th grade}   &\makecell{At least \\ 9th grade}   &\makecell{At least \\ 12th grade}   \\
 \hline  \textbf{Disabled} & & & & & \\
Over 25 average&      -0.120   &      -0.112   &       0.006** &       0.011** &      -0.006   &      -0.010   \\
            &     (0.088)   &     (0.072)   &     (0.003)   &     (0.005)   &     (0.010)   &     (0.013)   \\
Under 25 average&       0.076   &       0.130   &      -0.005   &      -0.008   &       0.014   &       0.022*  \\
            &     (0.106)   &     (0.083)   &     (0.003)   &     (0.007)   &     (0.013)   &     (0.012)   \\
Callaway \& Sant'Anna average&       0.086   &       0.118*  &      -0.002   &      -0.007   &       0.021*  &       0.012   \\
            &     (0.085)   &     (0.070)   &     (0.003)   &     (0.006)   &     (0.011)   &     (0.010)   \\
Age 6 to 25 average&      -0.035   &       0.018   &       0.003   &       0.001   &       0.012   &       0.002   \\
            &     (0.064)   &     (0.049)   &     (0.003)   &     (0.005)   &     (0.008)   &     (0.008)   \\
Under age 6 average&       0.173   &       0.228*  &      -0.011***&      -0.015*  &       0.016   &       0.039** \\
            &     (0.157)   &     (0.123)   &     (0.003)   &     (0.009)   &     (0.018)   &     (0.017)   \\
Age -10 to 5 average&       0.245*  &       0.250** &      -0.010***&      -0.018** &       0.032*  &       0.024   \\
            &     (0.130)   &     (0.109)   &     (0.003)   &     (0.008)   &     (0.017)   &     (0.015)   \\
\hline
Observations&      481559   &      481559   &      481559   &      481559   &      481559   &      481559   \\
Pre-mandate mean&       10.28   &        9.74   &        0.05   &        0.09   &        0.74   &        0.51   \\
\hline\hline
 \\ \hline \hline
		\textbf{Non-disabled} & & & & & \\
Over 25 average&      -0.051   &      -0.067   &      -0.000   &       0.001   &      -0.009   &      -0.018** \\
            &     (0.054)   &     (0.041)   &     (0.000)   &     (0.001)   &     (0.006)   &     (0.008)   \\
Under 25 average&       0.172***&       0.031   &      -0.000   &       0.001   &       0.005   &       0.009   \\
            &     (0.057)   &     (0.042)   &     (0.000)   &     (0.001)   &     (0.007)   &     (0.010)   \\
Callaway \& Sant'Anna average&       0.158***&       0.043   &      -0.001** &      -0.000   &       0.007   &       0.010   \\
            &     (0.049)   &     (0.035)   &     (0.000)   &     (0.001)   &     (0.006)   &     (0.008)   \\
Age 6 to 25 average&       0.098***&       0.031   &      -0.000** &       0.000   &       0.006   &       0.008   \\
            &     (0.035)   &     (0.024)   &     (0.000)   &     (0.000)   &     (0.004)   &     (0.006)   \\
Under age 6 average&       0.236***&       0.031   &       0.000   &       0.001   &       0.004   &       0.010   \\
            &     (0.083)   &     (0.059)   &     (0.000)   &     (0.001)   &     (0.009)   &     (0.014)   \\
Age -10 to 5 average&       0.247***&       0.059   &      -0.001***&      -0.000   &       0.009   &       0.014   \\
            &     (0.078)   &     (0.053)   &     (0.000)   &     (0.001)   &     (0.008)   &     (0.012)   \\
\hline
Observations&     7298493   &     7298493   &     7298493   &     7298493   &     7298493   &     7298493   \\
Pre-mandate mean&       12.04   &       11.12   &        0.00   &        0.01   &        0.90   &        0.72   \\
\\ \hline \hline 
    \end{tabular}
    }
    \caption*{\footnotesize Note: The table shows difference-in-difference estimates of the impacts of the mandates on educational attainment at age 25-35 using the Census and ACS data. Column (1) shows impacts on years of education, estimated from the data based on \textcite{jaegerEstimatingReturnsEducation2003} and column (2) years of education, top-coded at 12. Column (3) shows effects on the probability of having no schooling, column (4) less than 4th grade education, column (5) at least 9th grade education, and column (6) at least 12th grade education. Over 25 average and under 25 average refer to a simple average of event-study coefficients for individuals above and below age 25 at the time of a mandate, respectively. Callaway \& Sant'Anna average refers to a weighted average of estimated impacts, with weights given by the share belonging to each treated cohort in the sample. Age 6 to 25 average, under age 6 average, and age -10 to 5 average refer to simple averages of event-study coefficients for those ages at the time of the mandate's implementation. Standard errors clustered at the state level shown in parentheses. \\
* p \textless 0.1, ** p \textless 0.05, *** p \textless 0.01
    }
\end{table}

At what margin did these increases in education occur? Column (2) of Table \ref{tab:census_edu} highlights that this increase is driven by increases through 12th grade, the maximum grade covered by the mandates. In this column, years of education are top-coded at 12 (the end of high school), given that the mandates covered public education through high school and did not apply to higher education. Estimated impacts for disabled individuals are larger, indicating that most of the positive impact of the mandates was concentrated on school years before the end of high school. The estimates suggest that disabled individuals under age 6 at the time of the mandates attained an additional 0.23 years of education before the end of high school as a result of the mandates. Again, the event-study coefficients as shown in Figure \ref{fig:census_yr_edu_bHS} highlight that effects are larger for those who were younger at the time of the mandate's implementation.

More than just the number of years of education, the mandates reduced the probability of very poor education outcomes, such as having no education at all, among disabled people. Columns (3) and (4) of Table \ref{tab:census_edu} study these outcomes. Among disabled people under age 6 at the time of the law, a simple average of event study coefficients suggests a 1.1pp decline in having no schooling at all, representing a 22\% fall relative to the pre-mandate disabled mean of 5\% of individuals having no schooling. At the same time, the coefficients also suggest a 1.5pp decline in having less than 4th grade education, which is large relative to a pre-mandate disabled mean of 9\%. In both cases, although the estimates for those over age 25 are also statistically significant, the event study plot in Figure \ref{fig:census_no_school} and null estimates for those ages 6-25 help minimize concerns about pre-trends. 

Beyond very poor education outcomes, the mandates also improved high school education for disabled individuals. Column (5) of Table \ref{tab:census_edu} shows that, for disabled individuals under age 25 at the time of the law, there is an average of 1.6pp increase in finishing 9th grade (although imprecisely estimated), with significant effects as large as 3.2pp for those age -10 to 5 at the time of the mandate. This effect is large, representing about a 4.3\% increase over the pre-mandate disabled mean of 74\%. Column (6) shows that increases in the probability of finishing 12th grade have a similar magnitude. Again, there are no impacts for those who were older than school age at the time of the mandates' enactment. The event study plot in Figure \ref{fig:census_9} again shows the expected dynamics. 

Turning to spillovers for non-disabled individuals, expected effects are ambiguous. If the mandates meant that fewer resources were available for non-disabled students, they may have had negative spillovers. On the other hand, positive spillovers may be generated by increased school resources, improved class environments, and improved peer effects.

I find evidence of positive spillovers in educational attainment. The results in Column (1) of Table \ref{tab:census_edu} suggest that non-disabled individuals who were under age 25 at the time of the mandate experienced an increase in educational attainment of about 0.17 years, with impacts up to 0.24 years for those under age 6 at the time of the mandate. These effects are large and on par with the estimated impacts for disabled individuals, although they are more precisely estimated due to the larger number of non-disabled individuals in the sample. As with the effects for disabled individuals, Figure \ref{fig:census_yr_edu} shows that increases were concentrated among those who were young at the time of the mandates. 

However, the impacts for non-disabled individuals occurred at different margins than those for disabled individuals. While most of the positive impact in years of education for disabled individuals occurred before the end of high school, estimates in column (2) of Table \ref{tab:census_edu} indicate that much of the increase for non-disabled individuals occurred after high school. While disabled individuals experienced large declines in poor schooling outcomes like having no school or less than 4th grade education, for non-disabled individuals, these outcomes were very rare even before the mandates. Accordingly, there is very little estimated impact in columns (3) and (4). Similarly, nearly 90\% of non-disabled individuals finished 9th grade even before the mandates, and any positive impacts on this outcome in column (5) are small and not significant. Increases in educational attainment for non-disabled individuals are instead driven by increases in education beyond high school, perhaps due to better educational preparation in high school. 

How big are these estimates relative to previous work? I find that the mandates increased disabled people's probability of finishing 12th grade by 3.9pp if they were under age 6 when the mandate was enacted. As discussed in more detail in Section \ref{acs_disability}, not every disabled adult in the Census had a disability since childhood but the mandate's impacts are likely concentrated among those who did. Although the Census does not collect information on the duration of disability in most years, the presence of this information in the 1970 Census allows us to estimate that approximately 42\% of these adults had their disabilities since childhood. Further, using the estimates from Table \ref{tab:nhes_rsce} in Section \ref{section:nhes_education} above, we might also suppose that these effects are concentrated among the approximately 18\% of those who began receiving services for a disability following the mandates. Assuming no effect for all other disabled people, this estimate corresponds to an increase of $3.9/.42/.18 = $ 51.6pp increase in finishing 12th grade. Previous work by \textcite{ballisLongRunImpactsSpecial2021} considers the impact of an opposite reform, which resulted in removals from special education in Texas. They find that, for the marginal removed student, removal at age 10 decreases the probability of graduating high school by 51.9pp, directly in line with this back-of-the-envelope calculation. 

The work by \textcite{ballisLongRunImpactsSpecial2021} also documents negative spillovers of removal from special education on non-disabled students. Consistent with my results, they find no impact of peers' special education removal on high school completion. They also find that students' removal from special education hurts their non-disabled peers' college enrollment. They study effects among students who were not receiving services for a disability (ie, general education students) at age 10 when their peers lost access to these services. They find being in an average district, in which 3.7\% of special education students were removed from receiving services, made general education students 0.9 percentage points less likely to enroll in college. Scaling this impact to match the 18\% increase in services in my results, this translates to a 4.4 percentage point increase in college enrollment, or, assuming that college enrollment translates to 3 additional years of education,  0.13 more years of education for a non-disabled student. This is smaller than my estimate of a 0.24 increase in years of education. However, the results from \textcite{ballisLongRunImpactsSpecial2021} must be interpreted in light of the context of school funding in this setting and the fact that only selected marginal students lost their services.

Consistent with these spillovers, contemporary reports highlighted the potential for non-disabled students to benefit as more services were provided to disabled students for at least three reasons. First, instructional methods might have improved as teachers learned to work with a variety of students. For example, commenting on disabled students who were mainstreamed, or brought into classrooms with their non-disabled peers, \textcite{johnsonSocialStructureSchool1982} note, ``The instructional procedures needed for constructive mainstreaming also benefit nonhandicapped students: the shy student sitting in the back of the classroom, the overaggressive student who seeks acceptance through negative behaviors, the bright but socially inept students, and the average student who does his or her work but whom the teacher never seems to notice.'' At the same time, resources for disabled students, such as additional teachers, aides, psychologists, and teacher training, could also benefit non-disabled students by improving their access to teachers and staff. As explored further in the following section, investments in these resources may have been significant. For example, in the first two years following its mandate, Boston increased the number of resource room teachers, who would provide part-time additional instruction to disabled children, by 60\% \parencite{rafteryLETTERSEDITORHow1976}. How this impacted a classroom was described by one teacher in the 1990s: ``The assistant and therapists work with all the children in the class, not just the disabled. [...] `It gives me many more hands, so all the children benefit' '' \parencite{holtSegregatedEducationNot1994}. Finally, assistive technology designed to assist disabled students might benefit all students, as pointed out by one school principal who observed, ``We had to get computers that talk for our visually impaired kids. [...] Well, it turns out those help other kids learn to read, too'' \parencite{buzbeeSPECIALKIDSSPECIAL1995}.

\subsection{Impacts on labor market outcomes}

\begin{figure}[H]
\caption{Effects on labor force participation and employment}
\label{fig:census_emp}
    \begin{subfigure}{.5\textwidth}
        \caption{Labor force participation}
        \centering
        \includegraphics[width=\textwidth]{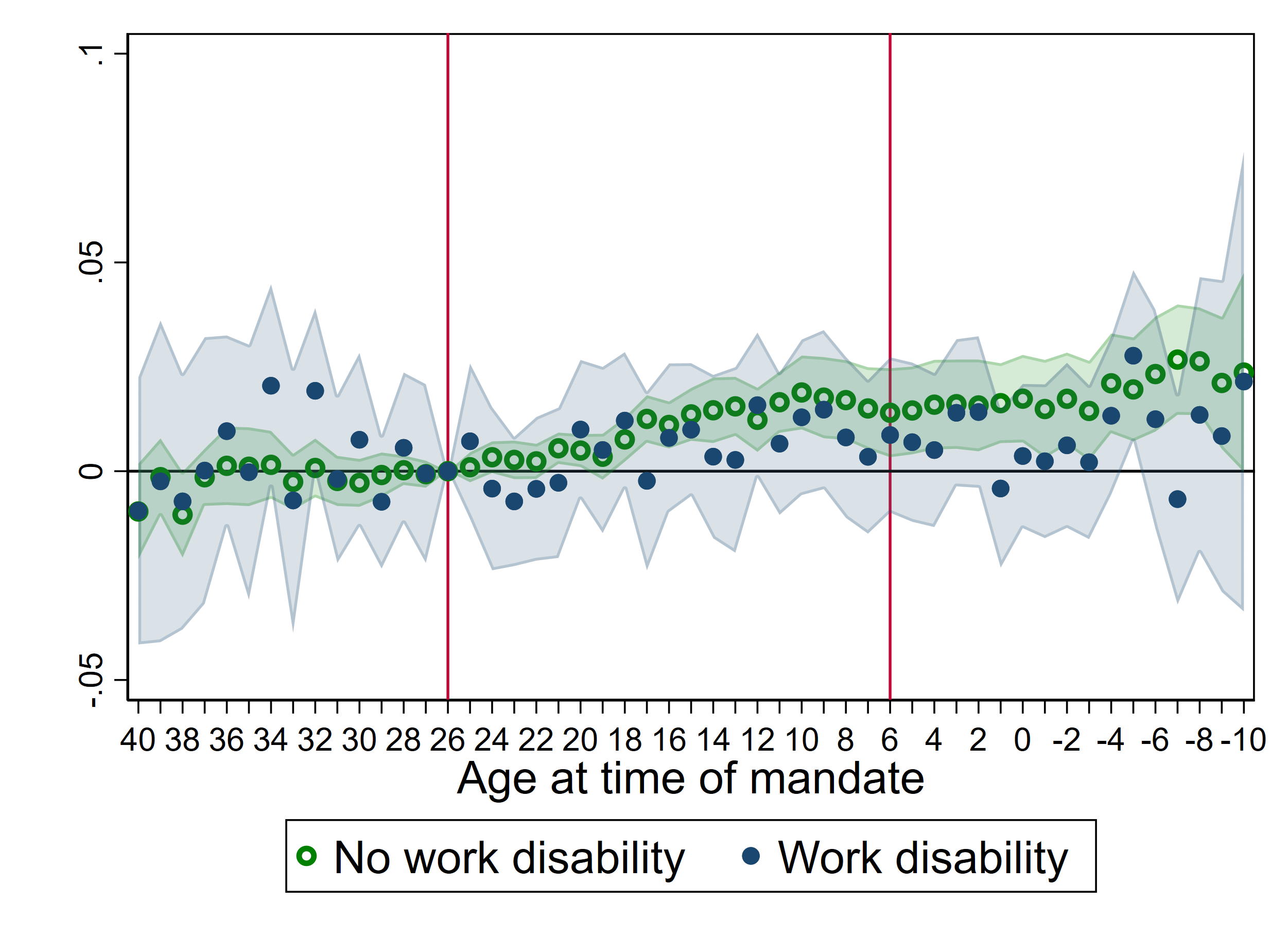}
        \label{fig:census_lf}  
    \end{subfigure}
    \begin{subfigure}{.5\textwidth}
        \caption{Employment}
        \centering
        \includegraphics[width=\textwidth]{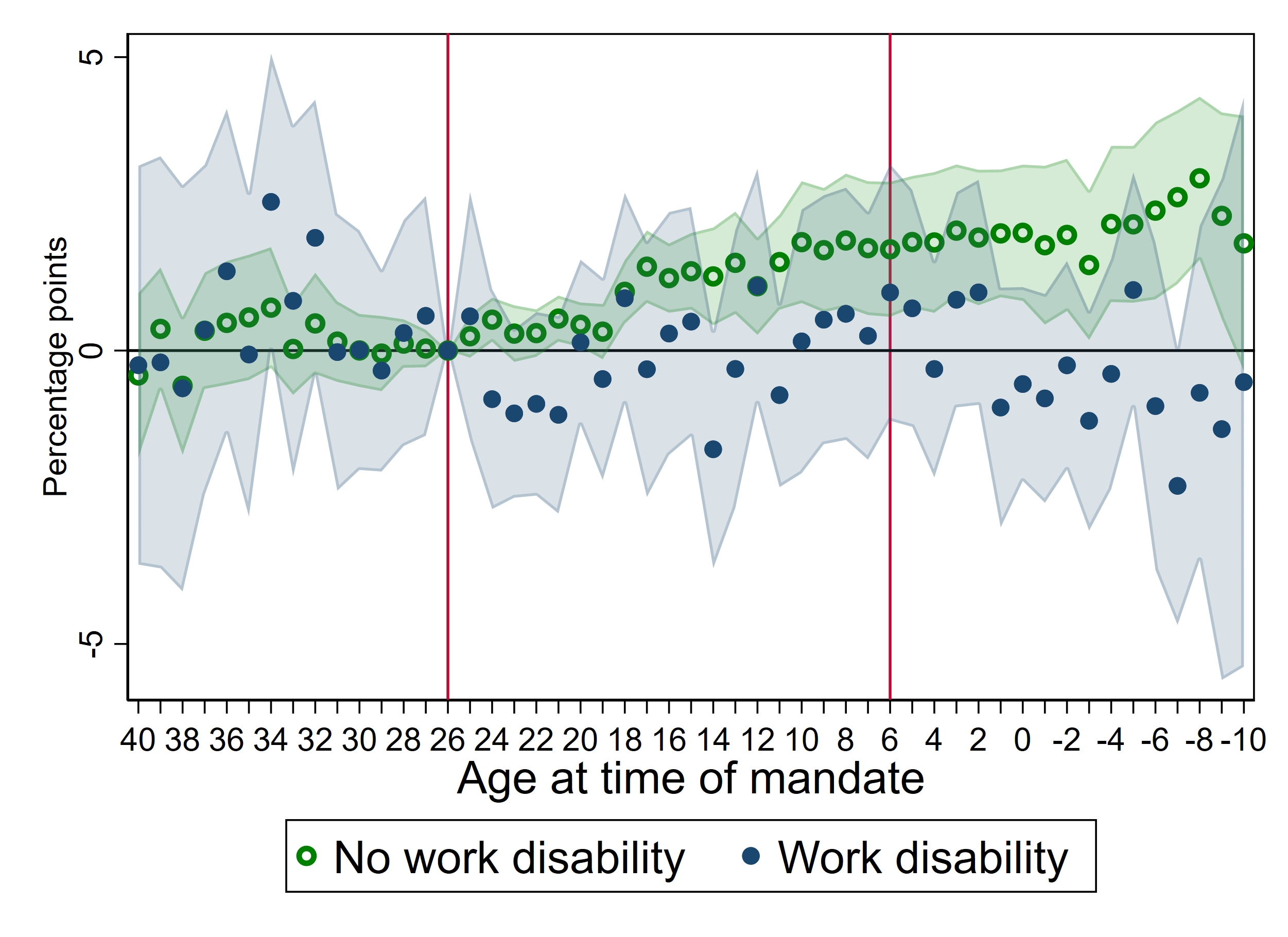}
        \label{fig:census_employed}  
    \end{subfigure}
    \begin{subfigure}{.5\textwidth}
        \caption{Recent employment}
        \centering
        \includegraphics[width=\textwidth]{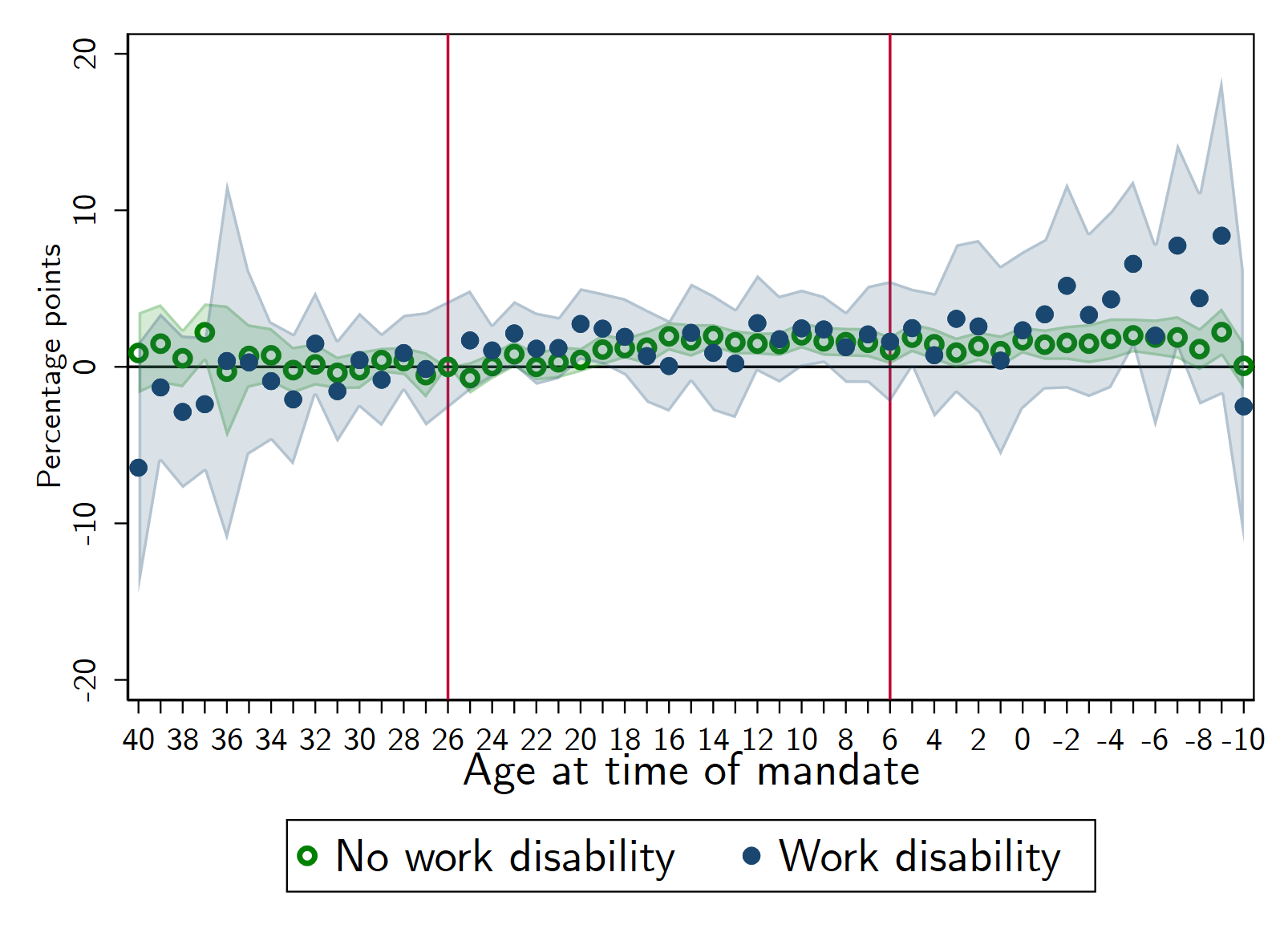}
        \label{fig:census_nm_occ}  
    \end{subfigure}
    \begin{subfigure}{.5\textwidth}
        \caption{Social security receipt}
        \centering
        \includegraphics[width=\textwidth]{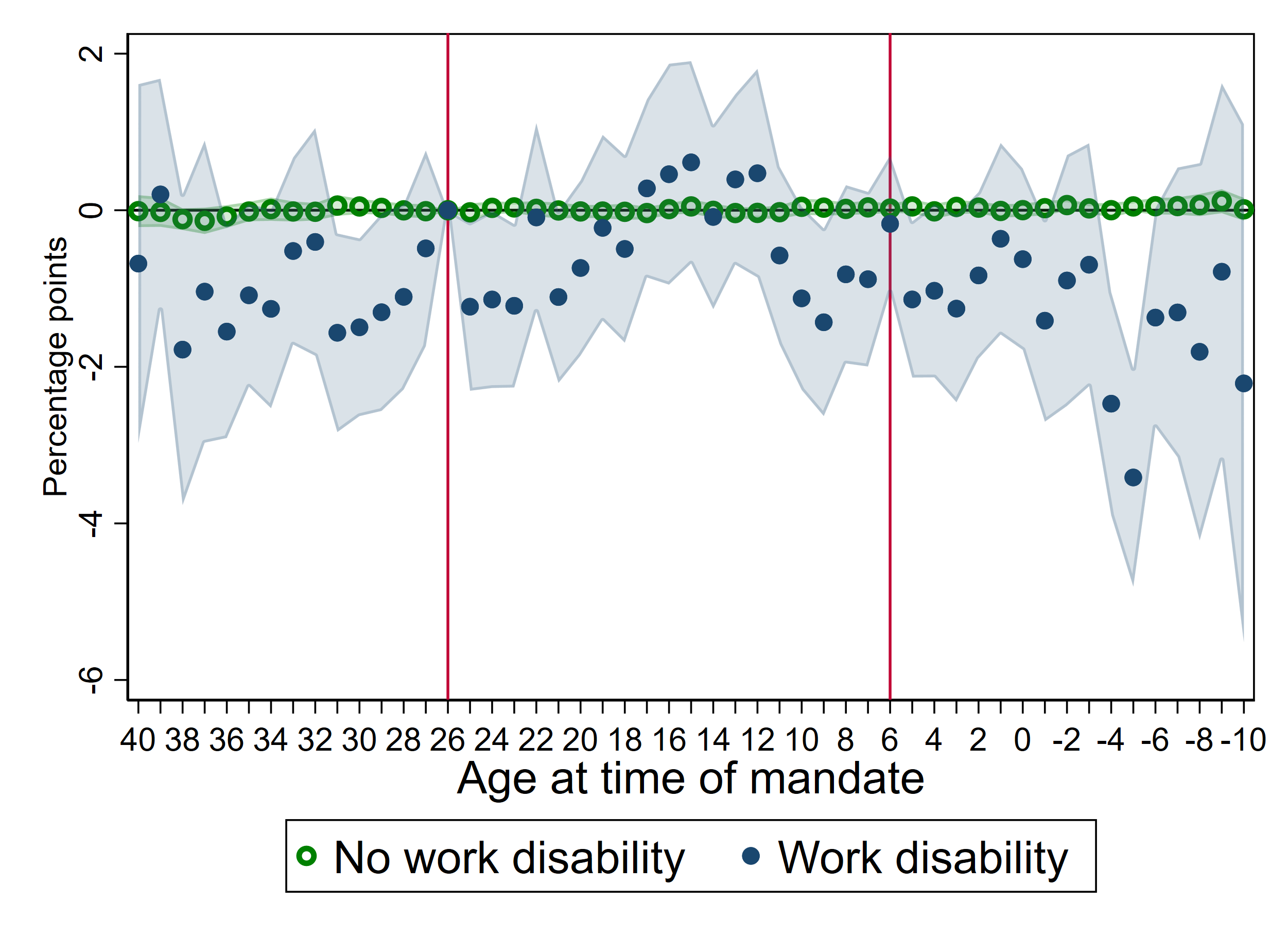}
        \label{fig:census_incss}  
    \end{subfigure}
        \caption*{\footnotesize Note: 95\% confidence intervals shown, standard errors clustered at the state level. The figures plot difference-in-difference event-study estimates of the impact of the mandates on labor force participation, employment, recent employment, and social security receipt with the sample split between disabled and non-disabled individuals age 25-35 in the Census and ACS.}
\end{figure}

With improvements in education for disabled students, we might also expect increased labor market engagement in adulthood, and I find some evidence of this. Table \ref{tab:census_emp} and Figure \ref{fig:census_emp} explore labor market outcomes. Columns (1) and (2) show no significant impact on labor force participation or employment for disabled individuals, and the magnitudes are small. This null effect persists even when including the possibility of being either employed or in school, as in column (3). However, column (4) examines whether the individual has an occupation listed in the Census data, a question which is asked of individuals who have worked in the last 5-10 years. Thus, this is an indicator of having some work experience in the recent past. Here, there is stronger evidence of an increase in employment of disabled individuals. Among those under age 6 at the time of the mandate, there is a 2.9pp increase in the probability of being employed recently as defined by this measure. Column (5) measures impacts on the inverse hyperbolic sine of wage income (measured in constant 1990 dollars) and finds an insignificant increase of about 0.05 or 5\%.\footnote{Results are virtually identical when using a log+1 transformation instead of inverse hyperbolic sine.} However, these employment results should be interpreted in light of the Census disability definition only including those who report limitations in or inability to work and are likely an underestimate of the true impact of the mandates on disabled people.

Consistent with increased labor market engagement and the avoidance of very poor education outcomes, the mandates also resulted in disabled individuals being less likely to receive Social Security payments, which include insurance for permanent disability. For young adults, these payments likely largely comprise payments from the Supplemental Security Income (SSI) program. SSI was created in 1972 and provided benefits to disabled adults who were ``unable to engage in any substantial gainful activity'' \parencite{socialsecurityadministrationSSIHistoryProvisions2000}. Thus, a decline in Social Security receipt would suggest a decline in individuals being unable to participate in any kind of work or ``gainful activity''. 

As expected given that increased education and work experience would result in fewer people being completely unable to work, Figure \ref{fig:census_incss} and Table \ref{tab:census_emp} show that the mandates reduced Social Security receipt among disabled individuals below schooling age at the time of their implementation. For disabled individuals under age 6 when the mandate was implemented, there is a 2.9pp decline in the probability of receiving these payments. This is quite large given that, before the mandate, 7\% of disabled individuals reported receiving these benefits.  

For non-disabled individuals, consistent with their increases in education, I find positive effects on employment, labor force participation, and income (Table \ref{tab:census_emp}). For non-disabled individuals who were under age 6 at the time of the mandate, I find a 3.0pp increase in labor force participation and a 2.8pp increase in employment. Studying impacts on wage incomes, I find large positive results, with an increase of 0.27 in the inverse hyperbolic sine (IHS) of income, or about 27\%, for those under age 6 at the time of the mandates. Looking at those age -10 to 5 (that is, removing people who were born more than 10 years following the mandate, for whom there are relatively few observations) results in a slightly smaller effect of 0.18 IHS points, or about 18\%.

\begin{table}[H]
    \centering
        \caption{Effects on labor market outcomes}
    \label{tab:census_emp}
    \resizebox{\textwidth}{!}{%
    \begin{tabular}{lcccccccc}
    \hline \hline \\
                 &\multicolumn{1}{c}{(1)}   &\multicolumn{1}{c}{(2)}   &\multicolumn{1}{c}{(3)}   &\multicolumn{1}{c}{(4)}   &\multicolumn{1}{c}{(5)}   &\multicolumn{1}{c}{(6)}   \\
            &         LFP   &  Employment   &\makecell{Employed \\ or in school}   &\makecell{Employed \\ recently}   &\makecell{IHS \\ wage income}   &\makecell{Social security \\receipt}   \\
 \hline  \textbf{Disabled} & & & & & \\
Over 25 average&      -0.001   &       0.003   &       0.005   &      -0.017*  &      -0.079   &      -0.008*  \\
            &     (0.008)   &     (0.008)   &     (0.008)   &     (0.009)   &     (0.080)   &     (0.004)   \\
Under 25 average&       0.005   &      -0.002   &       0.000   &       0.018***&       0.022   &      -0.018***\\
            &     (0.006)   &     (0.006)   &     (0.005)   &     (0.006)   &     (0.058)   &     (0.004)   \\
Callaway \& Sant'Anna average&       0.007   &      -0.001   &      -0.003   &       0.014** &       0.007   &      -0.008*  \\
            &     (0.006)   &     (0.007)   &     (0.006)   &     (0.007)   &     (0.071)   &     (0.004)   \\
Age 6 to 25 average&       0.005   &      -0.001   &      -0.002   &       0.006   &      -0.012   &      -0.005   \\
            &     (0.007)   &     (0.007)   &     (0.007)   &     (0.007)   &     (0.074)   &     (0.004)   \\
Under age 6 average&       0.004   &      -0.003   &       0.002   &       0.029***&       0.052   &      -0.029***\\
            &     (0.007)   &     (0.006)   &     (0.006)   &     (0.007)   &     (0.053)   &     (0.003)   \\
Age -10 to 5 average&       0.009   &      -0.004   &      -0.007   &       0.026***&       0.048   &      -0.014***\\
            &     (0.007)   &     (0.007)   &     (0.007)   &     (0.007)   &     (0.074)   &     (0.005)   \\
\hline
Observations&      481559   &      481559   &      481559   &      481559   &      481559   &      481559   \\
Pre-mandate mean&        0.54   &        0.50   &        0.52   &        0.79   &        5.78   &        0.07   \\
\hline\hline
 \\
		\textbf{Non-disabled} & & & & & \\
Over 25 average&      -0.001   &       0.003   &       0.005   &      -0.001   &      -0.002   &       0.000   \\
            &     (0.003)   &     (0.004)   &     (0.004)   &     (0.002)   &     (0.036)   &     (0.000)   \\
Under 25 average&       0.021***&       0.020***&       0.015***&       0.008***&       0.194***&       0.001*  \\
            &     (0.004)   &     (0.004)   &     (0.004)   &     (0.003)   &     (0.037)   &     (0.000)   \\
Callaway \& Sant'Anna average&       0.014***&       0.015***&       0.012***&       0.008***&       0.136***&       0.000   \\
            &     (0.004)   &     (0.004)   &     (0.004)   &     (0.003)   &     (0.037)   &     (0.000)   \\
Age 6 to 25 average&       0.010***&       0.011***&       0.010***&       0.006** &       0.109***&       0.000   \\
            &     (0.003)   &     (0.003)   &     (0.003)   &     (0.002)   &     (0.026)   &     (0.000)   \\
Under age 6 average&       0.030***&       0.028***&       0.020***&       0.010***&       0.268***&       0.001***\\
            &     (0.005)   &     (0.006)   &     (0.005)   &     (0.003)   &     (0.049)   &     (0.000)   \\
Age -10 to 5 average&       0.019***&       0.021***&       0.016***&       0.010***&       0.179***&       0.000   \\
            &     (0.006)   &     (0.006)   &     (0.005)   &     (0.003)   &     (0.053)   &     (0.000)   \\
\hline
Observations&     7298493   &     7298493   &     7298493   &     7298493   &     7298493   &     7298493   \\
Pre-mandate mean&        0.71   &        0.68   &        0.69   &        0.91   &        7.54   &        0.01   \\
\\ \hline \hline 
    \end{tabular}
    }
    \caption*{\footnotesize Note: The table shows difference-in-difference estimates of the impacts of the mandates on labor market outcomes at age 25-35 using the Census and ACS data. Column (1) shows effects on labor force participation, column (2) on employment, column (3) on whether the individual is either employed or in school, column (4) on whether the individual has a non-missing occupation, that is, whether they have been employed in the last 5-10 years, and column (5) on whether they report receiving Social Security income. Over 25 average and under 25 average refer to a simple average of event-study coefficients for individuals above and below age 25 at the time of a mandate, respectively. Callaway \& Sant'Anna average refers to a weighted average of estimated impacts, with weights given by the share belonging to each treated cohort in the sample. Age 6 to 25 average, under age 6 average, and age -10 to 5 average refer to simple averages of event-study coefficients for those ages at the time of the mandate's implementation. Standard errors clustered at the state level shown in parentheses. \\
* p \textless 0.1, ** p \textless 0.05, *** p \textless 0.01
    }
\end{table}

This effect on incomes is quite large, but not surprising. First, the magnitude corresponds closely with the increase in employment. Among those who were non-disabled and employed before the mandates, the mean IHS income was 9.79, which, when multiplied by a 2.8pp increase in employment, translates directly into a 0.27 IHS income increase. At the same time, as discussed above, the largest increases in education among non-disabled individuals occurred after high school (that is, on the margin of having at least some college). Research on the returns to education has estimated that, in this period in the US, the returns to having at least some college were about 15-20\% \parencite{jamesCollegeWagePremium2012}. 

As would be expected, virtually no non-disabled people age 25-35 report receiving Social Security payments, and there are very small effects on this outcome among non-disabled individuals.

\FloatBarrier

\subsection{Concerns with disability measures}\label{acs_disability}
In this section, I address concerns about the results discussed above. I show that they are not likely driven by changes in disability identification in the surveys and I discuss the overlap between disabilities measured in childhood and adulthood. 

One possible concern with this analysis is that the mandates changed who was identified or identified themselves as disabled, meaning that improvements in outcomes could be driven by changes in composition of the group rather than by meaningful changes in the outcomes of particular individuals. However, two pieces of evidence suggest that this is not the case. 

First, the probability of identifying as disabled did not change meaningfully in response to the mandates either in the NHES or in the Census datasets. Using the difference-in-differences approach to test whether there is any impact of the law on the probability of a child being identified by their parents as disabled in the NHES survey, I find no evidence of such an effect. Appendix Figure \ref{fig:prob_disabled} shows the event study estimates for the overall probability of being identified as disabled and the effects for each type of disability. Although the overall results show a negative and significant effect 3 years after the law, the aggregate point estimate for post period is insignificant. Further, there do not appear to be clear trends in the type of disability. There is also little evidence of impacts on disability identification in the Census data. Appendix Figure \ref{fig:prob_disabled} plots the event study coefficients. The magnitudes of the estimated impacts are very small, with the overall average for individuals under age 25 at the time of the mandate being only 0.2 percentage points, and with a trend that differs from the trend of estimated impacts. Thus, there is little evidence that the mandates substantially changed how disability was identified in these surveys. 

At the same time, the main education and employment effects are robust to studying impacts on the whole sample, rather than splitting the sample by disability. 
This means that these positive impacts cannot be driven solely by a change in composition of the two groups. 

Another concern with interpreting this analysis is the disability measure used in the Census, which is measured in adulthood only and cannot give a complete picture of a person's disabilities as a child, when they would have received services. In the Census sample, the available definition of disability is the response to a question which asks whether the person has a health or physical condition which limits the kind or amount of work the person can do. This definition of disability is only available for those who are ages 16 and above. More detail on the text of the question in each year can be found in Appendix \ref{appendix_censusdata}. I shed light on the ability of this definition to capture childhood disability experiences by examining comparisons between this definition and that in the NHES, additional descriptive statistics from the Census data, and another data source which includes disability measures both in adulthood and childhood. 

As noted above, the definition of disability available in the Census is an imperfect measure of the disabilities an individual may have had as a child and differs from the measure of child disability in the NHES. However, both definitions are based on an individual's functioning and ability to perform major life tasks. It seems likely than an individual who is categorized as disabled in the NHES due to having difficulty with walking, talking, hearing, exercising, or moving a limb is also likely to have limitations in the amount or type of work they can do. Appendix Figure \ref{fig:acs_vs_nhes} shows that, when studying young adults, patterns of race and gender by disability are similar between the Census and NHES. 

Some additional descriptive statistics from the Census also help to understand the difference between disabilities measured in childhood and adulthood. In the 1970 Census data only, the data contain a follow-up question which asks those who are disabled about the duration of their disability. Appendix Figure \ref{fig:disabdur_by_age} highlights that the share of disabled people with long-term disabilities is high and persistent across ages. Of those ages 25-35, 42\% report having had their disability for more than 10 years (the highest value recorded by the Census), that is, since childhood.

An additional data source containing questions both about childhood and adulthood disabilities can also shed light on the comparison between the two. The National Health Interview Survey (NHIS) is a national survey with detailed information on individuals' health status and history. Using data from 1997-2018, this dataset contains both a question on whether an individual's health limits or prevents them from working (as in the Census) and detailed questions about the person's abilities (similar to the NHES) and the duration of their disabilities. I define an individual as having a childhood disability if they have any functional limitation (eg, difficulty walking, lifting an object, or climbing stairs) or activity limitations (eg, in bathing, self care, or working) due to a condition that began when the person was 21 years old or younger. This allows for comparison between the two definitions: childhood limitations similar to those in the NHES and adulthood work disability as defined in the Census. However, a limitation of this analysis is that it may not include adults who had disabilities as children and no longer have them as adults. 

The results show that, among adults age 25-35 who report having a work disability by the Census definition, 45.6\% also had this disability as a child (Appendix Figure \ref{fig:nhis_disability_duration}). In contrast, this is true for only 2.5\% of those who do not report having a work disability. In this way, the Census disability measure is effective in differentiating individuals with childhood disabilities from those without. However, the results using the Census definition should be considered a lower bound because they include individuals who did not have a childhood disability.

\FloatBarrier

\section{The role of school resources}
\label{section:financial}
To understand the mechanisms for positive effects on adulthood outcomes, I consider the finances of school districts around the time of the mandates. A serious concern about the implementation of the  mandates was a lack of funding for their provisions and the potential that they would redirect scarce resources away from non-disabled students. Given this lack of mandated funding, as expected, I find no evidence of changes in spending by state governments. However, school districts raised substantial funds and increased expenditures and employment, and these may explain a large portion of the positive spillovers on non-disabled students. 

\subsection{Impacts on state and school district finances}

In this section, I use data from the Historical Finance Data Base of Individual Local Governments (IndFin), produced by the Census Bureau. This dataset contains information on the finances of cities, towns, states, and school districts in the 1967 fiscal year and annually from 1970-2008. In years ending in a 2 or 7, the data come from the Census of Governments, and data from other years come from the Annual Survey of Governments combined with other compilations of financial data. As a result, in most years over this period, nearly all school districts are represented. Appendix \ref{appendix_indfin} further describes the cleaning and coverage of the database. 

Increases in the provision of services for disabled students would have required substantial increases in the financial resources needed to educate disabled students, but this funding was not provided by state governments. The mandates were unfunded by states, meaning that state governments largely did not commit additional funds to providing services or change their funding formulas (that is, the formulas determining the funding for each school district based on the number of students enrolled and other factors) at the time of the mandates' implementation. Even though ``special education'' was often considered in state governments' funding formulas, states still did not provide enough funds to cover the rapidly increasing costs of services over this period.\footnote{For example, in Michigan, the \textit{Durant vs. State of Michigan} lawsuit filed in 1980 and decided in 1997 found that the state was only responsible for less than 30\% of the cost of ``special education'' and that this funding had been severely underprovided \parencite{clearyDurantWhatHappened1999}.} 

Instead, local school districts would bear the responsibility for paying for these mandates. As the headline of an article in the \textit{Boston Globe} published in 1974 about the passing of Massachusetts's mandate wrote, ``Education law on handicapped will hike taxes'' \parencite{thorntonEducationLawHandicapped1974}. The article noted, ``The biggest problem with the new law is that it requires the entire first year of funding, estimated at \$50 million to \$100 million, to come from local property taxes, with a promise of partial reimbursement from the state at the end of fiscal 1975.''\footnote{For a further discussion of local school district financial responsibility for Massachusetts's mandate, see \textcite{SPECIALCHILDChapter1975}. As one education commissioner reflected on his state mandate, he noted, ``while special education does add to the already heavy burden on the property tax, I don't consider it a heavy price to pay when you consider the improvement it has brought about for children with handicaps'' \parencite{paveChapter766Progress1980}. 
} 
If schools did not receive additional funding, this might have meant fewer resources for non-disabled students, potentially harming their educational outcomes. Thus, finances at the local school district level are crucial to understanding the outcomes for students. 

\begin{figure}[htb]
        \caption{Share of school district revenues by source, 1967}
        \centering
        \includegraphics[width=.5\textwidth]{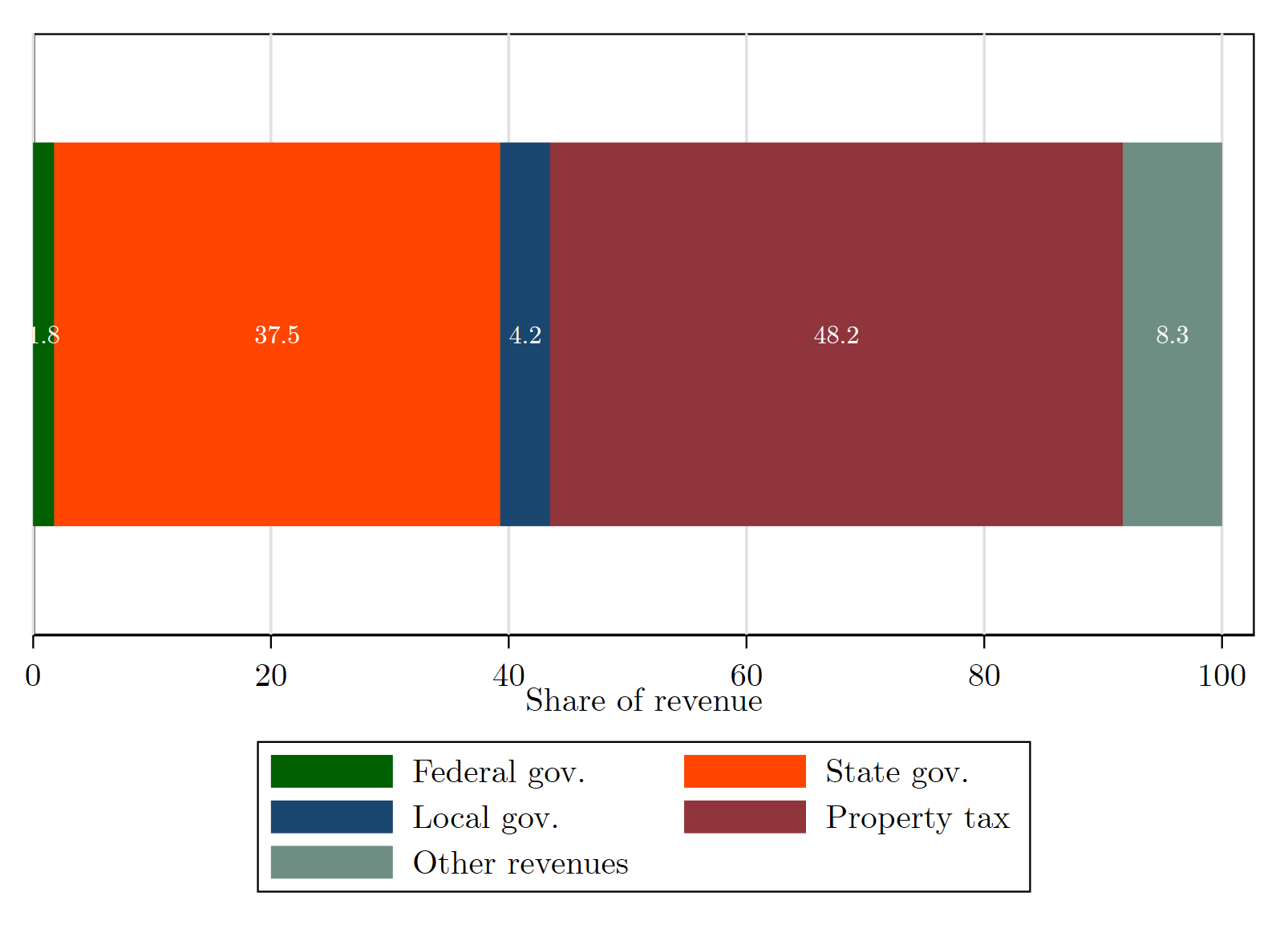}
        \label{fig:indfin_descriptive}  
            \caption*{\footnotesize Note: The figure plots average revenue per student according to revenue source for school districts in 1967. Most revenues come from state government and property taxes, with federal government, local government, and other sources each comprising a small share of total revenues.}

\end{figure}

To understand the funding sources available to school districts in this period, Figure \ref{fig:indfin_descriptive} highlights the two important sources of revenues for school districts in 1967: property taxes and state government funds. The figure shows, for school districts in 1967, the share of revenues according to their source. Property taxes provided 48\% of a school district's revenue per student and state funds provided another 37.5\%. In contrast, funds from the federal government and other local governments (such as counties, cities, or towns) were very small. Other sources of revenues include interest on retirement investments, local income taxes, and other charges. 

\begin{figure}[H]
        \caption{State spending per child, in thousands of 1990 dollars}
        \centering
        \includegraphics[width=.5\textwidth]{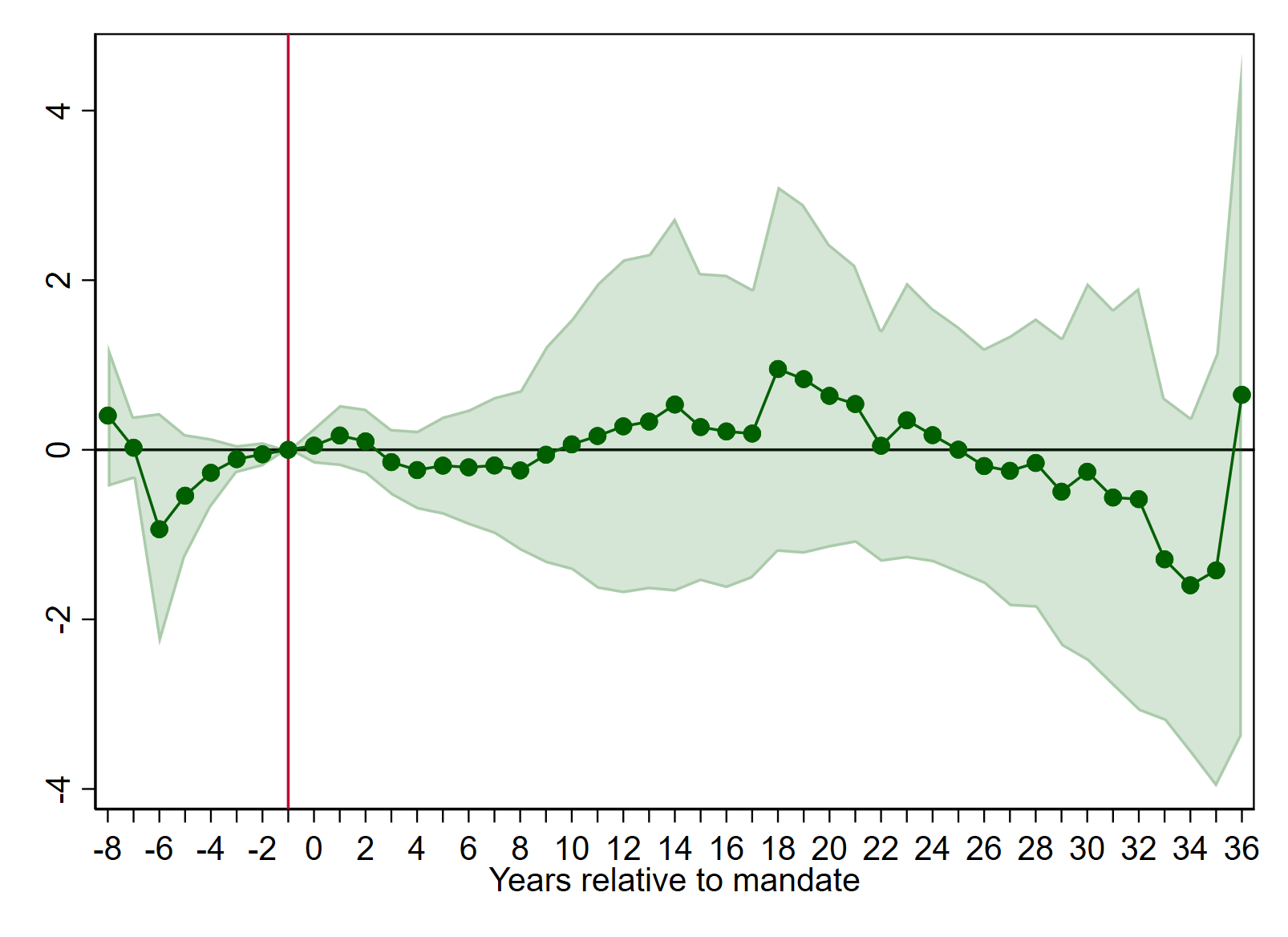}
        \label{fig:indfin_state_exp}  
        \caption*{\footnotesize Note: 95\% confidence intervals shown, standard errors clustered at the state level. The figure plots difference-in-difference event study estimates of the impact of the mandates on state government spending per child, in thousands of constant 1990 dollars.}
\end{figure}

Consistent with the unfunded nature of the mandates, the mandates did not increase education spending at the state government level. Table \ref{tab:indfin_state_exp_rev} shows the impacts of the mandates on overall state expenditures, state expenditures on education, and direct state expenditures on education (that is, excluding transfers to other governments). As in the other analyses, the table presents aggregations of the estimated treatment effects for different periods: the pre-mandate period, the post-period (using both a simple average and the Callaway and Sant'Anna average), years 0-9 following the mandate (the short term), years 10-19 following the mandate, and years 20+ after the mandate (the long term). The results in column (1) show that there is no increase in state spending in response to the mandates at any time period. The event study coefficients appear in Figure \ref{fig:indfin_state_exp} and also show no discernible impact of the mandates on state government spending. Further, columns (2)-(5) of Table \ref{tab:indfin_state_exp_rev} show no clear impact or a slight decline in education spending, no effect on direct education spending, and no evidence of impacts on tax revenues or borrowing.

\begin{table}[h]
    \centering
    \caption{Financial effects of the mandates at the state level}
    \label{tab:indfin_state_exp_rev}
        \resizebox{\textwidth}{!}{%
    \begin{tabular}{lccccc} 
        \hline \hline 
                    &\multicolumn{1}{c}{(1)}   &\multicolumn{1}{c}{(2)}   &\multicolumn{1}{c}{(3)}   &\multicolumn{1}{c}{(4)}   &\multicolumn{1}{c}{(5)}   \\
            &  Total exp.   &Education exp.   &Direct ed. exp.   &    Tax rev.   &   Borrowing   \\
\hline
Pre-period average&      -0.165   &       0.078   &       0.037   &       0.123*  &      -0.038   \\
            &     (0.175)   &     (0.060)   &     (0.042)   &     (0.065)   &     (0.039)   \\
Post-period average&      -0.073   &      -0.273*  &      -0.084   &       0.135   &      -0.087   \\
            &     (0.755)   &     (0.163)   &     (0.107)   &     (0.472)   &     (0.191)   \\
Callaway \& Sant'Anna average&      -0.012   &      -0.263   &      -0.087   &       0.056   &      -0.086   \\
            &     (0.674)   &     (0.163)   &     (0.098)   &     (0.370)   &     (0.191)   \\
Years 0-9 average&      -0.095   &      -0.146   &      -0.026   &      -0.092   &      -0.019   \\
            &     (0.300)   &     (0.137)   &     (0.045)   &     (0.207)   &     (0.103)   \\
Years 10-19 average&       0.319   &      -0.181   &      -0.063   &       0.279   &      -0.010   \\
            &     (0.758)   &     (0.202)   &     (0.095)   &     (0.439)   &     (0.269)   \\
Years 20+ average&      -0.111   &      -0.353*  &      -0.116   &       0.264   &      -0.167   \\
            &     (1.028)   &     (0.197)   &     (0.162)   &     (0.659)   &     (0.214)   \\
\hline
Observations&        1782   &        1782   &        1782   &        1782   &        1782   \\
Pre-mandate mean&        7.81   &        2.31   &        1.36   &        4.00   &        0.50   \\
 \\ 
        \hline
    \end{tabular} 
            } \\
    \caption*{\footnotesize Note: All variables are expressed in thousands of dollars per child age 5-19, in constant 1990 dollars. That is, a mean of 1 indicates 1000 1990 dollars per child. The table shows difference-in-difference estimates of the impacts of the mandates on expenditures and revenues using the state finance data. Column (1) shows effects on total spending per child, column (2) on education spending per child, column (3) on direct education spending per child, column (4) on tax revenues per child, and column (5) on borrowing per child. Pre-period average and post-period average refer to a simple average of event-study coefficients before and after the implementation of a mandate, respectively. Callaway \& Sant'Anna average refers to a weighted average of estimated impacts, with weights given by the share belonging to each treated cohort in the sample. Years 0-9 average, Years 10-19 average, and Years 20+ average refer to simple averages of event-study coefficients for those years. Standard errors clustered at the state level shown in parentheses. \\
    * p \textless 0.1, ** p \textless 0.05, *** p \textless 0.01
    }
\end{table}

This analysis serves as a useful placebo test. If a concern is that the timing of the mandates correlates with whether states were already expanding the capacity of their education systems, the results should show evidence of an increasing trend in state education spending. More generally, this also suggests that the mandates do not correlate with large expansions of state government capacity in general. The lack of evidence of impacts at the state level suggests that the mandates' largest effects may have been on local school districts, and that they were likely not correlated with other major shifts in state policy. 

\begin{figure}[H]
\caption{Mandate effects on school district expenditures and revenues}
\centering
    \begin{subfigure}{.45\textwidth}
        \caption{Education expenditures (in thousand 1990 dollars per student)}
        \centering
        \includegraphics[width=\textwidth]{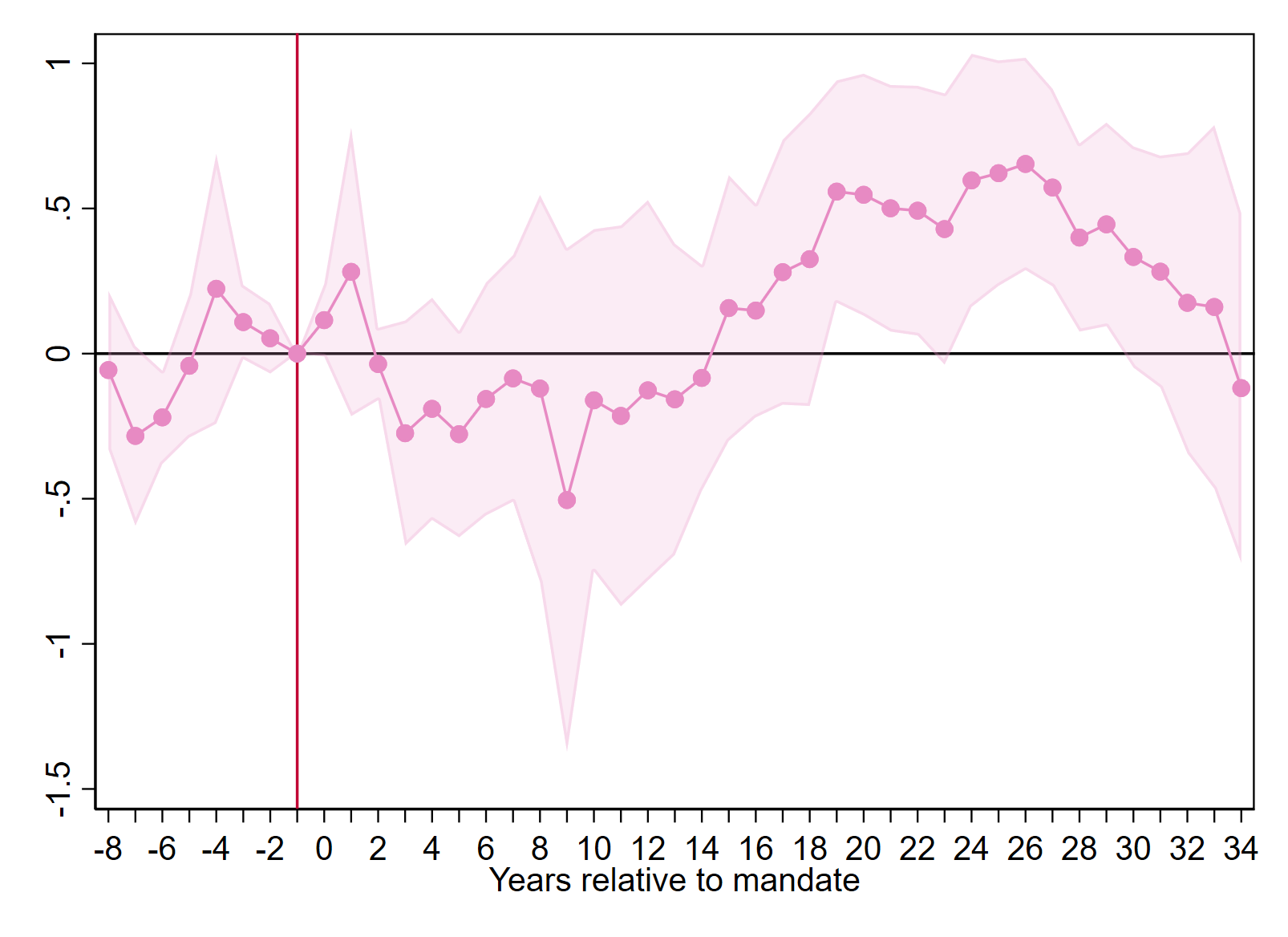}
    \end{subfigure}
    \hfill
    \begin{subfigure}{.45\textwidth}
        \caption{Tax revenues (in thousand 1990 dollars per student)}
        \centering
        \includegraphics[width=\textwidth]{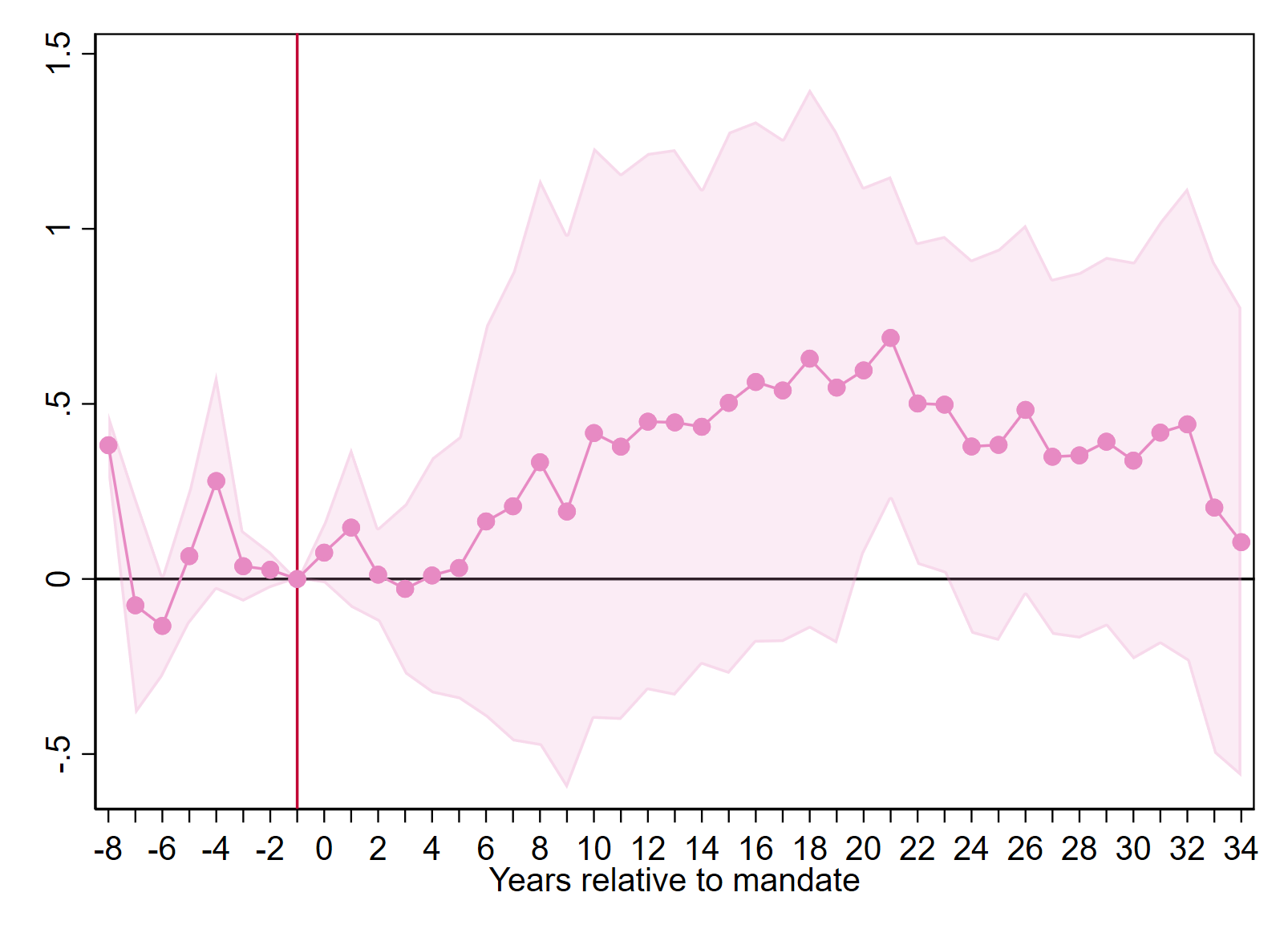}
    \end{subfigure}

    \begin{subfigure}[b]{.45\textwidth}
        \caption{Education expenditures (in thousand 1990 dollars per student), weighted by 1967 enrollment}
        \centering
        \includegraphics[width=\textwidth]{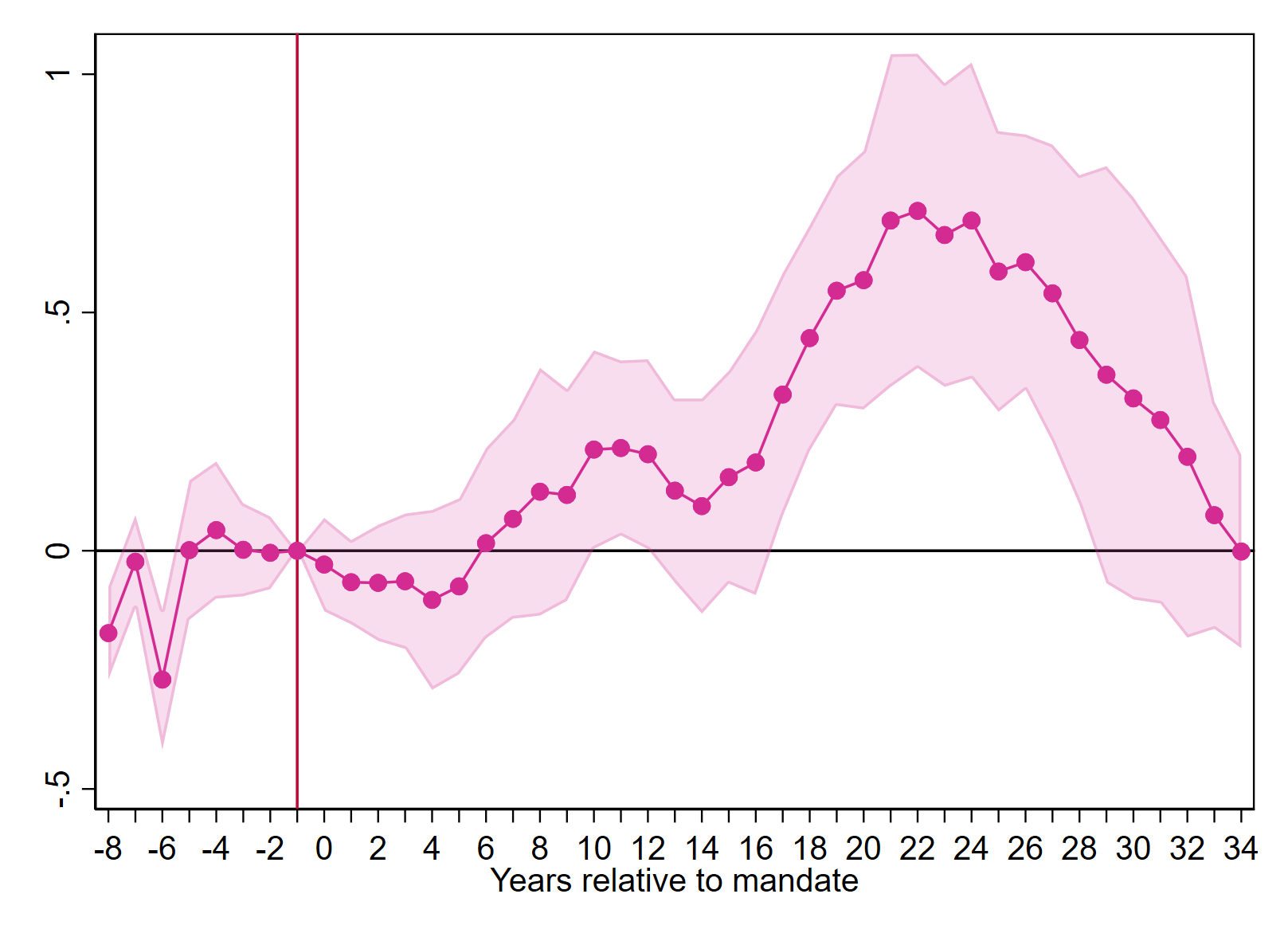}
    \end{subfigure}
    \hfill
    \begin{subfigure}[b]{.45\textwidth}
        \caption{Tax revenues (in thousand 1990 dollars per student), weighted by 1967 enrollment}
        \centering
        \includegraphics[width=\textwidth]{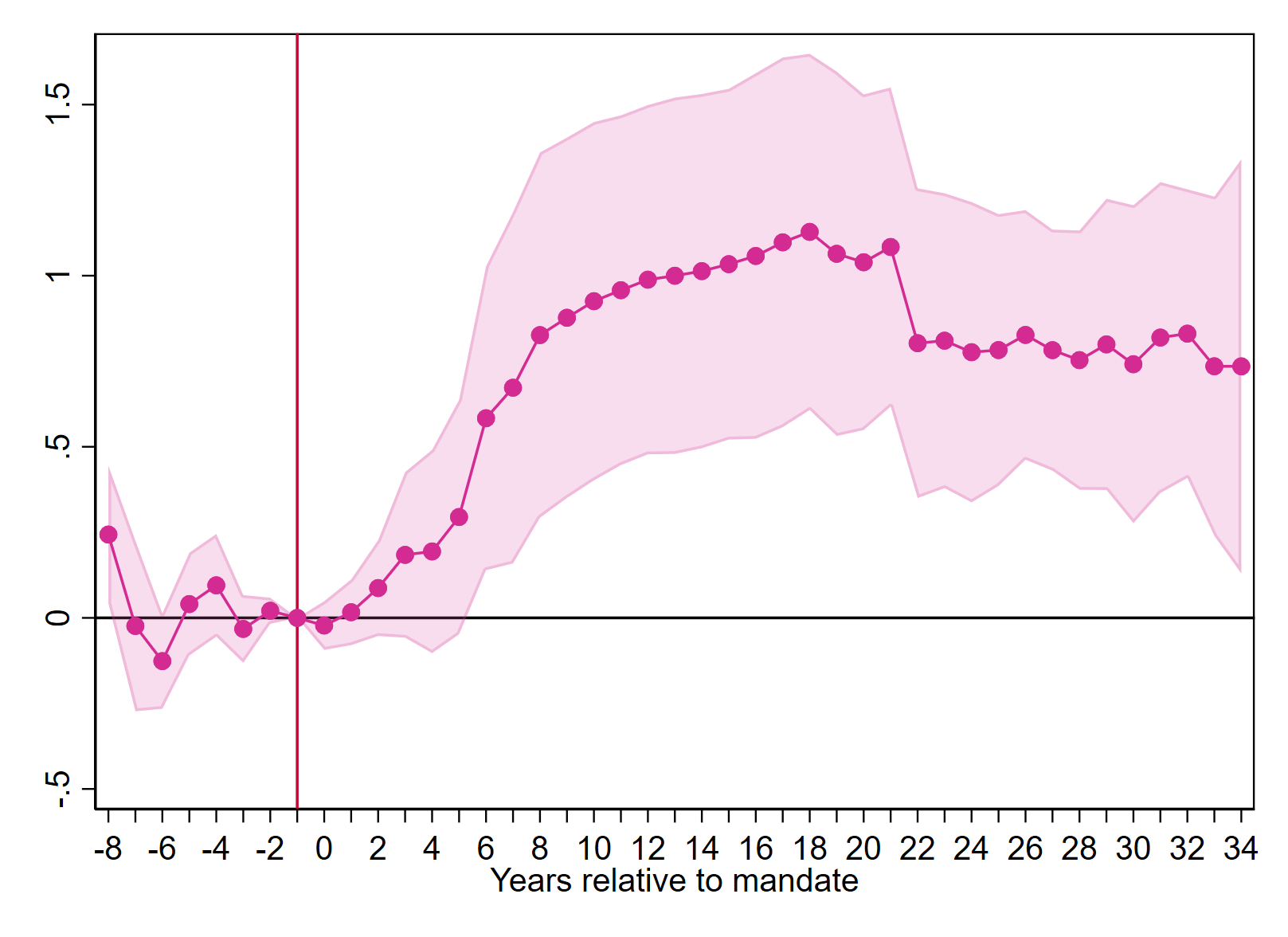}
    \end{subfigure}
        \label{fig:indfin_exp_rev}  
        \caption*{\footnotesize Note: 95\% confidence intervals shown, standard errors clustered at the state level. The figure plots difference-in-difference event study estimates of the impact of the mandates on school district spending per student and tax revenues per student, in thousands of constant 1990 dollars.}
\end{figure}

When looking at impacts on local school district revenues and expenditures, funding per student held steady even as enrollments increased, and increased in the long term. These effects can be seen clearly in Figure \ref{fig:indfin_exp_rev}, which presents the event study plots studying expenditures per student and tax revenues per student at the school district level. These plots highlight that there is little concerning evidence of pre-trends. On average across school districts, education spending per student held constant over the first few years following the mandates, and increased in the long term.

The magnitudes of these increases in expenditures and revenues per student are large, up to 15\% in the long term. Table \ref{tab:indfin_dist_exp_rev} quantifies the size of these impacts. Averaging all post-period observations, column (1) shows that the mandates increased education expenditures by about \$191 per student in 1990 dollars. This effect is not significant but is quite large, representing a 6.7\% increase over the pre-mandate average. Although there is little evidence of an increase in the first 10 years following the mandates, this occurs in the context of increasing school enrollments, as documented above. This means that even constant spending per student represents an increase in total expenditures. Expenditures per student increase by more in the long term, with spending per student increasing by \$464.

\begin{table}[thb]
    \centering
    \caption{Financial effects of mandates on school districts}
    \label{tab:indfin_dist_exp_rev}
        \resizebox{\textwidth}{!}{%
    \begin{tabular}{lccccccc} 
        \hline \hline 
                    &\multicolumn{1}{c}{(1)}   &\multicolumn{1}{c}{(2)}   &\multicolumn{1}{c}{(3)}   &\multicolumn{1}{c}{(4)}   &\multicolumn{1}{c}{(5)}   &\multicolumn{1}{c}{(6)}   &\multicolumn{1}{c}{(7)}   \\
            &Education exp.   &\makecell{Education exp.\\(weighted)}   &Capital outlay   &Operating exp.   &    Tax rev.   &\makecell{Tax rev.\\(weighted)}   &Intergovernmental rev.   \\
\hline
Pre-period average&      -0.124*  &      -0.105***&      -0.080** &      -0.039   &       0.026   &      -0.003   &      -0.112** \\
            &     (0.068)   &     (0.037)   &     (0.039)   &     (0.047)   &     (0.039)   &     (0.036)   &     (0.048)   \\
Post-period average&       0.234   &       0.359***&      -0.025   &       0.206   &       0.381*  &       0.820***&      -0.133   \\
            &     (0.156)   &     (0.099)   &     (0.040)   &     (0.154)   &     (0.231)   &     (0.197)   &     (0.145)   \\
Callaway \& Sant'Anna average&       0.191   &       0.287***&      -0.022   &       0.168   &       0.404*  &       0.770***&      -0.199   \\
            &     (0.154)   &     (0.080)   &     (0.036)   &     (0.153)   &     (0.243)   &     (0.184)   &     (0.141)   \\
Years 0-9 average&      -0.056   &      -0.008   &      -0.009   &      -0.056   &       0.156   &       0.371***&      -0.255*  \\
            &     (0.117)   &     (0.070)   &     (0.038)   &     (0.092)   &     (0.197)   &     (0.143)   &     (0.140)   \\
Years 10-19 average&       0.091   &       0.209***&      -0.031   &       0.106   &       0.491   &       0.855***&      -0.369   \\
            &     (0.178)   &     (0.080)   &     (0.031)   &     (0.151)   &     (0.320)   &     (0.220)   &     (0.255)   \\
Years 20+ average&       0.464** &       0.623***&      -0.027   &       0.397** &       0.391*  &       0.955***&       0.108   \\
            &     (0.188)   &     (0.171)   &     (0.094)   &     (0.198)   &     (0.234)   &     (0.255)   &     (0.229)   \\
\hline
Observations&      402721   &      402721   &      402787   &      402733   &      402772   &      402772   &      402752   \\
Pre-mandate mean&        2.83   &        2.83   &        0.30   &        2.47   &        1.42   &        1.42   &        1.31   \\
 \\ 
        \hline
    \end{tabular} 
            } \\
    \caption*{\footnotesize Note: All variables are expressed in thousands of dollars per enrolled student, in constant 1990 dollars. That is, a mean of 1 indicates 1000 1990 dollars per student. The table shows difference-in-difference estimates of the impacts of the mandates on expenditures and revenues using the school district finance data. Column (1) shows effects on education spending per enrolled student, column (2) the same with results weighted by school district size in 1967, column (3) on capital outlays per student, column (4) on operating expenditures per student, column (5) on tax revenues per student, column (6) the same with results weighted by school district size in 1967, and column (7) on intergovernmental transfers per student, that is, transfers received from federal, state, or local governments. Pre-period average and post-period average refer to a simple average of event-study coefficients before and after the implementation of a mandate, respectively. Callaway \& Sant'Anna average refers to a weighted average of estimated impacts, with weights given by the share belonging to each treated cohort in the sample. Years 0-9 average, Years 10-19 average, and Years 20+ average refer to simple averages of event-study coefficients for those years. Standard errors clustered at the state level shown in parentheses. \\
    * p \textless 0.1, ** p \textless 0.05, *** p \textless 0.01

    }
\end{table}

The estimated impact on education expenditures appears more quickly when school districts are weighted according to their size. The results in column (2) of Table \ref{tab:indfin_dist_exp_rev} are weighted according to a school district's population in 1967, the first year of data. By giving more weight to school districts with more students, these results give a better picture of the experience of the average student rather than the average school district. The results indicate large and significant increases in school spending beginning as soon as 10 years following the mandate, with a long term increase of \$623 per student, or about 22\%. Although there is a significant estimate for the pre-period average coefficient, the event study plot which appears in Figure \ref{fig:indfin_exp_rev} highlights that this is largely driven by one pre-period estimate, rather than indicative of a concerning pre-trend.

These increases in expenditures were largely driven by operating expenditures rather than capital outlays. Columns (3) and (4) of Table \ref{tab:indfin_dist_exp_rev} split expenditures into capital outlays such as construction of schools or facilities and operating expenses. There is no significant increase in capital outlays and, instead, increases in spending are largely driven by operating expenditures. 

To pay for these additional costs, school districts began to raise revenues via increased taxes within the first few years following the mandates, with increases remaining high in the long term. Column (5) of Table \ref{tab:indfin_dist_exp_rev} shows that tax revenues rose within 10 years of the mandates and remained high in the long term, with school districts raising, on average, an additional \$391 per student in the long term, a 28\% increase over the pre-mandate mean of \$1420 per student, although somewhat imprecisely estimated. The timing of this increase is shown in Figure \ref{fig:indfin_exp_rev}, which plots the corresponding event study coefficients. As with expenditures, the increases occur more quickly and are larger when results are weighted by school district enrollment (column (6)). 

Meanwhile, column (7) of Table \ref{tab:indfin_dist_exp_rev} shows no increase in the funds provided by other governments, mainly state governments. This is consistent with states shifting responsibility for the education of disabled students onto local districts. 

\begin{figure}[htb]
\caption{Mandate effects on expenditures and revenues, by 1967 district size}
\label{fig:indfin_exp_rev_het_size}
    \begin{subfigure}{.45\textwidth}
        \caption{Education expenditures (in thousand 1990 dollars per student)}
        \centering
        \includegraphics[width=\textwidth]{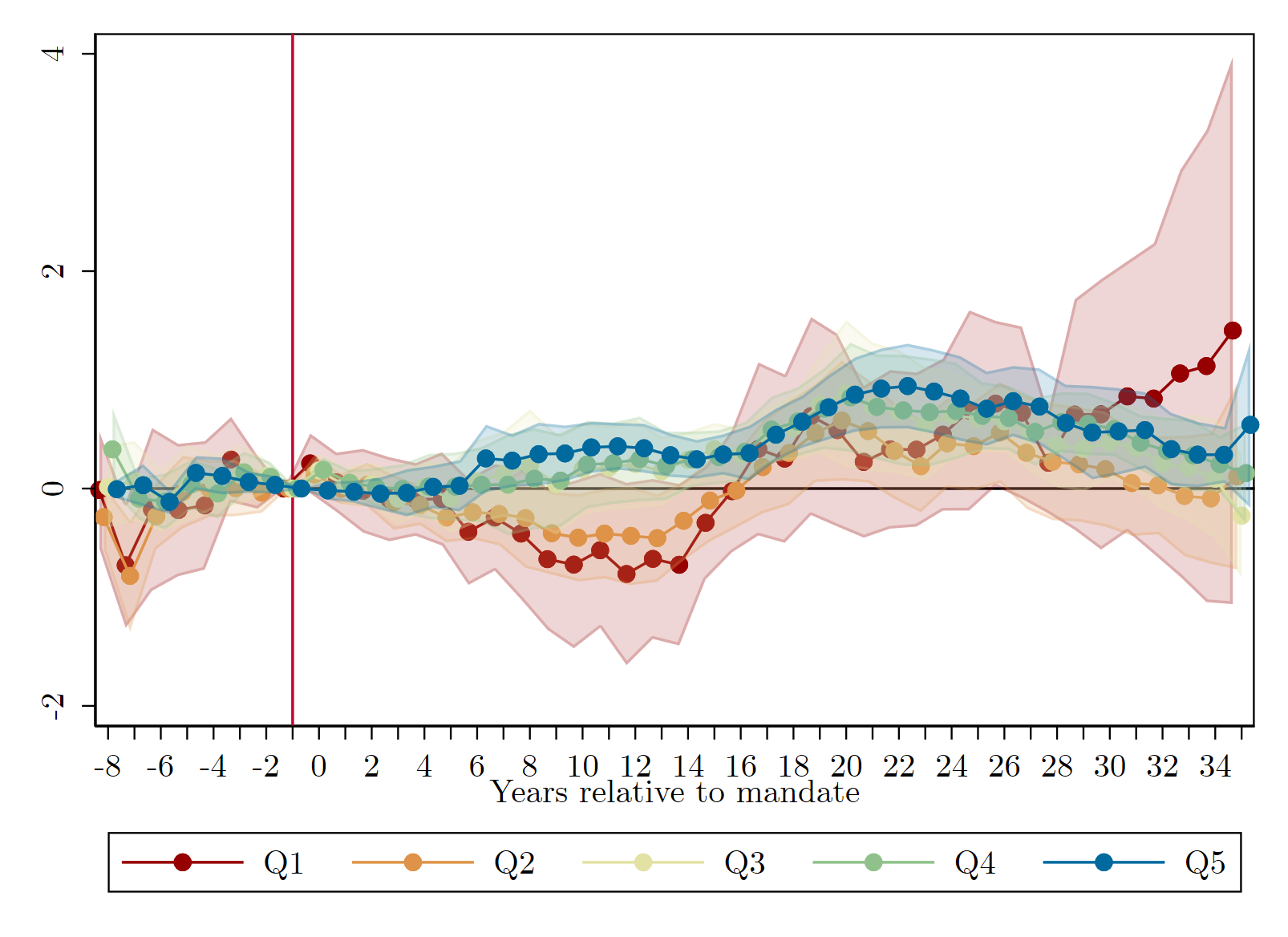}
    \end{subfigure}
    \begin{subfigure}{.45\textwidth}
        \caption{Tax revenues (in thousand 1990 dollars per student)}
        \centering
        \includegraphics[width=\textwidth]{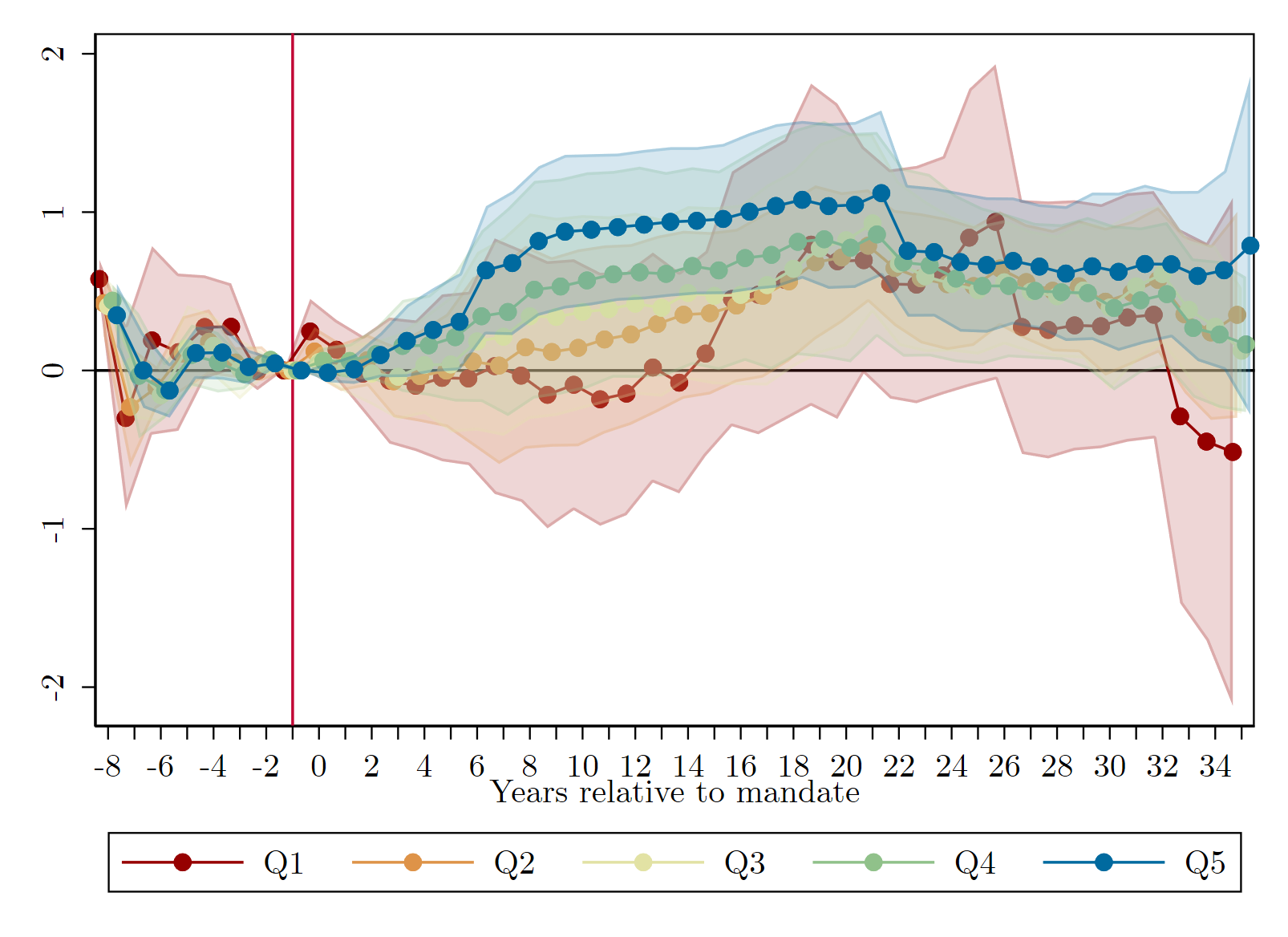}
    \end{subfigure}
        \caption*{\footnotesize Note: 95\% confidence intervals shown, standard errors clustered at the state level. The figure plots difference-in-difference event study estimates of the impact of the mandates on school district spending per student and tax revenues per student, in thousands of constant 1990 dollars. The sample is split according to the quintile of enrollment in 1967, with Q1 representing the smallest districts and Q5 representing the largest.}
\end{figure}

As suggested by the difference between the weighted and unweighted estimates, this analysis masks substantial heterogeneity in school districts' responses. Figure \ref{fig:indfin_exp_rev_het_size} shows the impacts on expenditures and revenues according to school district size, as determined by the quintile of their enrollment in 1967, the first year available in the data. The figure plots the event study estimates showing the impacts on school districts' education spending and tax revenues, split by quintile of school district size. In the short term, smaller school districts (ie, those in Q1 and Q2) struggled to increase property taxes and education spending per student fell. On the other hand, large school districts (ie, those in Q4 and Q5) were able to increase both spending and taxes in the short term, as quickly as 5 or 6 years after the mandates. In the long term, the gaps according to district size were largely closed. This is consistent with the fact that small school districts are likely to be rural school districts which have lower property values and thus more difficulty raising local funds \parencite{gutierrezSmallSparseDefining2023}. Disparities between school districts were emphasized in a 1978 \textit{Washington Post} article describing Virginia's mandate, with more urban areas in Northern Virginia leading implementation and other areas lagging behind \parencite{boodmanHandicapSchoolingState1978}.

\begin{figure}[hbpt]
\caption{Mandate effects on expenditures and revenues, by 1967 district spending}
\label{fig:indfin_exp_rev_het_exp}
    \begin{subfigure}{.45\textwidth}
        \caption{Education expenditures (in thousand 1990 dollars per student)}
        \centering
        \includegraphics[width=\textwidth]{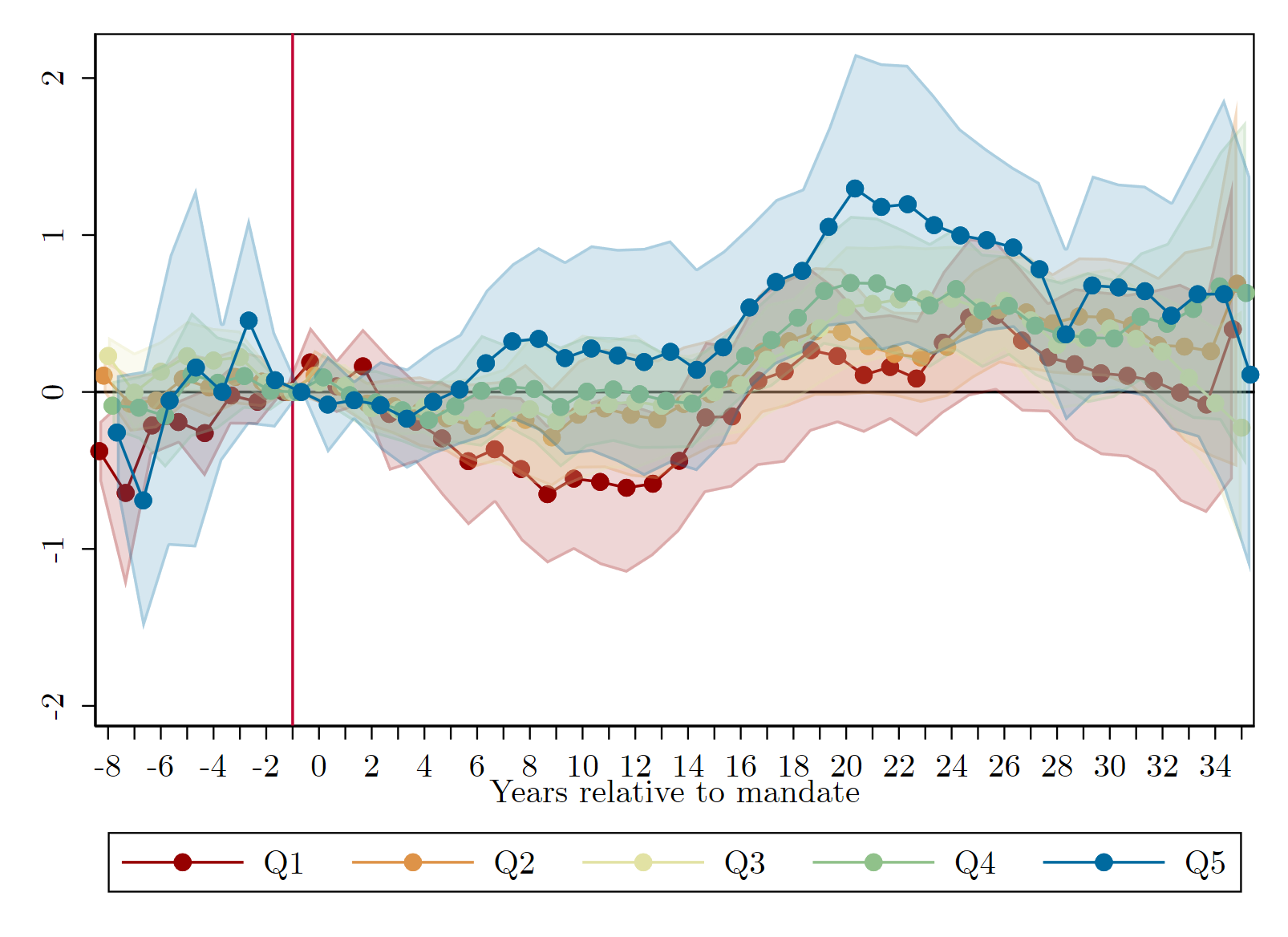}
    \end{subfigure}
    \begin{subfigure}{.45\textwidth}
        \caption{Tax revenues (in thousand 1990 dollars per student)}
        \centering
        \includegraphics[width=\textwidth]{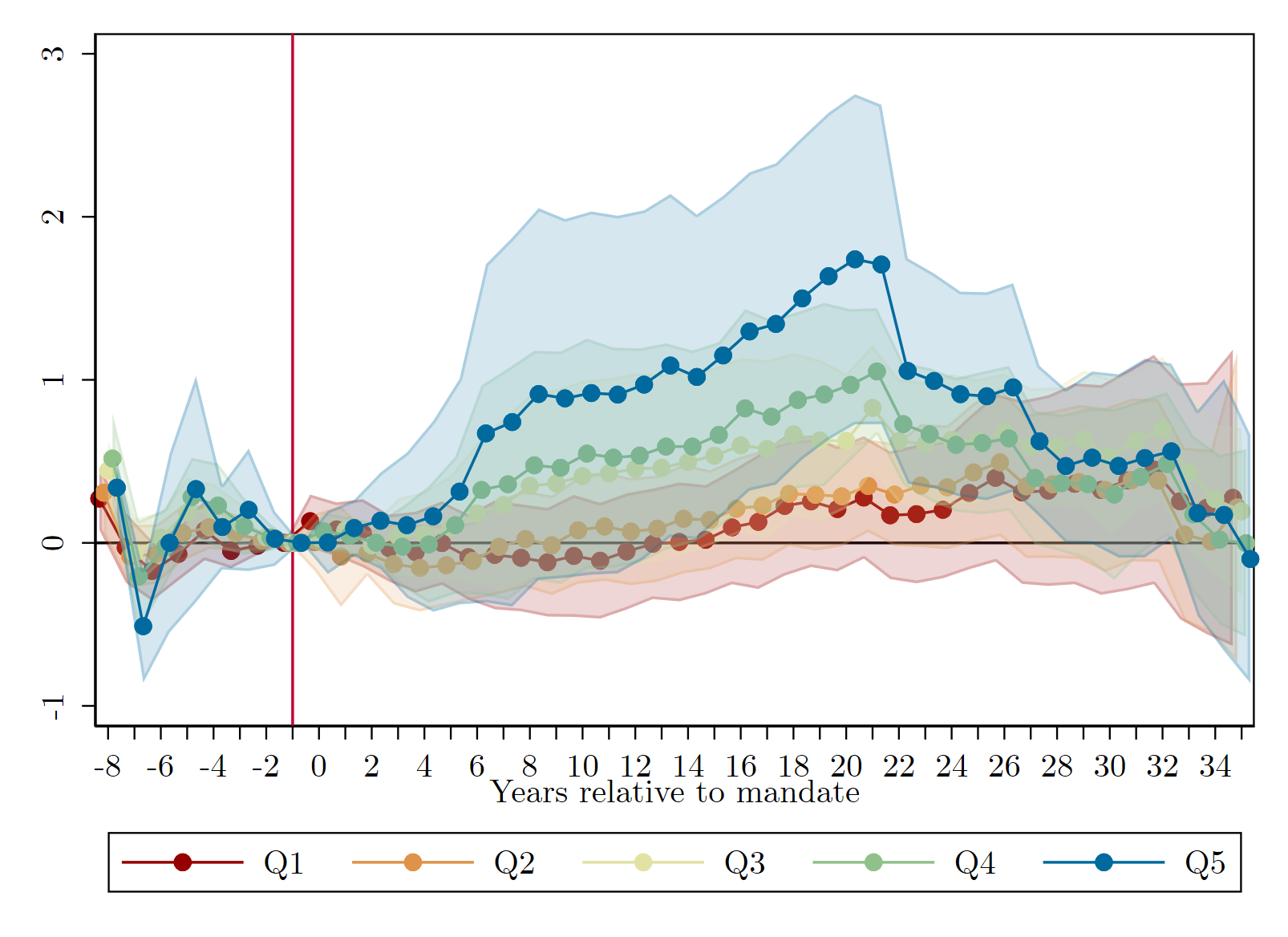}
    \end{subfigure}
        \caption*{\footnotesize Note: 95\% confidence intervals shown, standard errors clustered at the state level. The figure plots difference-in-difference event study estimates of the impact of the mandates on school district spending per student and tax revenues per student, in thousands of constant 1990 dollars. The sample is split according to the quintile of expenditures per student in 1967, with Q1 representing the lowest expenditure districts and Q5 representing the highest.}
\end{figure}

The patterns are similar for the gaps between more and less wealthy school districts, as shown in Figure \ref{fig:indfin_exp_rev_het_exp}, which plots event study estimates according to quintile of district spending per student in 1967. In this sense, the mandates may have emphasized existing inequalities between school districts in terms of spending.

How large are these increases in expenditures relative to the costs of providing services to disabled students? Table \ref{tab:cost_estimates} compares the increased funding with multiple estimates of the marginal cost of providing services to disabled students under the mandates. In each column, the cost of providing services is compared with the estimated increase in long-term funding. The costs are drawn from two contemporary sources, which both estimate the added cost to a school of providing services to a disabled student \parencite{hartmanEstimatingCostsEducating1981,chaikindSpecialEducationCosts1990}. These sources provide costs averaged across type of disability, based on calculations from actual school spending in this period. My calculations use \textcite{hartmanEstimatingCostsEducating1981} to assume 10.1\% of students were disabled. 

The marginal cost of the mandates depends on the increase in the share of students receiving services caused by the mandate. Table \ref{tab:cost_estimates} presents three scenarios. The first uses the marginal increase in services provided from the estimates generated in Section \ref{section:nhes_education} using the NHES data (Table \ref{tab:nhes_rsce}). However, a possible concern with this scenario is that the number of children receiving services may have increased over time but the NHES results offer a picture only in the short term. The second scenario takes a more extreme view of marginal increase by using the dataset compiled in Section \ref{section:num_served} to note that, averaging across states, the share of children receiving services increased from 1.6\% in 1957 to 8.6\% in 1990. Assuming that rates of service provision were otherwise at least not decreasing (and, potentially increasing for unrelated reasons), the maximum impact of the mandates is an increase of 7 percentage points overall (that is, 69.3 percentage points among disabled individuals). Finally, the third scenario is the most extreme, assuming that the full marginal cost of providing services for all 10.1\% of students assumed to be disabled was caused by the mandates. 

\begin{table}[h]
    \centering
    \caption{Cost estimates}
    \label{tab:cost_estimates}
        \resizebox{\textwidth}{!}{%
    \begin{tabular}{lccccc} 
        \hline \hline 
                                                                                                     & \multicolumn{1}{c}{(1)}                & \multicolumn{1}{c}{(2)}                   & \multicolumn{1}{c}{(3)}                \\
                                                                                             & \multicolumn{1}{c}{\makecell{NHES \\ marginal cost}} & \multicolumn{1}{c}{\makecell{Bounded \\ marginal cost}} & \multicolumn{1}{c}{\makecell{Full \\marginal cost}} \\ \hline
\multicolumn{4}{l}{\textbf{Using costs as in  \textcite{hartmanEstimatingCostsEducating1981}}}	 \\
\; Marginal number of disabled students receiving services	& 18.5 pp &	69.3 pp	& 100 pp \\
\; Estimated cost of learning support per recipient & 	\$3362	& \$3362	& \$3362 \\
 \quad   Marginal cost per student due to mandate &	 \$61.6	& \$230.7 &	\$332.9 \\
 \multicolumn{4}{l}{\;\underline{Individual born 10 years after the mandate:}}	 \\
 \quad Estimated spillover available & 278.1 & 109.0 & 6.8 \\
 \quad Implied increase in years of education & 0.27 & 0.10 & 0.01 \\
 \quad Percent of effect explained by funding & 112\% & 44\% & 3\% \\
\multicolumn{4}{l}{\;\underline{Individual born 14+ years after the mandate:}}	 \\
 \quad Estimated spillover available & 402.4 & 233.3 & 131.1 \\
 \quad Implied increase in years of education & 0.38 & 0.22 & 0.13 \\
\hline \hline \\

\multicolumn{4}{l}{\textbf{Using costs as in  \textcite{chaikindSpecialEducationCosts1990}}}			\\
\; Marginal number of disabled students receiving services &	18.5 pp & 	69.3 pp &	100 pp \\
\; Estimated cost of learning support per recipient	& \$4350 & \$4350 & \$4350 \\
 \quad  Marginal cost per student due to mandates & \$79.7 & \$298.5 & \$430.7 \\
 \multicolumn{4}{l}{\;\underline{Individual born 10 years after the mandate:}}	 \\
 \quad Estimated spillover available & 260.0 & 41.2 & -91.0 \\
 \quad Implied increase in years of education & 0.25 & 0.04 & -0.09 \\
 \quad Percent of effect explained by funding & 105\% & 17\% & -37\% \\
\multicolumn{4}{l}{\;\underline{Individual born 14+ years after the mandate:}}	 \\
 \quad Estimated spillover available & 384.3 & 165.5 & 33.3 \\
 \quad Implied increase in years of education & 0.37 & 0.16 & 0.03 \\


 \\ 
        \hline \hline
    \end{tabular} 
            } \\
        \caption*{\footnotesize Note: This table contains back-of-the-envelope calculations of the costs of providing services per student, increased funding per student, and the potential funding spillover for non-disabled students. Using estimates of the costs of providing services to disabled students from \textcite{hartmanEstimatingCostsEducating1981} and \textcite{chaikindSpecialEducationCosts1990}, the potential expenditure spillover for non-disabled students is estimated. Using the results in \textcite{jacksonEffectsSchoolSpending2016}, the implication of this funding for educational attainment is calculated and compared to the estimates of effects on educational attainment in the previous sections.}
\end{table}

The table shows that, in the long term (that is, 20+ years following the mandate), the increase in expenditures is in line with or exceeds the expected cost of providing services to disabled students in all three scenarios. In column (1), using the cost estimates from \textcite{hartmanEstimatingCostsEducating1981}, the provision of services for disabled students is assumed to cost \$3362 1990 dollars per recipient. Scaling this by the increase in services among disabled students and the share of disabled students in the population, the marginal cost per student due to the mandate is \$62 1990 dollars. 

In order to compare these revenues to these costs, the estimates from  Table \ref{tab:indfin_dist_exp_rev} showing the increase in expenditures per student are used. The estimates indicate an increase in expenditure per student of \$91 in years 10-19 after the mandates (insignificant) and \$464 in years 20+ after the mandates (significant). To align these with the timing of the education impacts, I consider the increased funding from the perspective of two potential individuals: one born 10 years following the mandate and one born 14+ years following the mandate (that is, whose entire school life occurs 20+ years after the mandate). For a student born 10 years after the mandate, averaging this funding increase over 12 years of education from age 6 to 18, and subtracting the marginal cost of services for disabled students, this creates an estimated spillover of \$278.1 1990 dollars per year. The spillover is assumed to be shared proportionally between disabled and non-disabled students. For a student born 14+ years after the mandate, the estimated spillover is larger due to the larger long-term increase in funding and would be \$402.4. As an alternative assessment of the costs of services for disabled students, using the larger cost estimates from \textcite{chaikindSpecialEducationCosts1990}, the spillovers are slightly smaller, \$260.0 and \$384.3, respectively. In columns (2) and (3), when assuming a larger marginal increase in the number of disabled students, the spillover is naturally smaller. However, there is a positive funding spillover for non-disabled students in nearly every scenario, except for that using the most extreme cost estimates. 

The size of this spillover can explain the magnitude of the increase in educational attainment of non-disabled students. The work in \textcite{jacksonEffectsSchoolSpending2016} suggests that, in a similar time period in the US, an increase in school spending of 10\% translated into a 0.27 increase in the years of education attained. This increase in educational attainment caused by funding can be compared to my estimated increase in educational attainment among non-disabled individuals under age 6 at the time of the mandates in Table \ref{tab:census_edu}, which was 0.24 years. Considering the perspective of an individual born 10 years after the mandate and the costs estimated by \textcite{hartmanEstimatingCostsEducating1981}, the increase in funding alone -- without considering, for example, peer effects between disabled and non-disabled students -- can explain 3\% to just over 100\% of the increase in educational attainment of the non-disabled students, depending on the scenario. In column (1), the scenario representing the largest spillover for non-disabled students, the increase in funding explains nearly exactly the estimated increase in educational attainment. However, as noted above, this scenario may underestimate the increase in service provision in the long term. In column (2), 17-44\% of the increase in educational attainment is explained, depending on which cost estimates are used. The most realistic scenario may be between these two, given that the NHES results may underestimate the costs of the mandates by capturing only the short-term increase in services while the bounded cost scenario may overestimate the costs by attributing the entire increase in services over this period to the mandates. 

When considering a student who received the long-term increase in funding for all school years, as a student born 14 or more years after the mandate would, the potential increase in education explained by funding is even larger. For these students, the implied increase in years of education may be up to 0.38 years, which is larger than my estimated increase of 0.24.

\FloatBarrier
\subsection{Impacts on employment in education industry}
 
To understand the drivers of the increased spending per student, this section considers employment in the education industry using data from the Current Population Survey 1968-1989. I find evidence that the mandates resulted in increased employment of teachers and other public-sector education workers. This is in line with \textcite{jacksonEffectsSchoolSpending2016}, who study the impacts of increased school spending on educational attainment and find that positive effects are largely driven by increased employment of teachers and increases in teacher salaries. 

To study impacts on employment of teachers, I use data from the Current Population Survey (CPS) from 1968-1989 from IPUMS \parencite{floodIntegratedPublicUse2022}.\footnote{The March/Annual Social \& Economic Supplement (ASEC) data are used because this is the only month of data harmonized by IPUMS for 1968-1976.} The data provided by IPUMS include industry and occupation codes that are consistent over time, allowing me to identify whether individuals are employed in the education industry and in the public or private sector. I consider a sample of adults age 25-40 who were at least 25 at the time of the mandate's enactment in their state of residence in order to capture effects for those whose education was not affected by the mandates.

\begin{figure}[htb]
\caption{Mandate effects on employment in the education industry}
    \begin{subfigure}{.45\textwidth}
        \caption{Public-sector education employment}
            \label{fig:employed_pub_edu}  
        \centering
    \end{subfigure}
    \begin{subfigure}{.45\textwidth}
        \caption{Private-sector education employment}
            \label{fig:employed_priv_edu}  

        \centering
    \end{subfigure}
        \includegraphics[width=\textwidth]{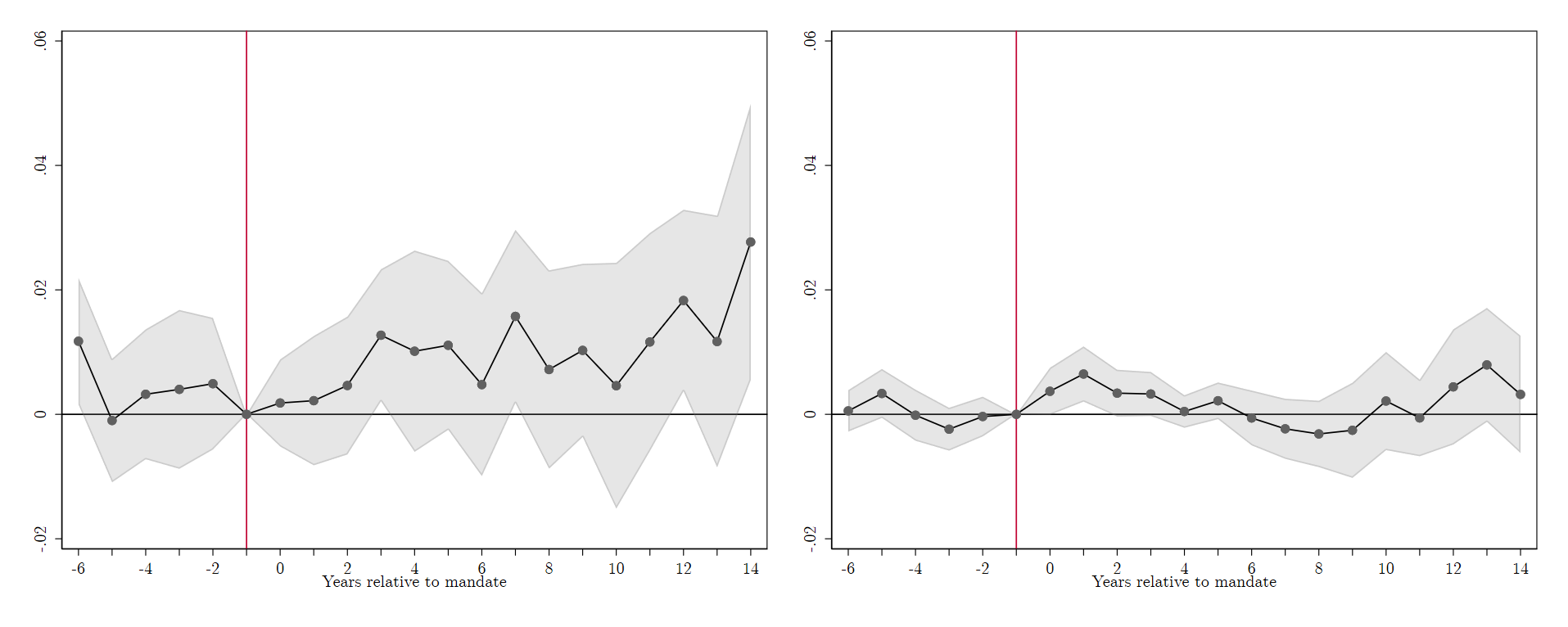}
        \caption*{\footnotesize Note: 95\% confidence intervals shown, standard errors clustered at the state level. The figure plots difference-in-difference event study estimates of the impact of the mandates on employment in public sector education and private sector education using the CPS data.}
\end{figure}

Following the mandates, employment in public education increased. Figure \ref{fig:employed_pub_edu} plots the event-study coefficients studying the probability that a given individual is employed in both the education industry and the public sector. The figure shows that, following the mandates, the probability of being employed in public education increased by about 1 percentage point over the first 15 years after the mandate's implementation, and up to 1.5 percentage points in the long term, relative to a pre-mandate mean of 5\%. Although the estimates are noisy, the effects are quite large. 

In contrast, there is no evidence of such an increase in employment of workers in education in the private sector. The null effects appear in Figure \ref{fig:employed_priv_edu}. This serves as a placebo test: if a concern is that a general expansion in education in these states is correlated with the mandates, then employment should expand in both the public and private sector.

\FloatBarrier

\section{Family and social impacts}

Given the large impacts of these laws, they likely also had impacts on the families of disabled individuals. In this section, I show evidence that the mandates increased the employment of mothers of disabled children and that they increased the likelihood of disabled individuals heading their own households and becoming parents.

\subsection{Impacts on parents' employment}

To the extent that parents, especially mothers, may have had to care for their children in the absence of suitable public schooling, and in line with the reduction in school absences, we should expect increases in parents' labor supply as a result of these laws. 

Returning to the NHES data used in Section \ref{section:nhes_education} and a study of the short-term impacts of the mandates, Table \ref{tab:nhes_parent} shows that the mandates caused large increases in the employment of mothers of children with disabilities. In column (1), effects on whether the mother reported that their main activity was employment range from 17.3-25.7 percentage points, depending on aggregation method used, relative to a mean of 26\%. In column (2), magnitudes are smaller and insignificant but still positive for the impact on the mother having any job (for example, doing housework as a main activity but working some of the time). At the same time, for non-disabled children, there is no evidence of any impact on mothers' employment. Columns (3) and (4) show that there is also little evidence of an impact on fathers' employment, although there is a pre-trend for disabled children. This pre-trend disappears when controlling for fathers' education (not shown). 

\begin{table}[htb]
    \centering
        \caption{Effects on parental employment}
    \label{tab:nhes_parent}
    \resizebox{\textwidth}{!}{%
    \begin{tabular}{lcccccccc}
    \hline \hline
                  &\multicolumn{1}{c}{(1)}   &\multicolumn{1}{c}{(2)}   &\multicolumn{1}{c}{(3)}   &\multicolumn{1}{c}{(4)}   \\
            &\makecell{Mother main act.\\ employed}   &Mother any job   &\makecell{Father main act. \\ employed}   &Father any job   \\
 \hline  \textbf{Disabled} & & & & & \\
Pre-period average&       0.021   &      -0.071   &      -0.072***&      -0.065***\\
            &     (0.050)   &     (0.104)   &     (0.020)   &     (0.022)   \\
Post-period average&       0.257***&       0.105   &       0.018   &       0.050   \\
            &     (0.093)   &     (0.138)   &     (0.065)   &     (0.065)   \\
Callaway \& Sant'Anna average&       0.173*  &       0.038   &      -0.020   &       0.021   \\
            &     (0.096)   &     (0.095)   &     (0.070)   &     (0.071)   \\
\hline
Observations&        1667   &        1667   &        1462   &        1462   \\
Pre-mandate mean&        0.26   &        0.37   &        0.92   &        0.94   \\
\hline\hline
\textbf{Non-disabled} & & & & & \\
Pre-period average&      -0.041   &      -0.019   &      -0.014   &       0.001   \\
            &     (0.031)   &     (0.028)   &     (0.015)   &     (0.017)   \\
Post-period average&       0.004   &       0.027   &       0.016   &       0.017** \\
            &     (0.052)   &     (0.053)   &     (0.010)   &     (0.007)   \\
Callaway \& Sant'Anna average&      -0.086   &      -0.056   &       0.010   &       0.017*  \\
            &     (0.065)   &     (0.073)   &     (0.012)   &     (0.009)   \\
\hline
Observations&       10411   &       10411   &        9386   &        9386   \\
Pre-mandate mean&        0.25   &        0.36   &        0.94   &        0.96   \\

    \\ \hline
    \end{tabular}
    }
    \caption*{\footnotesize Note: This table shows difference-in-difference estimates of the impacts of learning support mandates on parents' employment using the NHES data. Column (1) shows effects on the probability that a child's mother reports their main activity as employment, column (2) on whether the mother reports employment at all, column (3) on whether the child's father's main activity is employment, and column (4) on whether the father reports employment at all. Pre-period average and post-period average refer to a simple average of event-study coefficients before and after the implementation of a learning support mandate, respectively. Callaway \& Sant'Anna average refers to a weighted average of estimated impacts, with weights given by the share belonging to each treated cohort in the sample. Standard errors clustered at the state level shown in parentheses. \\
    * p \textless 0.1, ** p \textless 0.05, *** p \textless 0.01
    }
\end{table}

The magnitudes of these estimates are in line with previous literature examining the impact of public kindergarten and preschools on mothers' labor supply. Estimates from this literature suggest that, for single mothers with no other children younger than preschool age, eligibility for public kindergarten increased employment by 15-20 percentage points \parencite{fitzpatrickRevisingOurThinking2012}, with larger effects when considering take-up of the program \parencite{cascioMaternalLaborSupply2009}. Other authors estimate that enrollment in preschool or subsidized childcare increased mothers' employment by 1.8-7.7 percentage points \parencite{bakerUniversalChildCare2008,olivettiEconomicConsequencesFamily2017}. My estimates, which include both single and married mothers with and without other children, of a 10 percentage point increase in employment (although insignificant) are on par with these. Because effects are larger when considering whether employment was the mother's main activity (the closest available measure to full-time employment), my results also point to intensive-margin changes in the amount of work done by mothers of disabled children as a result of the mandates. 
 
\subsection{Impacts on households and parenthood}

An important motivation for the provision of educational services for disabled children was to provide them with the opportunity to live independent lives and to participate in their communities. The implementation of the mandates may have influenced social participation outcomes by enabling greater independence from one's parents or relatives, increased social interaction during school years, and changing social norms around disability. In this section, I find some evidence that the mandates increased the probability of disabled individuals heading their own households and becoming parents in adulthood.

\begin{figure}[htb]
\caption{Effects on social outcomes}
    \begin{subfigure}{.45\textwidth}
        \caption{Probability of heading household (men)}
        \centering
        \includegraphics[width=\textwidth]{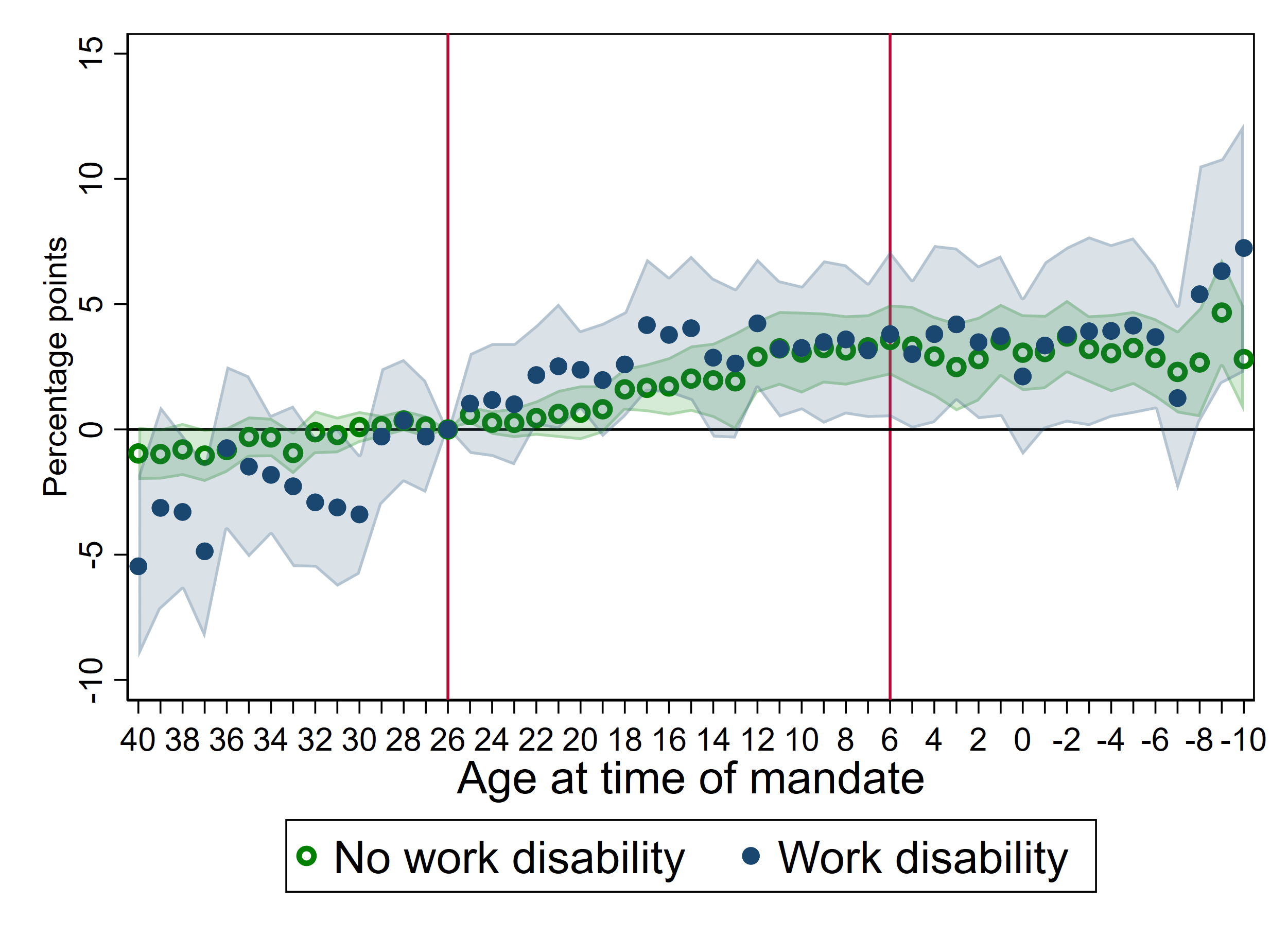}
    \end{subfigure}
    \begin{subfigure}{.45\textwidth}
        \caption{Probability of being a parent (age 30-40)}
        \centering
        \includegraphics[width=\textwidth]{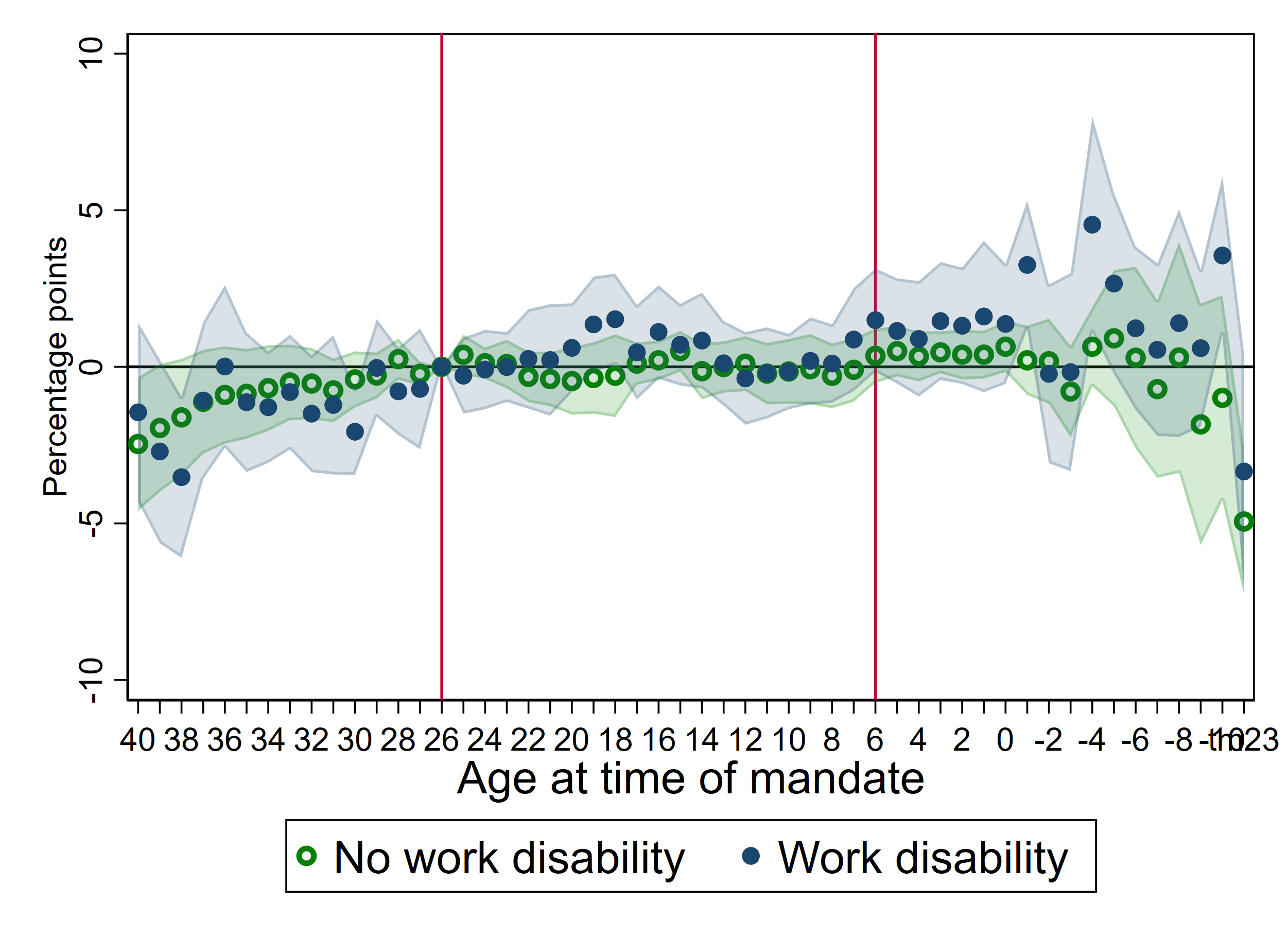}
    \end{subfigure}
        \label{fig:census_social}  
        \caption*{\footnotesize Note: 95\% confidence intervals shown, standard errors clustered at the state level. The figure plots difference-in-difference event-study estimates of the impact of the mandates on the probability of being a household head (among men age 25-35) and the probability of being a parent (at age 30-40), with the sample split between disabled and non-disabled individuals in the Census and ACS.}
\end{figure}

Consistent with increased education and labor force participation resulting in greater independence, I show that the mandates increased the probability of heading one's own household among both disabled and non-disabled individuals. Returning to the Census data and design used in Section \ref{section:census_education}, Figure \ref{fig:census_social} and columns (1) and (2) of Table \ref{tab:census_social} present impacts on the probability of being registered in the Census as a head of household. The probability of heading a household is studied overall in column (1) and for male respondents only in column (2) given the Census definitions of household heads.\footnote{Since 1980, the head of the household was determined to be ``any household member in whose name the property was owned or rented''. Before this, ie, in the 1970 Census, households with a married couple were determined to be headed only by a male respondent \parencite{rugglesIPUMSUSAVersion2024}.} Although there is some evidence of a pre-trend in this outcome (as measured by the pre-period average effect), for disabled people under age 6 at the time of the mandate, the mandates increased the probability of heading one's own household by 3.4pp overall and 3.1pp among men. For non-disabled people, also consistent with their increased education and employment, the probability of heading one's own household increased by 2.2pp and there is much less evidence of a concerning pre-trend. 

Considering later-in-life family outcomes, I show no impact on marriage rates for and a slight increase in the probability of disabled individuals becoming parents. I study parenthood at age 30-40 to attempt to better capture the age at which many individuals have already had children. Figure \ref{fig:census_social} and column (4) of Table \ref{tab:census_social} show that disabled people also became more likely to become parents by age 30-40 (1.5pp more likely for those under age 6 at the mandate's enactment). This is consistent with the mandates leading to greater independence, changing norms around parenthood and disability, and potentially greater prospects for the children of those with hereditary disabilities. Meanwhile, there is no effect on parenthood for non-disabled people. 

\begin{table}[htb]
    \centering
        \caption{Effects on social outcomes}
    \label{tab:census_social}
    \resizebox{.8\textwidth}{!}{%
    \begin{tabular}{lcccccccc}
    \hline \hline \\
                 &\multicolumn{1}{c}{(1)}   &\multicolumn{1}{c}{(2)}   &\multicolumn{1}{c}{(3)}   &\multicolumn{1}{c}{(4)}   \\
            &        Head   & Head (male)   &     Married   &      Parent   \\
 \hline  \textbf{Disabled} & & & & & \\
Over 25 average&      -0.017***&      -0.035***&      -0.014   &      -0.015*  \\
            &     (0.006)   &     (0.012)   &     (0.010)   &     (0.008)   \\
Under 25 average&       0.034***&       0.031***&       0.008   &       0.012** \\
            &     (0.009)   &     (0.011)   &     (0.007)   &     (0.005)   \\
Callaway \& Sant'Anna average&       0.034***&       0.032***&       0.006   &       0.007   \\
            &     (0.009)   &     (0.012)   &     (0.008)   &     (0.005)   \\
Age 6 to 25 average&       0.027***&       0.029***&       0.008   &       0.004   \\
            &     (0.008)   &     (0.010)   &     (0.008)   &     (0.004)   \\
Under age 6 average&       0.040***&       0.033** &       0.008   &       0.015*  \\
            &     (0.012)   &     (0.013)   &     (0.008)   &     (0.008)   \\
Age -10 to 5 average&       0.042***&       0.040***&       0.002   &       0.016*  \\
            &     (0.011)   &     (0.015)   &     (0.009)   &     (0.008)   \\
\hline
Observations&      481559   &      262627   &      481559   &      566624   \\
Pre-mandate mean&        0.45   &        0.67   &        0.62   &        0.68   \\
\hline\hline
 \\
		\textbf{Non-disabled} & & & & & \\
Over 25 average&      -0.001   &      -0.004   &      -0.014** &      -0.012*  \\
            &     (0.002)   &     (0.004)   &     (0.006)   &     (0.007)   \\
Under 25 average&       0.022***&       0.021***&      -0.003   &       0.000   \\
            &     (0.005)   &     (0.006)   &     (0.005)   &     (0.006)   \\
Callaway \& Sant'Anna average&       0.020***&       0.024***&       0.005   &       0.001   \\
            &     (0.004)   &     (0.006)   &     (0.004)   &     (0.004)   \\
Age 6 to 25 average&       0.013***&       0.018***&       0.008***&      -0.000   \\
            &     (0.003)   &     (0.005)   &     (0.003)   &     (0.004)   \\
Under age 6 average&       0.030***&       0.023***&      -0.012*  &       0.003   \\
            &     (0.007)   &     (0.007)   &     (0.007)   &     (0.008)   \\
Age -10 to 5 average&       0.032***&       0.031***&       0.002   &       0.001   \\
            &     (0.006)   &     (0.007)   &     (0.006)   &     (0.008)   \\
\hline
Observations&     7298493   &     3550424   &     7298493   &     7224379   \\
Pre-mandate mean&        0.48   &        0.87   &        0.83   &        0.84   \\
\\ \hline \hline 
    \end{tabular}
    }
    \caption*{\footnotesize Note: The table shows difference-in-difference estimates of the impacts of the mandates on household formation outcomes at age 25-35 using the Census and ACS data. Column (1) shows effects on whether the individual is a household head, column (2) on whether the individual is a household head among men, column (3) on whether the individual is married, and column (4) on whether the individual is a parent at age 30-40. Over 25 average and under 25 average refer to a simple average of event-study coefficients for individuals above and below age 25 at the time of a mandate, respectively. Callaway \& Sant'Anna average refers to a weighted average of estimated impacts, with weights given by the share belonging to each treated cohort in the sample. Age 6 to 25 average, under age 6 average, and age -10 to 5 average refer to simple averages of event-study coefficients for those ages at the time of the mandate's implementation. Standard errors clustered at the state level shown in parentheses. \\
* p \textless 0.1, ** p \textless 0.05, *** p \textless 0.01
    }
\end{table}

\FloatBarrier

\section{Cost-benefit analysis and marginal value of public funds}
\label{section:mvpf}
In this section, I quantify the monetary costs and benefits of the mandates and assess their cost effectiveness. I find that the monetary benefits of the mandate are large. Due to the large increases in income they generate, the mandates pay for themselves by raising more public revenues than they cost. Even so, I caveat this analysis by highlighting that the non-monetary benefits of the mandates are possibly multiple times larger than the monetary ones.

I assess the total costs and benefits of the mandates over in the long term using the marginal value of public funds (MVPF) framework as outlined by \textcite{hendrenUnifiedWelfareAnalysis2020}. The MVPF is given by the ratio of benefits received by beneficiaries of the mandates (their willingness to pay) to the net cost to the government in the long-term. This gives a sense of the value received for each dollar of public funds spent. I consider a hypothetical individual born in 1990, with present values discounted to the individual's birth at a rate of 3\% per year and an assumed marginal tax rate of 20\%, consistent with the parameters used by \textcite{hendrenUnifiedWelfareAnalysis2020}. 

To calculate total willingness to pay, I consider the benefits due to increased incomes for both disabled and non-disabled individuals, as well as the reduction in private tuition costs due to the shift from private to public education. 
I draw estimates from the adulthood impacts for those under age 6 at the time of the mandate as examined in section \ref{section:census_education}.

To quantify the increase in income for disabled individuals, I use the estimated small but insignificant increase in wage income for those under age 6 at the time of the mandate from Table \ref{tab:census_emp}, which is a 0.052 increase in IHS income (about 5.2\%). In Table \ref{tab:mvpf}, I transform this into levels using the pre-mandate mean wage income among disabled individuals and applying the assumed marginal tax rate of 20\%. The increase amounts to \$423 higher post-tax income per year, in 1990 dollars. Considering this income increase to be constant from age 25-67 and discounting to the individual's birth year, I find a lifetime value of this increased income of \$4,784 for each disabled individual. 

\begin{table}[htb]
    \centering
        \caption{Cost-benefit analysis}
    \label{tab:mvpf}
    \resizebox{\textwidth}{!}{%
    \begin{tabular}{lcccccccc}
    \hline \hline \\
     
&\multicolumn{1}{c}{(1)}   &\multicolumn{1}{c}{(2)}   &\multicolumn{1}{c}{(3)}   &\multicolumn{1}{c}{(4)}   &\multicolumn{1}{c}{(5)}     &\multicolumn{1}{c}{(6)} \\
 & \makecell{Base case} & \makecell{Best-case} & \makecell{Worst-case} & \makecell{More conservative\\ income} & \makecell{More conservative\\costs} & 	Disabled only \\
\hline
\textbf{Benefits to disabled} & & & & & & \\
\; Increased income & 0.052 & 0.156 & -0.052 & 0.048 & 0.052 & 0.052 \\
\quad Annual \$ amount, after tax & \$423 & \$1,272 & (\$422) & \$390 & \$423 & \$423 \\
\quad Lifetime PV & \$4,784 & \$14,394 & (\$4,773) & \$4,416 & \$4,784 & \$4,784 \\
\textbf{Benefits to non-disabled} & & & & & & \\
\; Increased income as adult (IHS points) & 0.268 & 0.364 & 0.172 & 0.179 & 0.268 & - \\
\quad Annual \$ amount, after tax & \$3,422 & \$4,695 & \$2,180 & \$2,271 & \$3,422 & - \\
\quad Lifetime PV & \$38,738 & \$53,152 & \$24,682 & \$25,703 & \$38,738 & - \\
\textbf{Benefits to all} & & & & & & \\
\; Reduced cost of private education & & & & & &\\
\quad Annual \$ amount, elementary & \$1,270 & \$1,270 & \$1,270 & \$1,270 & \$1,270 & \$1,270 \\
\quad Lifetime PV, elementary & \$186 & \$393 & (\$22) & \$186 & \$186 & \$3,003 \\
\quad Annual \$ amount, secondary & \$2,432 & \$2,432 & \$2,432 & \$2,432 & \$2,432 & \$2,432 \\
\quad Lifetime PV, secondary & \$74 & \$156 & (\$9) & \$74 & \$74 & \$5,750 \\
\quad Total lifetime PV & \$259 & \$549 & (\$31) & \$259 & \$259 & \$8,753 \\
\textbf{Total willingness to pay} & \$36,899 & \$51,306 & \$22,831 & \$24,647 & \$36,899 & \$13,537 \\
\hline
\textbf{Net cost to government} & & & & & & \\
\; Marginal cost per student & (\$464) & (\$96) & (\$832) & (\$464) & (\$623) & (\$464) \\
\quad Lifetime PV & (\$3,868) & (\$796) & (\$6,940) & (\$3,868) & (\$5,194) & (\$3,868) \\
\; Reduction in disability benefits & & & & & & \\
\quad Among disabled, IHS points & 0.280 & 0.337 & 0.223 & 0.280 & 0.280 & 0.280 \\
\quad Annual \$ amount & \$84 & \$102 & \$67 & \$84 & \$84 & \$84 \\
\quad Lifetime PV & \$59 & \$71 & \$47 & \$59 & \$59 & \$953 \\
\; Increased tax revenue, disabled & & & & & & \\
\quad Annual \$ amount & \$106 & \$318 & (\$105) & \$98 & \$106 & \$106 \\
\quad Lifetime PV & \$74 & \$222 & (\$74) & \$68 & \$74 & \$1,196 \\
\; Increased tax revenue, non-disabled & & & & & & - \\
\quad Annual \$ amount & \$856 & \$1,174 & \$545 & \$568 & \$856 & - \\
\quad Lifetime PV & \$9,086 & \$12,467 & \$5,789 & \$6,029 & \$9,086 & - \\
\textbf{Total net gov. budget} & \$5,351 & \$11,964 & (\$1,178) & \$2,288 & \$4,025 & (\$1,719) \\
\hline
\textbf{Net present value} & \$42,250 & \$63,270 & \$21,653 & \$26,935 & \$40,924 & \$11,818 \\
\textbf{MVPF} & $\infty$ & $\infty$ & $19.39$ & $\infty$ & $\infty$ & $7.87$ \\
\hline \hline

 
    \end{tabular}
    }
    \caption*{\footnotesize Note: This table presents a cost-benefit analysis of the mandates under various scenarios. The methodology is described in section \ref{section:mvpf}. Negative dollar amounts are shown in parentheses. 
    }
\end{table}

For non-disabled individuals, I find large lifetime benefits driven by increased incomes. Among this group, the results in Table \ref{tab:census_emp} showed an increase of 0.268 in IHS income for those under age 6 at the time of the mandate. Thus, for a non-disabled individual, the estimated annual post-tax increase in income is \$3,422 in 1990 dollars. This corresponds to a lifetime value of \$38,738 in 1990 dollars.

I also include the benefits to individuals due to reduced private schooling costs. I use the estimated impacts on private school enrollments from Table \ref{tab:cps_enroll}, which showed a 2.8pp shift from private to public school enrollments in the long term. To estimate the reduction in tuition costs due to this shift, I use the median private school tuition for elementary and secondary education in 1985, converted to 1990 dollars \parencite{williamsPrivateSchoolEnrollment1987}.\footnote{Although the hypothetical individual in this analysis is born in 1990, these are some of the closest available contemporary estimates of private school tuition.} I assume that individuals attend elementary school from age 6-13 and secondary school from age 14-15.7, corresponding to the average pre-mandate years of education among disabled individuals. On average across individuals, this translates to savings of \$259 per year. 

Summing up the benefits across these groups, and assuming that disabled individuals are 6.2\% of the population, as in the Census data, I find a total willingness to pay for the mandates of \$36,899 in 1990 dollars (equivalent to over \$90,000 in 2025). 

To calculate net costs from the perspective of the government (aggregated across all levels of government), I consider the increases in the cost of education, the reduction in disability benefit payments, and the increase in tax revenues coming from higher incomes. For the increased education cost, I use the long-run estimated expenditure impact of the mandate from Table \ref{tab:indfin_dist_exp_rev}, which is \$464 per student per year. Assuming 12 years of compulsory education, this amounts to a present value of \$3,868 per student. At the same time, the government experienced savings in the form of reduced disability benefits, the value of which fell by 0.28 IHS points among disabled individuals, giving a present value of \$59 per person overall. The government also received increased tax revenue due to the higher earnings of both disabled and non-disabled individuals. Once scaled by their shares in the population, the increase amounts to an average increase of \$74 per person due to the increase in income for disabled individuals and \$9,086 due to the increase for non-disabled individuals.    

These increased tax revenues exceed the cost of the mandates, that is, the mandates pay for themselves from the perspective of public funds. In fact, over an individual's lifetime, the benefits to government revenue are 2.4 times larger than the increased educational costs. In the terminology used in the MVPF framework, the MVPF is infinite. Beyond MVPF, even a simple cost-benefit analysis indicates that the program returns more than 9.5x in benefits to the recipients compared to its costs. Very high or infinite MVPFs are relatively common for interventions that affect children due to their large, lifelong benefits and increases in government tax revenues (eg, \textcite{hendrenUnifiedWelfareAnalysis2020, ganimianAugmentingStateCapacity2024a, hojmanPublicChildcareBenefits2022}).

To understand the robustness of this estimate, I follow the approach in \textcite{hendrenUnifiedWelfareAnalysis2020} to estimate 95\% confidence intervals for the MVPF. This is done by bootstrapping, drawing each estimate used in the calculation of the MVPF from an independent random normal distribution with mean given by its estimated value and standard deviation given by its standard error. I perform 10,000 bootstrap replications. I use two methods to construct confidence intervals. First, I consider the simple percentile confidence intervals, given by the 2.5th and 97.5th percentiles of the bootstrapped distribution of the MVPF. Second, given the non-normality of the distribution of the bootstrapped MVPFs, I use the bias-corrected confidence intervals given by \textcite{efronJackknifeBootstrapOther1982}. In both cases, the lower bound of the 95\% confidence intervals is still infinite. In other words, the MVPF is only less than infinite in approximately 1\% of bootstrap iterations; with about 99\% confidence, it is infinite.

I further test robustness of the MVPF to the estimated values by calculating it using the most extreme plausible values. First, I construct a best-case estimate by using the upper bound of the 95\% confidence interval for benefits and the lower bound of the 95\% confidence interval for costs to obtain the most favorable estimates of cost effectiveness implied by these measures of uncertainty. As shown in column (2) of Table \ref{tab:mvpf}, these estimates again imply an infinite MVPF, with the increase in government revenues 12x larger than the increase in costs. Next, in column (3), I use the other bound of the confidence intervals to construct the worst-case cost-effectiveness implied by these measures. Since the marginal costs for students are substantially higher and the marginal increase in incomes is smaller, the mandates no longer pay for themselves. However, the MVPF is still quite large, over 19, meaning that each dollar of public funds spent returned over \$19 of value to beneficiaries in the long term. In column (4), I use the smaller estimate of the increase in income among individuals age -10 to 5 at the time of the mandate from Table \ref{tab:census_emp}. Again, the MVPF is infinite. Finally, column (5) shows that the MVPF remains infinite when using the larger cost estimated when weighting school districts by pre-period enrollment in Table \ref{tab:indfin_dist_exp_rev}.

I also test robustness of the analysis to the choice of the discount rate. Since the benefits of mandates occur later in a person's life than the expenses, the MVPF will decrease with higher discount rates. Appendix Figure \ref{fig:mvpf_discount} shows that the MVPF remains infinite when assuming any discount rate lower than approximately 6.2\%. This discount rate is substantially higher than the discount rate of 3\% used by prior literature \parencite{hendrenUnifiedWelfareAnalysis2020} and the 2\% rate used by the federal Office of Management and Budget to prepare cost-benefit analyses for federal regulations \parencite{usombCircularNoA42023}. It is also higher than discount rates developed by the Congressional Budget Office based on Treasury yields over the next 30 years, which are expected to be no higher than 4.4\% \parencite{ashExploringEffectsMedicaid2023}. However, it is slightly lower than peak estimates of fair value discount rates based on investment yields over the next 30 years, which are expected to be as large as 7.3\% by 2054 \parencite{ashExploringEffectsMedicaid2023}. Using this largest estimate of the discount rate, the MVPF is 13.2.  

Given that a primary motivation of the mandates was to benefit disabled individuals, in column (6), I consider the beneficiaries to be only disabled individuals. In this case, I assume that the shift to public from private school is concentrated entirely among disabled individuals. Since the benefits are concentrated among a smaller group, the average lifetime present value of these savings is higher. Even so, when considering disabled beneficiaries, the mandates do not pay for themselves, because the estimated increase in income among this group is small. Still, the benefits are large relative to the costs, with an MVPF of 7.87. This analysis is necessarily limited by the limited income effects detectable for disabled people, which are limited by the definition of disability available in the Census. 

It should also be noted that this analysis quantifies only the measurable monetary benefits from these policies, which are likely only a small fraction of their overall benefits. In particular, policymakers should not conclude from this analysis that providing services for disabled students is only cost-effective when it generates large spillovers for non-disabled individuals. The benefits to disabled individuals of better functioning in everyday life, greater independence, and greater involvement in one's community are likely very large but difficult, if not impossible, to quantify. Some modern research hints at the quality of life improvements that these services can have: for example, a study of German children age 12-18 with dyslexia found that, following an individualized treatment for dyslexia, they experienced substantially lower levels of anxiety and depression and that their emotional wellbeing and relationships with friends and family improved \parencite{mollEconomicEvaluationDyslexia2023}. Further, higher educational attainment has been linked to large non-monetary benefits in the form of improved health and longevity \parencite{kruegerEconomicValueEducation2019}. It is likely that some of the largest impacts of the mandates were improvements in everyday life, like those pointed out by one \textit{Boston Globe} headline, referring to the mandate in Massachusetts: ``David, at 7, can crawl -- because of Ch. 766'' \parencite{cohenDavid7Can1975}. As another young person who received services under this mandate commented, ``I (Steven) would have died without it.'' \parencite{howardImpactChapter7661981}.

\section{Conclusion}
\begin{quote}
    Of a child in Massachusetts 3 years after its mandate: ``Chris has learned to read, add, subtract, multiply, divide, knit, and crochet. This is the boy who spoke not at all until he was nine.'' \parencite{cohenFocusSpecialChild1975}
\end{quote}

This paper has documented and analyzed the large, broad, and long-term positive impacts of the implementation of mandates for states to provide educational services for disabled students. These positive impacts include benefits for the disabled individuals impacted, as well as positive spillovers for their parents and non-disabled peers. This work provides new evidence on one of the largest education reforms in recent US history and some of the most comprehensive evidence on the long-term impacts of educational services for disabled students. 

My results show that the mandates drove rapid increases in the probability of being recommended to receive services for a disability and actually using them. These increases are quite large: I find an increase of 22.5-25.3 percentage points in the probability of disabled students being recommended to receive educational services (relative to a mean of 32\% before the mandates). I also find large increases of 18.3-18.5 percentage points in the probability of actually receiving these services, relative to a baseline mean of 16\% of disabled students receiving services in states with no mandate.  The magnitude of the increase in services in line with estimates of an increase in the number of students receiving services using state-level data from a newly-constructed series. The mandates also increased disabled students' probability of being transferred to ``special education'' classes, reduced the probability that they were frequently absent from school, and increased the probability of repeating a grade. Meanwhile, I find no strong effects of the mandates on these outcomes for non-disabled students.

As the scope and services offered by public education expanded, I show evidence that the mandates increased school enrollments. Consistent with both the mandates requiring preschool to be provided for disabled students and with students staying longer in school, these increases occurred both among individuals of preschool age and above the age of compulsory schooling (ie, age 16-20). The mandates also caused a substantial shift from private to public education.

Studying educational attainment in adulthood, I find that the mandates substantially increased total education attainment for disabled individuals. For those who were below school age at the time of the mandates' implementation, effects were as large as an additional 0.23 years of education. These increases largely occurred before the end of high school and reduced the probability of extremely poor education outcomes for disabled individuals. 

With more education, disabled individuals also became more likely to have some work experience in adulthood and less likely to receive Social Security disability benefits. Although I do not find any impact on their labor force participation and employment rate at age 25-35, I show that the mandates made disabled individuals 2.9 percentage points more likely to have some work experience in adulthood. Given that the Census disability measure available over this period identifies only individuals whose disabilities limit or prevent them from working, these impacts are likely an underestimate of the true positive impacts of the mandates among disabled individuals. 

Despite the concern that these mandates could detract resources from non-disabled students, I document positive spillovers for non-disabled individuals. Among those who were below school age when the mandates were implemented, I find evidence that educational attainment increased by 0.25 years, with marginal years of education occurring at higher levels than among disabled individuals. I also find that the mandates led to large increases in labor force participation, employment, and wage income for non-disabled individuals. 

To better understand these positive effects, I use data on state and school district finances to show that the mandates caused long-term increases in education spending per student at the school district level. Using estimates from prior literature estimating increases in educational attainment driven by increased spending per student \parencite{jacksonEffectsSchoolSpending2016}, I show that this increased funding can explain a substantial portion of the positive spillover for non-disabled students. I show evidence that the mandates increased employment in public education, suggesting that funding increases were spent at least in part on teachers and staff. 

The mandates also had important effects on the families and social experiences of disabled individuals. Using data from the short term, I show that the mandates increased the probability that mothers of disabled children reported that their main economic activity was employment (as opposed to housework). For disabled individuals who were young at the time of the mandates, the mandates increased the probability of heading their own households and becoming parents themselves in adulthood. 

Analyzing the monetary costs and benefits of these mandates, I find that, under a variety of scenarios, the mandates pay for themselves by raising more government revenue than they cost. Since the mandates improved employment outcomes, they generate large increases in government revenues over an individual's lifetime. Even so, these monetary benefits are likely only a small fraction of the non-monetary benefits experienced by disabled individuals due to improved education and opportunities. 

This paper provides detailed and novel evidence on the numerous large, long-term, and important effects of these mandates on education, employment, and social outcomes for disabled individuals, and the positive spillovers for their parents and peers. It highlights how a large expansion in the availability of public education for disabled students can have net positive effects, with increases in overall education resources playing an important role. The results show that spending on educational services for disabled students is highly cost-effective and may even pay for itself given improved labor market outcomes.

Although current debates about special education policy take place in the context of a much more developed special education system, these debates still rely on estimates of the benefits of providing services to disabled students. This work quantifies the benefits that expansions in the services offered to disabled students can have. This work also has relevance for a number of countries around the world which still do not require individualized educational services for disabled students \parencite{waisathDismantlingBarriersAdvancing2024}. This evidence on the positive effects of providing services to disabled students can inform policymakers designing requirements for and funding for these programs.

\printbibliography

\appendix
\renewcommand*\thetable{\Alph{section}.\arabic{table}}
\renewcommand*\thefigure{\Alph{section}.\arabic{figure}}

\section{Regression discontinuity analysis and balance tests}\label{appendix_rdrobust}
\setcounter{figure}{0}    
\setcounter{table}{0}    
\FloatBarrier

This section addresses tests of the suitability of the comparison states for the main difference-in-difference analysis. Further details appear in Section \ref{section:num_served}. 

To test whether control states experienced any changes around the time of their state mandates, Table \ref{tab:rdrobust} contains regression discontinuity estimates studying the jump in the number of students receiving services for a disability around the time of a mandate's implementation in their state. The analysis is implemented using the \textbf{rdrobust} Stata package \parencite{calonicoRdrobustSoftwareRegressiondiscontinuity2017} which implements a local linear regression discontinuity design. The table presents an estimate using conventional estimators, an estimate with bias-corrected estimators and conventional confidence intervals, and an estimate with bias-corrected estimators and robust confidence intervals as developed by \textcite{calonicoRobustNonparametricConfidence2014}. Optimal bandwidths are selected by the procedures documented in \textcite{calonicoRobustNonparametricConfidence2014, calonicoRobustDataDrivenInference2014, calonicoOptimalBandwidthChoice2020}. 

This regression discontinuity analysis shows that treated states experienced substantial increases in the number of students receiving services at the time of their mandate's implementation, while control states did not. Columns (1) and (2) of Table \ref{tab:rdrobust} test for evidence of a jump in the year the mandate was implemented. Column (1) contains the results for control states, that is, states that implemented a mandate in 1978 and later. Column (2) contains the results for treated states, that is, states that implemented a mandate before 1978. The results show no significant evidence of such a jump, although estimates are positive for treated states and smaller and negative for control states. However, mandates may have been implemented only for a partial year in their year if implementation (eg, if they began in September). Because of this, columns (3) and (4) test for evidence of a jump in the following year. These highlight a substantial jump in service provision in treated states at the time of the mandate's implementation and no evidence of such a jump for the control states. The jump corresponds to a 1.15-1.23 percentage point increase in the number of students receiving services for a disability. At the same time, columns (5) and (6) test for a discontinuity in the slope (that is, a kink) around the year of the mandate's implementation. They show that there is also a (marginally insignificant) increase in the slope at the time of a mandate's implementation in the treated states. Again, there is no such increase in the control states. 

Although this setting does not correspond neatly to a regression discontinuity design because the mandates' impacts are expected to play out for many years rather than only at the point of discontinuity, this helps to reaffirm the suitability of the control states as a control group by showing that the mandates had no immediate impact in these states.

\begin{table}[htb]
    \centering
        \caption{Evidence of discontinuity in services for a disability around mandates}
    \label{tab:rdrobust}
    \resizebox{\textwidth}{!}{%
    \begin{tabular}{lcccccccc}
    \hline \hline
                  &\multicolumn{2}{c}{\underline{Level, time period 0}}&\multicolumn{2}{c}{\underline{Level, time period 1}}&\multicolumn{2}{c}{\underline{Slope, time period 0}}\\
            &\multicolumn{1}{c}{(1)}   &\multicolumn{1}{c}{(2)}   &\multicolumn{1}{c}{(3)}   &\multicolumn{1}{c}{(4)}   &\multicolumn{1}{c}{(5)}   &\multicolumn{1}{c}{(6)}   \\
            &Control states   &Treated states   &Control states   &Treated states   &Control states   &Treated states   \\
\hline
Conventional&    -0.00183   &     0.00645   &     0.00124   &      0.0123***&    -0.00130   &     0.00432   \\
            &   (0.00794)   &   (0.00444)   &   (0.00655)   &   (0.00426)   &   (0.00589)   &   (0.00279)   \\
[1em]
Bias-corrected&    -0.00370   &     0.00540   &   -0.000299   &      0.0115***&    -0.00194   &     0.00546*  \\
            &   (0.00794)   &   (0.00444)   &   (0.00655)   &   (0.00426)   &   (0.00589)   &   (0.00279)   \\
[1em]
Robust      &    -0.00370   &     0.00540   &   -0.000299   &      0.0115** &    -0.00194   &     0.00546   \\
            &   (0.00950)   &   (0.00527)   &   (0.00794)   &   (0.00525)   &   (0.00844)   &   (0.00399)   \\

    \\ \hline
    \end{tabular}
    }
    \caption*{\raggedright \footnotesize Note: Standard errors shown in parentheses, \\ * p \textless 0.1, ** p \textless 0.05, *** p \textless 0.01 }
\end{table}

\begin{figure}[htb]
\caption{Correlations between mandate timing and state characteristics} \label{fig:balance_coefplot}
    \begin{subfigure}{.5\textwidth}
        \caption{Correlation with year of state's mandate}
        \centering
        \includegraphics[width=\textwidth]{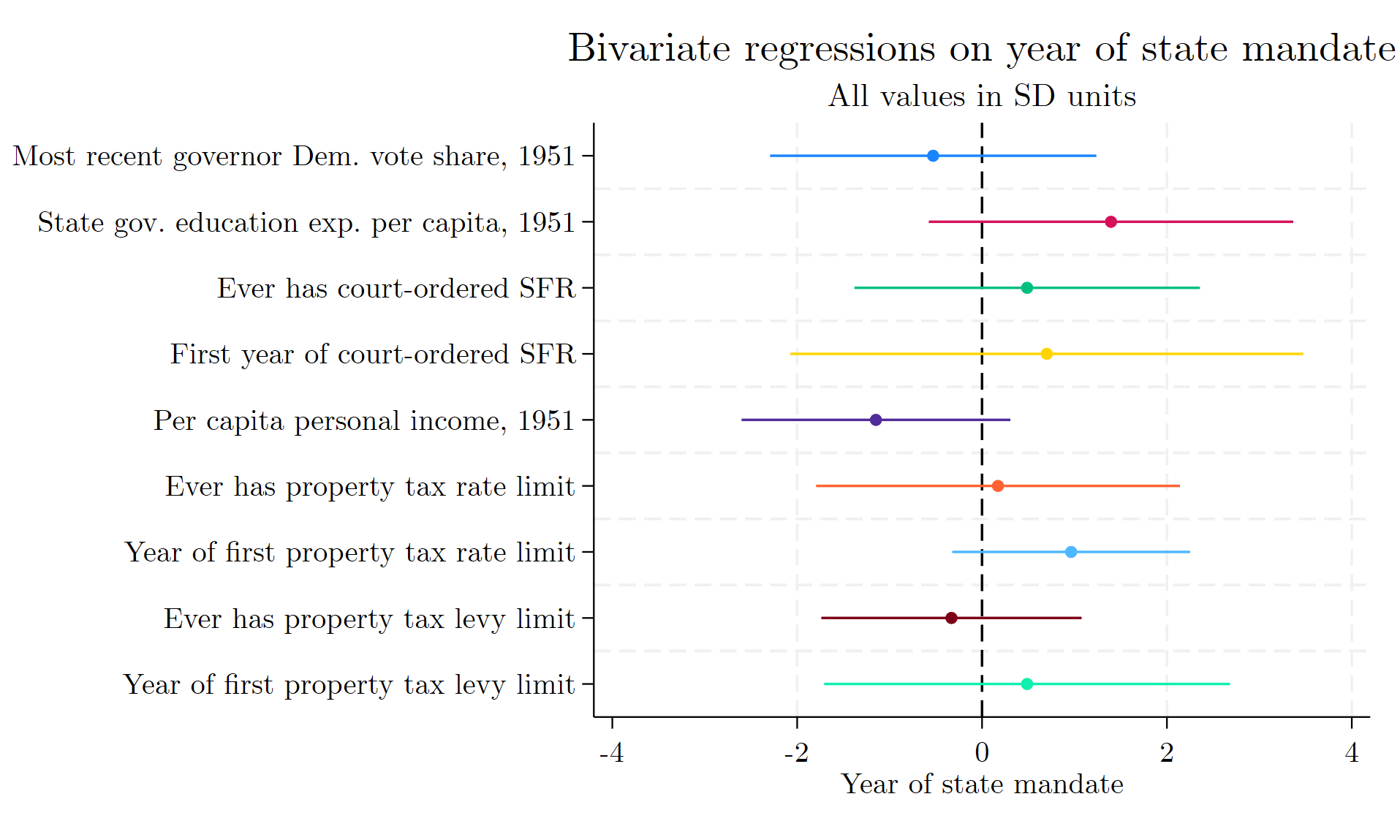}
    \end{subfigure}
    \begin{subfigure}{.5\textwidth}
        \caption{Correlation with being a control state}
        \centering
        \includegraphics[width=\textwidth]{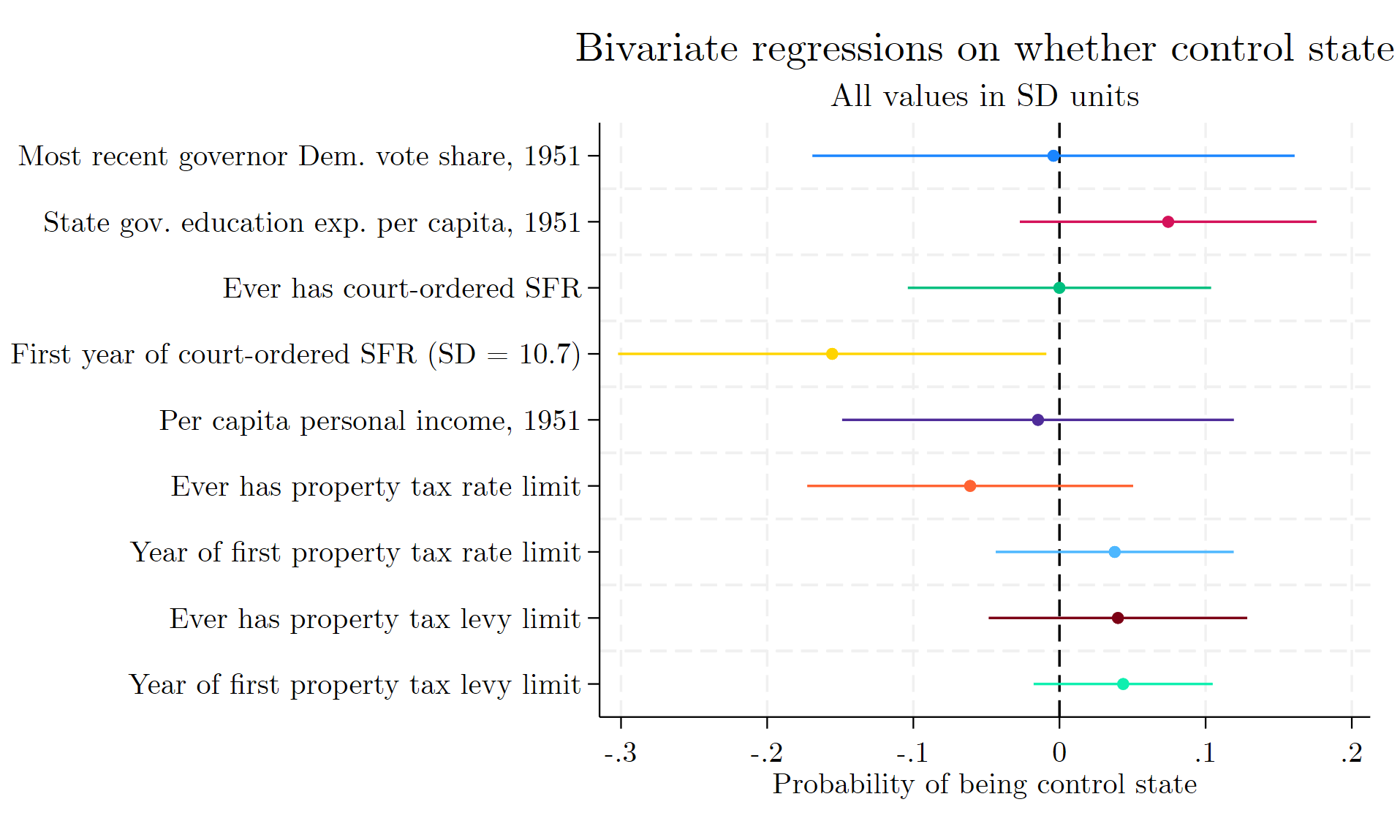}
    \end{subfigure}
            \caption*{\footnotesize Note: 95\% confidence intervals shown. The figures plot bivariate regressions of the year of a state's mandate (panel a) and whether it is a control state (panel b) on each state characteristic. }
\end{figure}

Another possible concern is that the implementation of a state mandate might be correlated with other characteristics of the state. This would be a concern to the extent that these characteristics are also correlated with trends in education outcomes. Figure \ref{fig:balance_coefplot} presents coefficients from bivariate regressions of the year of a state's mandate (panel a) or whether a state is one of the control states (panel b) on various state characteristics which might relate to trends in education in each state. 

One such possibility is that states that are otherwise more progressive were earlier to implement these mandates and also had faster growth in educational attainment, but there is no significant association between the timing of a state's mandate and its pre-period political tendencies. Data on Democrat share in the most recent governor election in 1951 are drawn from  \textcite{icpsrGeneralElectionData2013}. Figure \ref{fig:balance_coefplot} shows that states with higher pre-period Democrat shares in their governor elections did not pass mandates significantly earlier.

Another possible confounder is that states with more pre-period income or investment in education might pass mandates earlier and have better trends in education outcomes. I find little evidence of this.  Data on a state's per capita income and state-level education spending per capita are drawn from IndFin, described in Section \ref{section:financial}. There is no significant association between the timing of a state's mandate and its pre-period income or spending on education.

Both before and after this period, many states implemented other major education reforms which could confound the analysis. I use the compilation prepared by \textcite{jacksonEffectsSchoolSpending2016} to examine court-ordered school finance equalization reforms (SFR) in each state, which take place between 1971-2009. There is no significant correlation between the state's mandate and whether it ever had such a reform. Among states that had such a reform, there is a significant association between being a control state and the year of this order. However, this significance is not robust to a Bonferroni correction for multiple testing and is identified on only 4 control states which had such an order. 

Another important set of policies debated during this period was limits on property taxes, which might have limited states' ability to create new education initiatives while also harming education trends. I examine both limits on property tax rates and levies that either affect a state overall or with respect to its schools, as documented by \textcite{paquinChronicle161yearHistory2015}. I find little correlation between the timing of a state's mandate and limits on property taxes. 

\newpage 
\section{Supplementary tables and figures}\label{appendix_figures}
\setcounter{figure}{0}    
\setcounter{table}{0}    
\FloatBarrier
\begin{table}[htb]
    \centering
        \caption{Results on use of resources by type}
    \label{tab:nhes_rsce_by_type}
    \resizebox{\textwidth}{!}{%
    \begin{tabular}{lcccccccc}
    \hline \hline
                  &\multicolumn{1}{c}{(1)}   &\multicolumn{1}{c}{(2)}   &\multicolumn{1}{c}{(3)}   &\multicolumn{1}{c}{(4)}   \\
            &Intellectual disability   &Slow learner   &Speech therapy   &Emotional disability   \\
 \hline  \textbf{Disabled} & & & & & \\
Pre-period average&      -0.009   &       0.001   &       0.015   &      -0.005   \\
            &     (0.019)   &     (0.017)   &     (0.020)   &     (0.008)   \\
Post-period average&       0.048   &       0.105***&       0.001   &       0.003   \\
            &     (0.032)   &     (0.026)   &     (0.013)   &     (0.015)   \\
Callaway \& Sant'Anna average&       0.067** &       0.084***&       0.013   &      -0.001   \\
            &     (0.029)   &     (0.026)   &     (0.017)   &     (0.012)   \\
\hline
Observations&        1636   &        1636   &        1636   &        1636   \\
Pre-mandate mean&        0.03   &        0.06   &        0.08   &        0.01   \\
\hline\hline
\textbf{Non-disabled} & & & & & \\
Pre-period average&       0.001   &       0.016** &      -0.004   &      -0.005***\\
            &     (0.002)   &     (0.008)   &     (0.003)   &     (0.002)   \\
Post-period average&       0.003   &       0.008   &       0.000   &      -0.017***\\
            &     (0.004)   &     (0.013)   &     (0.002)   &     (0.003)   \\
Callaway \& Sant'Anna average&       0.004   &       0.026   &       0.001   &      -0.015***\\
            &     (0.004)   &     (0.026)   &     (0.002)   &     (0.003)   \\
\hline
Observations&       10270   &       10270   &       10270   &       10270   \\
Pre-mandate mean&        0.01   &        0.04   &        0.01   &        0.01   \\

    \\ \hline
    \end{tabular}
    }
    \caption*{\footnotesize Note: The table shows difference-in-difference estimates of the impacts of the mandates on the types of resources they used. The top panel shows results for disabled students and the bottom panel shows results for non-disabled students. Column (1) shows effects on the probability of using services for an intellectual disability, column (2) for slow learners, column (3) for speech therapy, and column (4) for an emotional disability. Pre-period average and post-period average refer to a simple average of event-study coefficients before and after the implementation of a mandate, respectively. Callaway \& Sant'Anna average refers to a weighted average of estimated impacts, with weights given by the share belonging to each treated cohort in the sample. Standard errors clustered at the state level shown in parentheses. \\
    * p \textless 0.1, ** p \textless 0.05, *** p \textless 0.01 }
\end{table}

\begin{figure}[htb]
\caption{Effects on school enrollments} \label{fig:enrolled615_1620}
    \begin{subfigure}{.5\textwidth}
        \caption{Age 6-15}
        \centering
        \includegraphics[width=\textwidth]{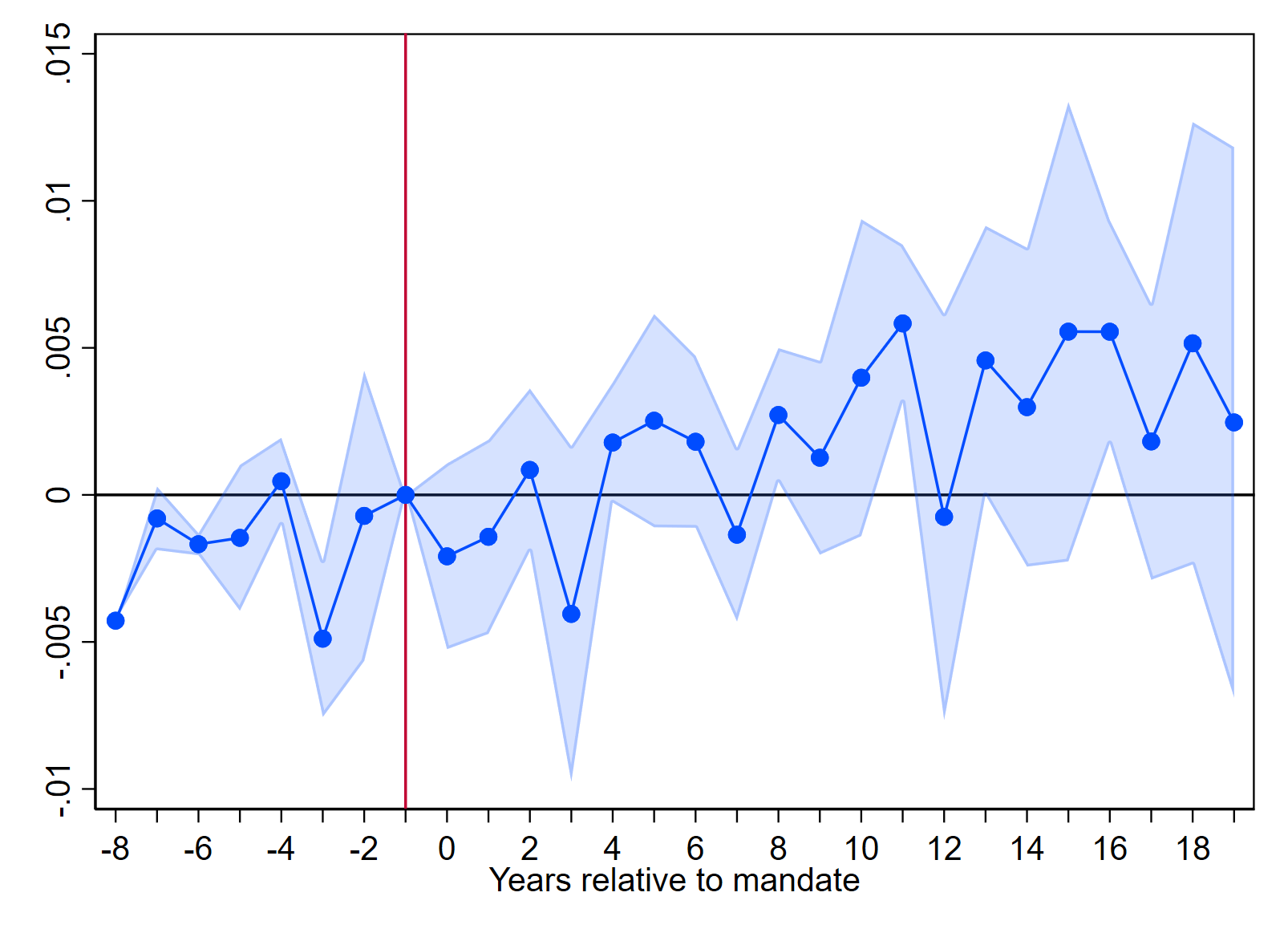}
    \end{subfigure}
    \begin{subfigure}{.5\textwidth}
        \caption{Age 16-20}
        \centering
        \includegraphics[width=\textwidth]{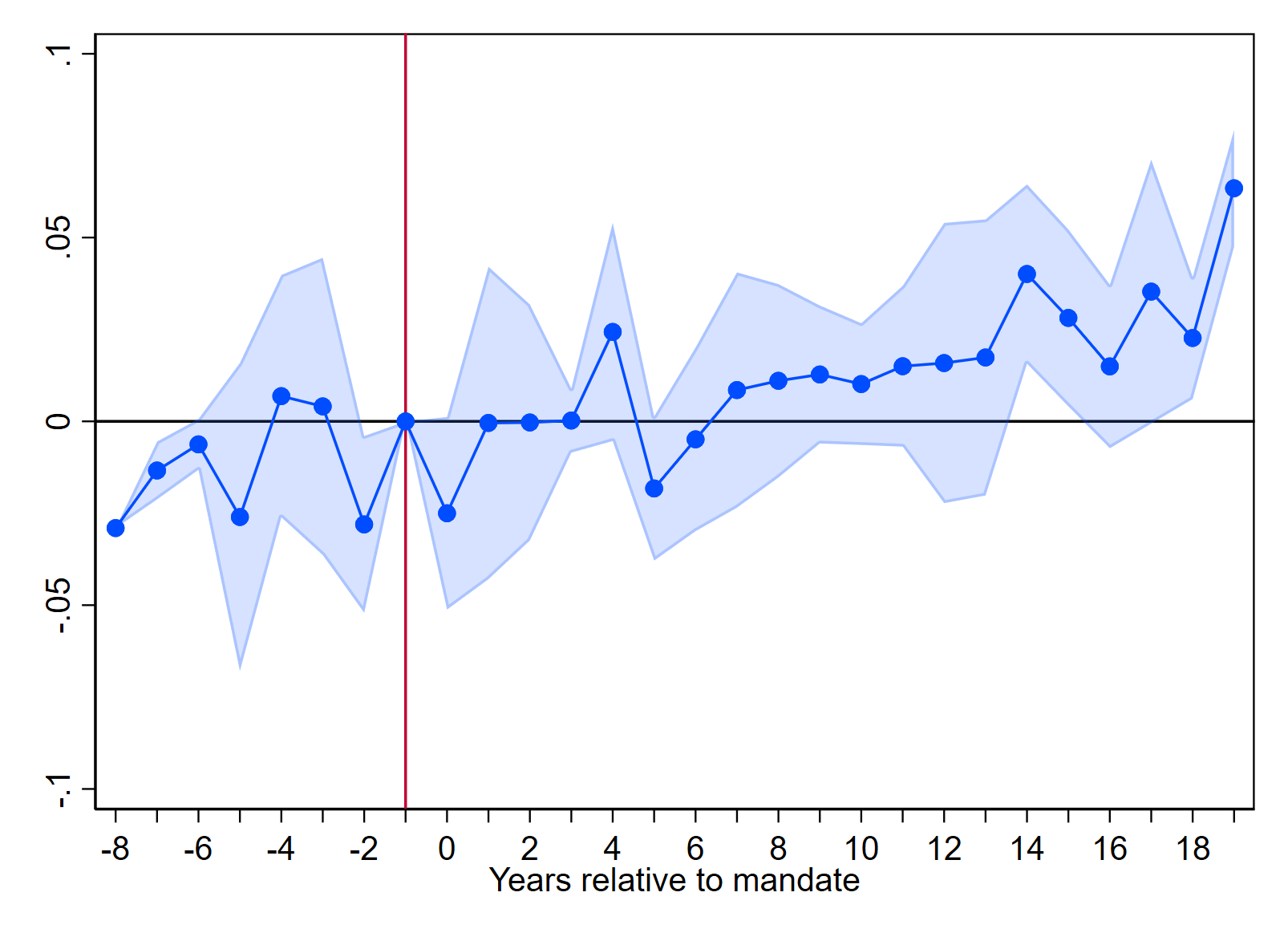}
    \end{subfigure}
            \caption*{\footnotesize Note: 95\% confidence intervals shown, standard errors clustered at the state level. The figures plot difference-in-difference event-study estimates of the impact of the mandates on the probability of being enrolled in school among individuals age 6-15 (panel (a)) and age 16-20 (panel (b)) in the CPS. }
\end{figure}

\begin{figure}[htb]
        \caption{Effects on probability of being identified as disabled}
    \begin{subfigure}{.45\textwidth}
        \caption{NHES: Effects on overall probability}
        \centering
        \includegraphics[width=\textwidth]{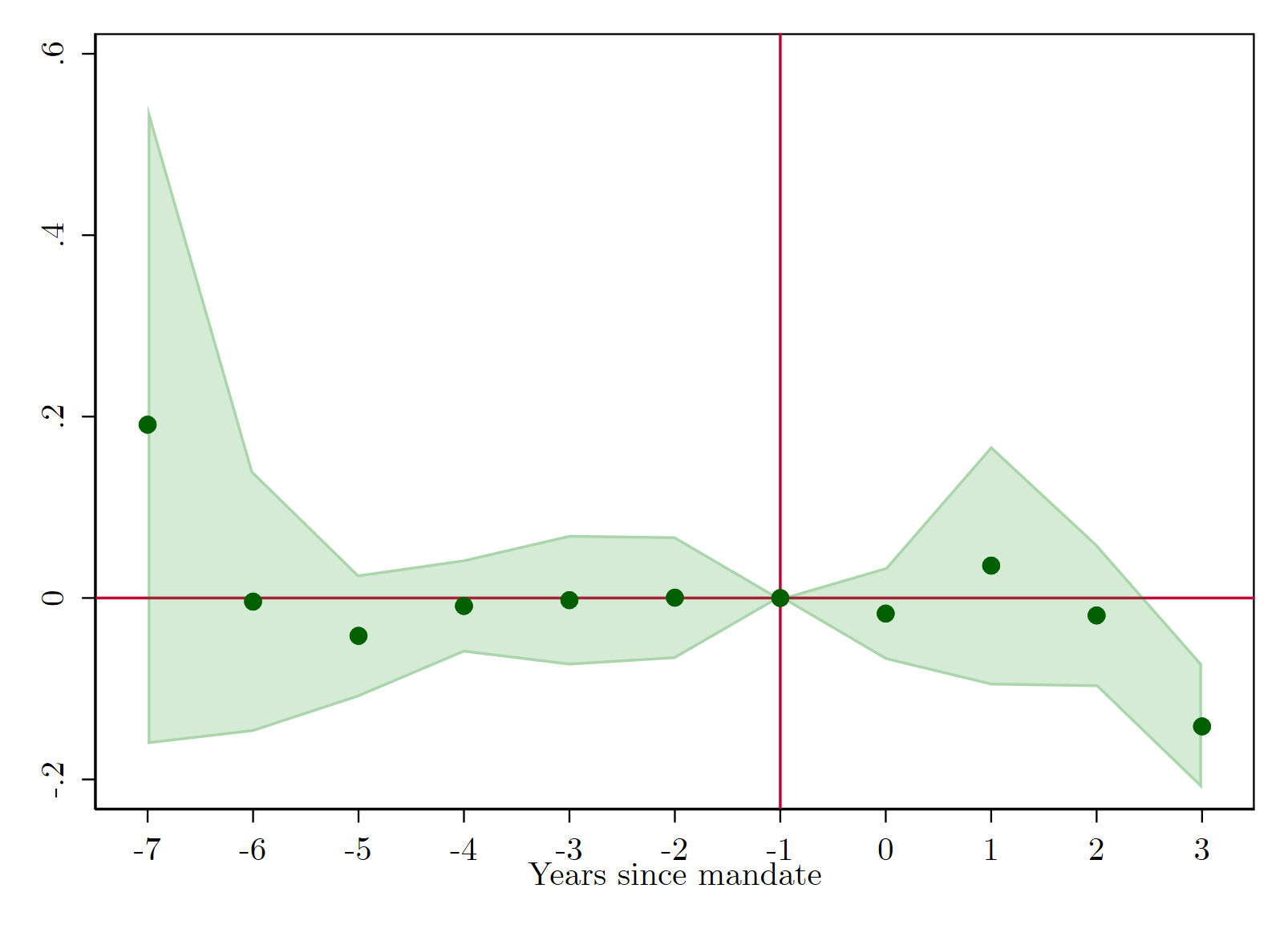}
    \end{subfigure} 
    \begin{subfigure}{.45\textwidth}
        \centering
        \caption{Census: Effects on overall probability}
        \includegraphics[width=\textwidth]{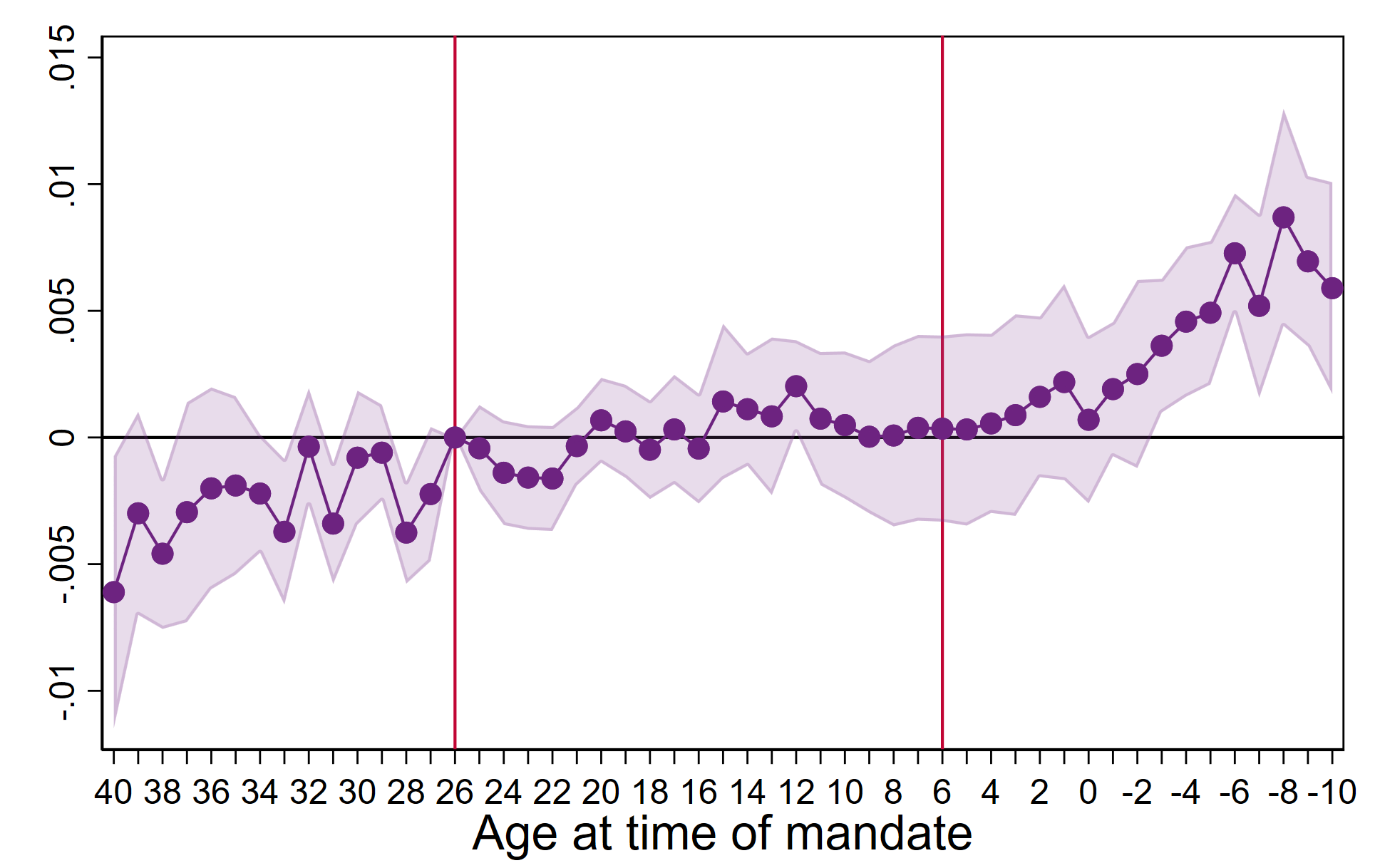}
    \end{subfigure}
        \label{fig:prob_disabled}  
        \caption*{\footnotesize Note: 95\% confidence intervals shown. The figure plots difference-in-difference event-study estimates of the impact of the mandates on the probability of being identified as disabled in NHES (panel a) and the Census data (panel b).}
\end{figure}
\begin{figure}[htb]
\caption{Descriptive statistics on Census disability measure}

    \begin{subfigure}{.5\textwidth}
        \caption{Duration of disability by age (Census 1970)}
        \centering
        \includegraphics[width=\textwidth]{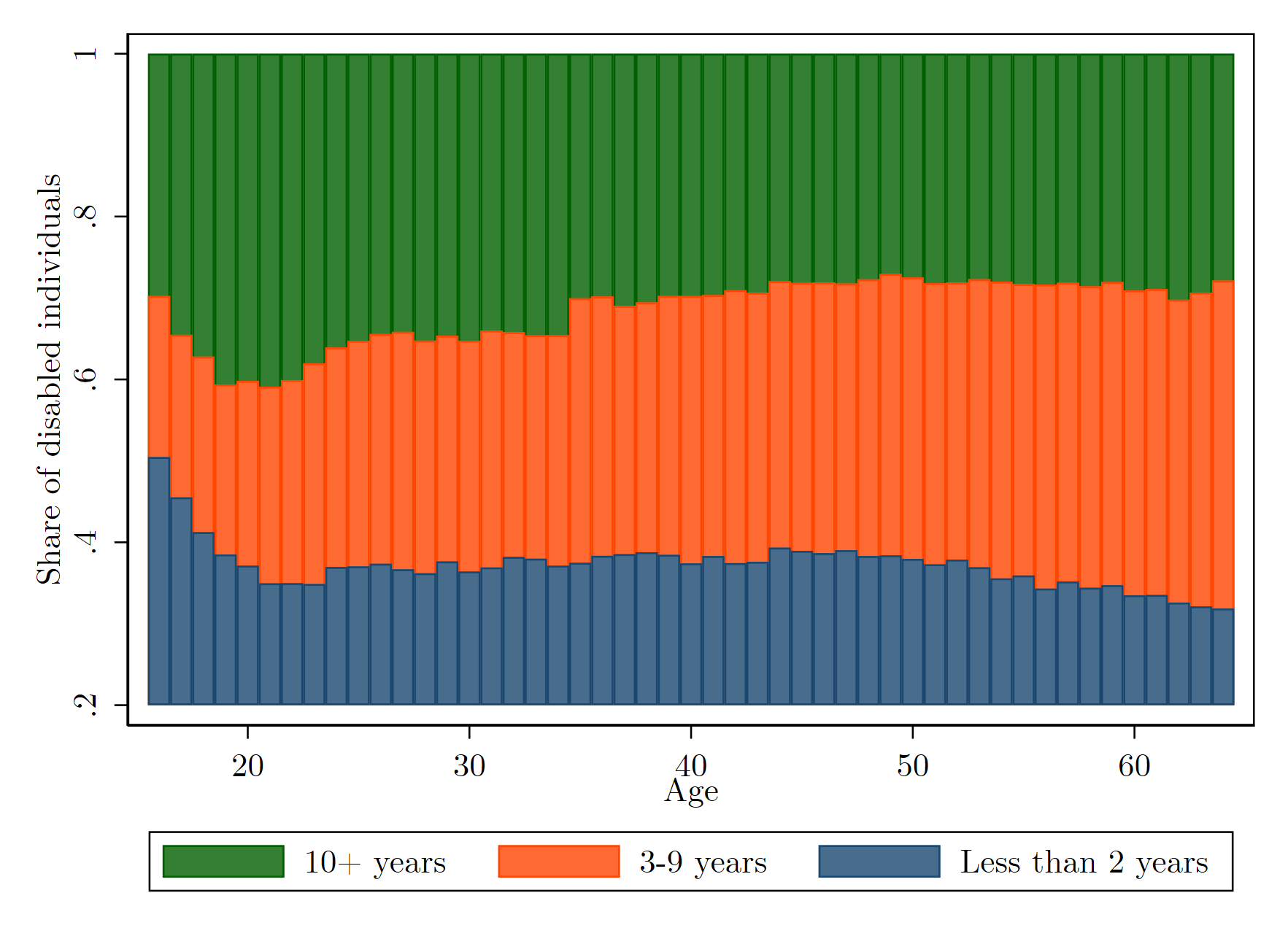}
        \label{fig:disabdur_by_age}  
    \end{subfigure}
    \begin{subfigure}{.5\textwidth}
        \caption{Comparison of disability definitions (NHIS)}
        \centering
        \includegraphics[width=\textwidth]{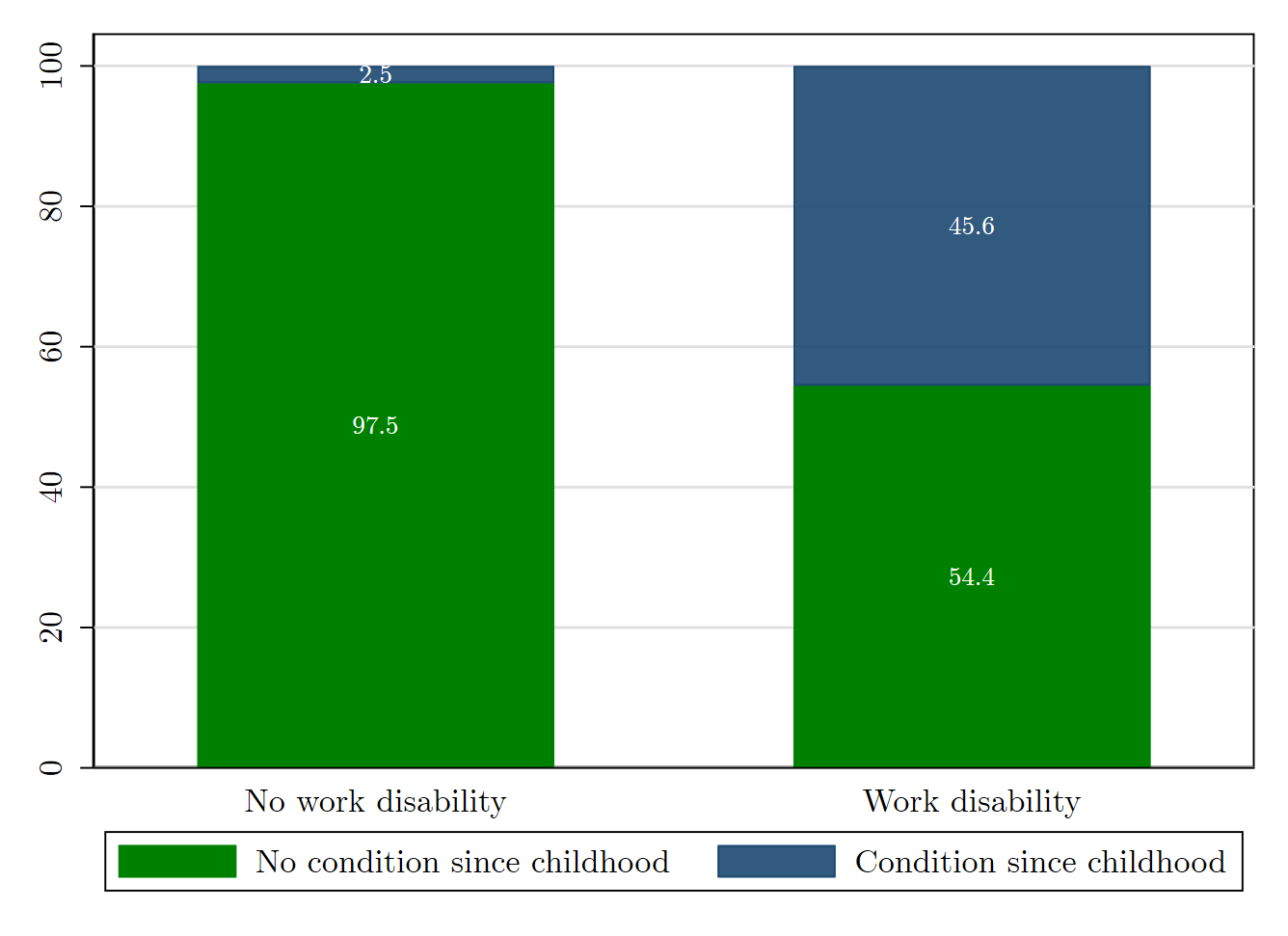}
        \label{fig:nhis_disability_duration}  
    \end{subfigure}
\end{figure}

\begin{figure}[htb]
\caption{Robustness of MVPF to discount rate} \label{fig:mvpf_discount}
        \centering
        \includegraphics[width=\textwidth]{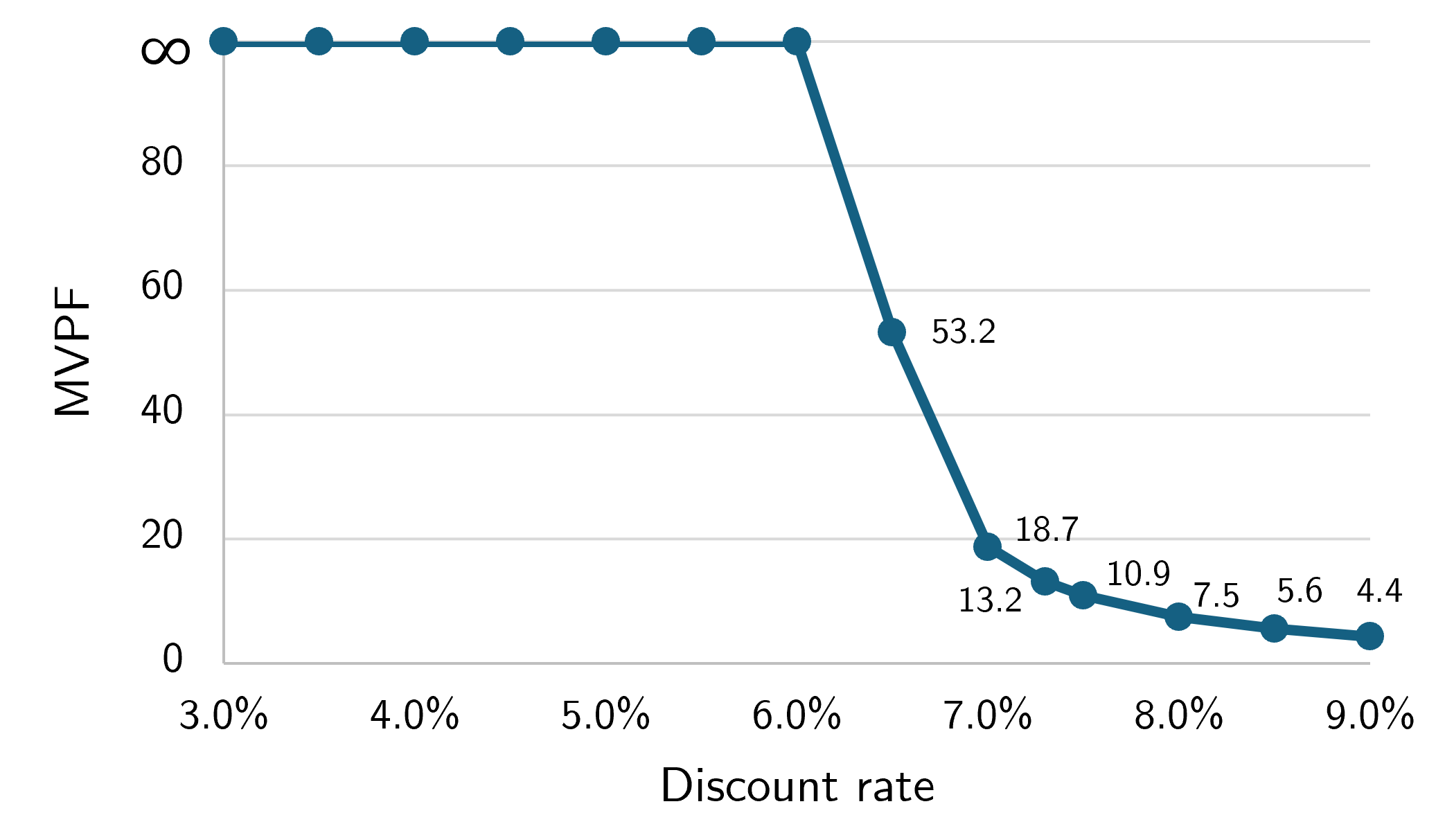}
    \caption*{\footnotesize Note: This figure plots the MVPF given different values of the discount rate used to compute present values.}
\end{figure}

\FloatBarrier

\newpage 

\section{Data appendix}

\subsection{State-level dataset on children receiving services}
\label{appendix_num_served}

Table \ref{tab:data_handicapped} presents the data sources compiled to construct the state-level series of children receiving services for a disability. 
\begin{table}[htb]
    \caption{Data sources on number of children receiving services for a disability}
    \centering
    \begin{tabular}{c|c}
    \hline \hline
         School year & Data source \\ \hline
         1952-53 & Biennial Survey of Education \\
         1957-58 & Biennial Survey of Education \\
         1965-66 & House Committee Report \parencite{perkinsChildrenSpecificLearning1969} \\
         1967-68 & \textcite{usofficeofeducationBetterEducationHandicapped1969} \\
         1968-69 & \textcite{grotbergDayCareResources1971}\\
         1971-72 & \textcite{uscongressCongressionalRecord1975} \\
         1972-73 & \textcite{uscongressCongressionalRecord1975} \\
         1976-forward & Annual reports on PL-142 implementation \\ \hline \hline
    \end{tabular}
    \label{tab:data_handicapped}
    
\end{table}

The number of children receiving services is normalized by estimates of the child population are drawn from the SEER database produced by the National Institutes of Health based on Census estimates for 1969-2023 \parencite{seerDownloadUSCounty2025}. For earlier years, Census data compiled by \textcite{hainesHistoricalDemographicEconomic2004} are used.

\subsection{Census data} \label{appendix_censusdata}
\setcounter{figure}{0}    

The relevant questions for defining disability in each Census sample can be found below: 
\begin{itemize}
    \item 1970 Form 1: ``Does this person have a health or physical condition which limits the kind or amount of work he can do at a job?''
    \begin{itemize}
        \item ``Health condition. This is a serious illness, or a serious handicap (impairment) affecting some part of the body or mind, which interferes with his ability to work at a job. Answer No for pregnancy, common colds, etc.''
    \end{itemize}
    \item 1980 1\% sample: ``Does this person have a physical, mental, or other health condition which has lasted for 6 or more months and which.... Limits the kind or amount of work this person can do at a job? Prevents this person from working at a job?''
    \begin{itemize}
        \item ``Mark Yes to part (b) if the health condition prevents this person from holding any significant employment.''
    \end{itemize}
    \item 1990 1\% sample: ``Does this person have a physical, mental, or other health condition that has lasted for 6 or more months and which Limits the kind or amount of work this person can do at a job? Prevents this person from working at a job?''
    \begin{itemize}
        \item ``The term `health condition' refers to any physical or mental problem which has lasted for 6 or more months. A serious problem with seeing, hearing, or speech should be considered a health condition. Pregnancy or a temporary health problem such as a broken bone that is expected to heal normally should not be considered a health condition.''
    \end{itemize}
    \item 2000 1\% sample and American Community Survey: ``Because of a physical, mental, or emotional condition lasting 6 months or more, does this person have any difficulty in doing any of the following activities: (Answer if this person is 16 YEARS OLD OR OVER.) Working at a job or business?''
\end{itemize}

\begin{figure}[htb]
\caption{Descriptive statistics on Census disability measure}
    \begin{subfigure}{.5\textwidth}
        \caption{Demographics of disability, Census 1970-1990}
        \centering
        \includegraphics[width=\textwidth]{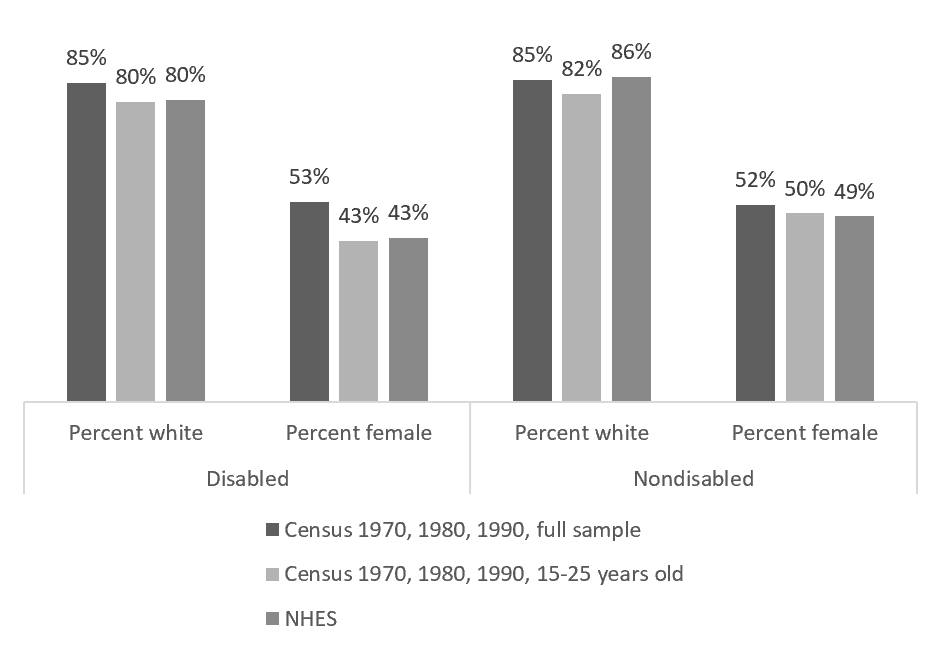}
        \label{fig:acs_vs_nhes}
    \end{subfigure}    
    \begin{subfigure}{.5\textwidth}
        \caption{Percent disabled by age, Census 1970-1990}
        \centering
        \includegraphics[width=\textwidth]{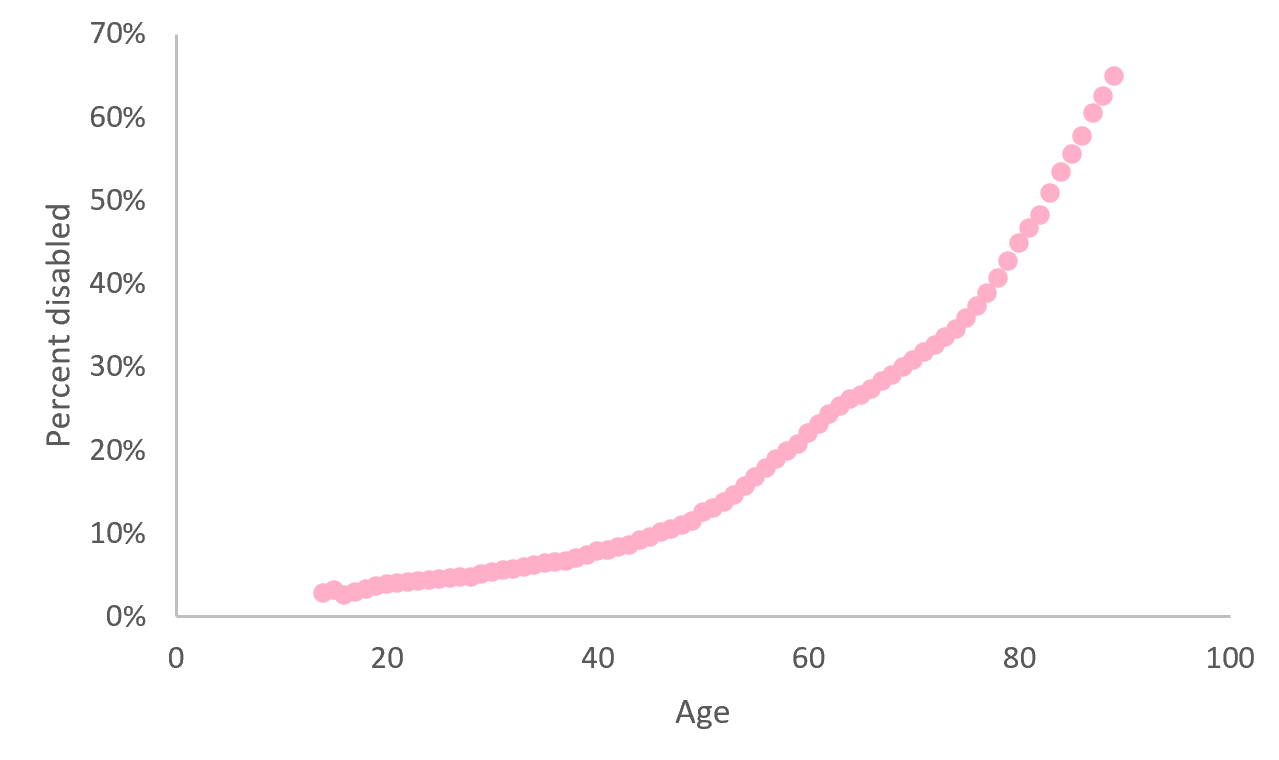}
        \label{fig:acs_disab_age}
    \end{subfigure}
\end{figure}
\FloatBarrier

\subsection{School district finances} \label{appendix_indfin}

I use data from the Historical Finance Data Base of Individual Local Governments (IndFin), produced by the Census Bureau. This dataset contains information on the finances of cities, towns, states, and school districts in the 1967 fiscal year and annually from 1970-2008. In years ending in a 2 or 7, the data come from the Census of Governments (census years), and data from other years come from the Annual Survey of Governments (sample years). However, even in sample years, additional data collected by states ensures that coverage is nearly complete. Still, for several years between 1970 and 1980 and 1993-1996, the sample consists of only a subsample of school districts.

To examine the coverage of this dataset, I compare the number of districts which appear in the IndFin dataset with the statistics reported by the National Center for Education Statistics in the \textit{Digest of Education Statistics} \parencite{ncesTable21410Number2022}\footnote{Although the data on the number of school districts is not annual, I carry forward estimates between years in order to compare annually. NCES also notes that statistics on the number of school districts are not directly comparable before/after 1980.} and total enrollment with \textit{State Comparisons of Education Statistics} \parencite{snyderStateComparisonsEducation1998}. 

\begin{figure}[htb]
\caption{Coverage of IndFin dataset}
    \centering
    \includegraphics[width=.7\textwidth]{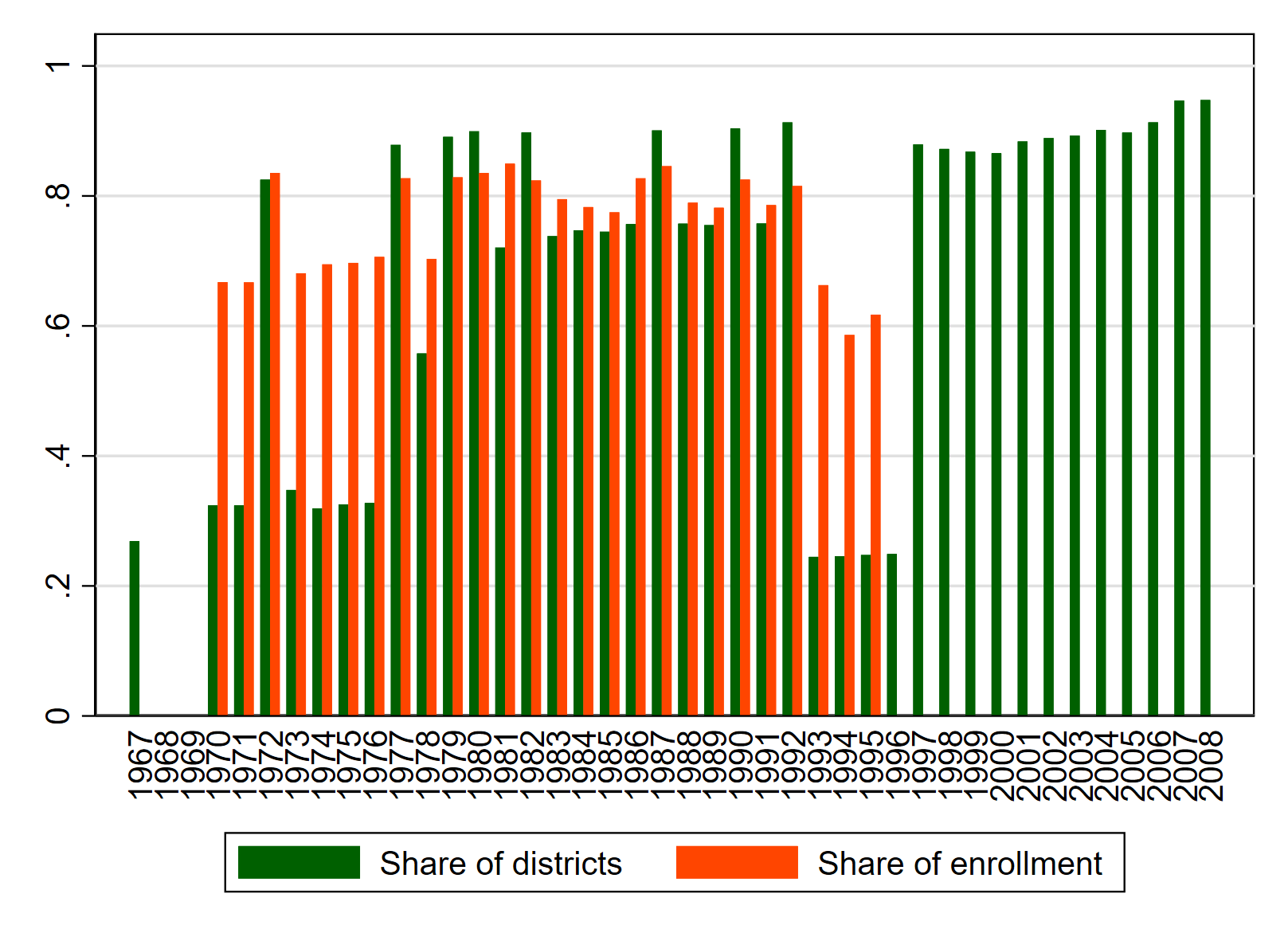}
    \label{fig:share_enroll_dist}
\end{figure}

Figure \ref{fig:share_enroll_dist} shows that the coverage of the dataset is high. In sample years, the IndFin dataset covers only about 1/3 of school districts, but these are weighted towards larger school districts, so that it represents more than 60\% of enrollment in almost every year. In census years, coverage is over 80\%. The coverage is not expected to be 100\%, as a number of students attend dependent or special school districts, which are not included in this analysis.

I use data from on the Consumer Price Index (CPIAUCSL) \parencite{usbureauoflaborstatisticsConsumerPriceIndex2025} in order to deflate the units in all analyses to constant 1990 dollars. To avoid implausible per-student estimates due to school districts with low enrollment, I remove any observations with per-student spending or revenues in 1990 dollars exceeding \$100,000 per student. 

\FloatBarrier

\section{State laws}\label{appendix_laws}

The timing of state mandates was compiled by manual review of the literature, in consultation with three data sources: \cite{nationalassociationofstatedirectorsofspecialeducationStateProfilesSpecial1977}, \cite{hensleyQuestionsAnswersEducation1975}, and \cite{uscongressCongressionalRecord1975}.

The table below summarizes the information for each state. In many states, several laws relating to a mandate to provide services for a disability (usually known as special education or services for exceptional children). In these cases, the year of the first mandate was taken to be the year of the earliest law which required states to provide educational services for a large group of disabled children (rather than those with one specific disability) and which established some standards or accountability for this. 

\begin{landscape}
\fontsize{9}{11}\selectfont

\begin{longtable}{lccPRccP}
 \multicolumn{1}{l}{State} & \multicolumn{1}{c}{Year of effect} & \multicolumn{1}{T}{Year of passage} & Name of relevant legislation & \multicolumn{1}{c}{Phrasing} & \multicolumn{1}{T}{Effective year\newline in NASDSE 1977} & \multicolumn{1}{T}{Year of passage \newline in Hensley et al. 1975}& Notes \\
 \hline \hline
\endfirsthead
\endhead
         Alabama & 1971 & 1971 & Alabama Laws and Joint   Resolutions of the Legislature of Alabama 1971, Act 106; S.13 & ``Each school board shall provide not less than twelve consecutive   years of appropriate instruction and special services for exceptional   children" & 1977 & 1971 & Law noted by Hensley is earlier version; 1977 version includes an   amendment to include ``profoundly retarded" students \\ \hline 
Alaska & 1971 & 1970 & Session Laws of Alaska, Chapter 144 & ``to provide competent education services for the exceptional   children of legal school age in the state for whom the regular school   facilities are inadequate or not available" & 1971 & 1974 & Earlier law from NASDSE used \\ \hline
Arizona & 1976 & 1973 & Session Laws of Arizona 1973 Chapter 181; House Bill 2256 & ``to guarantee equal educational opportunity to each handicapped   child in the state regardless of the schools, institutions or programs by   which such children are served" & 1976 & 1973 & Law passed in 1973, took effect in 1976 \\ \hline
Arkansas & 1979 & 1973 & Arkansas Act 102, the Handicapped Children's Act of 1973 & ``to provide as an integral part of the public schools, special   education sufficient to meet the needs and maximize the capabilities of   handicapped children" & 1979 & 1973 & Law passed in 1973, took effect in 1979 \\ \hline
California & 1978 & 1977 & Lanterman Act; Statutes of California 1977, Chapter 1247 (AB 1250) & ``all individuals with exceptional needs have a right to participate   in appropriate programs of publicly supported education and that special   educational programs and services for these persons are needed in order to   assure them of this right to an appropriate educational opportunity" & 1978 & 1974 & The law highlighted by Hensley et al. is only permissive of special   education, not a mandate \\ \hline
Colorado & 1973 & 1973 & Handicapped Children's Educational Act; Session Laws of Colorado 1973,   Ch. 354 & ``to provide means for education those children who are   handicapped" & 1973 & 1973 &  \\ \hline
Connecticut & 1967 & 1967 & Public Acts Passed by the General Assembly of the State of Connecticut   1967; An Act Concerning the Provision of Special Education; Public Act No.   627; SB No. 1788 of 1967 & ``The state board of education shall provide for the development and   supervision of the educational programs and services for children requiring   special education" & 1967 & 1966 & The act number in the two sources is the same; Hensley et al. may have a   typo in the year \\ \hline
Delaware & 1978 & 1977 & Laws of the State of Delaware of 1977, Chapter 190; Senate Bill No. 353 & ``that each handicapped person   as defined in this Chapter shall receive a free and appropriate public   education designed to meet his or her needs" & & 1935 & The law highlighted by Hensley et al. is only permissive of special   education, not a mandate; NASDSE was published in 1977, so does not include 1978 law \\ \hline
District of Columbia & 1975 & 1975 & District of Columbia Register, 1975, Rules of the Board of Education   Education of the Handicapped & ``shall provide an adequate free public education to handicapped   children for whom regular program of instruction is inadequate to meet their   special educational needs" & 1975 &  &  \\ \hline
Florida & 1973 & 1968 & Acts and Resolutions Adopted by the Legislature of Florida at   Extraordinary Sessions 1968, Chapter 68-24; Senate Bill 89-X & ``Each district school board shall provide an appropriate program of   special instruction, facilities, and related services for exceptional   children; such programs shall be implemented in annual increments so that all   exceptional children shall be served by 1973" & 1973 & 1968 & Law passed in 1968, fully took effect in 1973 \\ \hline
Georgia & 1977 & 1974 & Adequate Program for Education in Georgia Act; Acts and Resolutions of   the General Assembly of the State of Georgia 1974, No. 1242 (Senate Bill No.   672) & ``to assure that each Georgian has access to quality instruction   designed to develop his capacities to the maximum through programs that meet   his developmental and remedial educational needs" & 1977 & 1968 & Hensley et al. do not provide a citation for any 1968 law; earliest law   found is used \\ \hline
Hawaii & 1949 & 1949 & Series A-54: Act 29 & ``It is hereby declared to be of   vital concern to the Territory of Hawaii that all exceptional children   residing in the Territory of Hawaii be provided with instruction, special   facilities and special services for education, therapy and training to enable   them to live normal competitive lives. In order to effectively accomplish   such purpose the department of public instruction is authorized, and it shall   be its duty, to establish and administer instruction, special facilities and   special services for the education, therapy and training of exceptional   children...." & & 1949 &  \\ \hline
Idaho & 1972 & 1972 & General and Special Laws of the State of Idaho Passed by the Second   Regular Session of the Forty-First Idaho Legislature, Chapter 312; H.B. No.   754 & ``Each public school district is responsible for and shall provide   for the education and training of exceptional pupils resident   therein" & 1972 & 1973 & Earlier law from NASDSE used \\ \hline
Illinois & 1969 & 1967 & Illinois Public Acts, Regular Session 33, Public Act 76-27 & ``The Advisory Committee shall by July 1, 1967 complete and report to   the Superintendent of Public Instruction a comprehensive plan whereby all   handicapped children resident in the county may receive a good common school   education…. If any county fails to submit an acceptable plan by July 1, 1967,   then it shall be the duty of the Council to devise and recommend a   comprehensive plan for the education of handicapped children resident therein   prior to July 1, 1969" & 1969 & 1972 & Earlier law from NASDSE used \\ \hline
Indiana & 1973 & 1969 & Laws of the State of Indiana 1969, Chapter 396; H. 1071 & ``School boards of any school corporations that maintain a recognized   school may, until July 1, 1973, and shall thereafter, subject to any   limitation hereinafter specified, establish and maintain such special   educational facilities as may be needed…." & 1973 & 1969 & Law passed in 1969, took effect in 1973 \\ \hline
Iowa & 1975 & 1974 & Laws of the State of Iowa 1974 Session, Chapter 1172; S.F. 1163 & ``to provide an effective, efficient, and economical means of   identifying and serving children from under five years of age through grade   twelve who require special education" & 1975 & 1974 & Law passed in 1974, took effect in 1975 \\ \hline
Kansas & 1979 & 1974 & 1974 Session Laws of Kansas, Chapter 290; HB 1672 & ``The board of education of every school district shall provide   special education services for all exceptional children in the school   district and said special education services shall meet standards and   criteria set by the state board. Said special educations services shall be   planned and operative not later than July 1, 1979." & 1979 & 1974 & Law passed in 1974, took effect in 1979 \\ \hline
Kentucky & 1974 & 1970 & Acts of the General Assembly of the Commonwealth of Kentucky 1970,   Chapter 47; HB 256 & ``By July 1, 1974, all county and independent boards of education   shall operate special education programs to the extent required by, and   pursuant to, a plan which has been approved by the State Board of   Education" & 1979 & 1970 & Earlier law from Hensley et al. used \\ \hline
Louisiana & 1972 & 1972 & State of Louisiana Acts of the Legislature, Regular Session 1972; Act   368; House Bill No. 835 & ``It is and shall be the duty of the various branches and divisions   of the public school system of Louisiana, both state and local, to offer the   best available educational, learning, and training facilities, services,   classes, and opportunities to all children of school age within their   respective boundaries. This includes all children of school age whether   normal, exceptional, crippled, or otherwise either mentally or physically   handicapped, and whatever may be the degree of that handicap." & 1972 & 1972 &  \\ \hline
Maine & 1973 & 1973 & Acts, Resolves and Constitutional Resolutions as Passed by the One   Hundred and Sixth Legislature of the State of Maine, Chapter 609 & ``The commissioner shall provide or cause to be provided by   administrative units all regular and special education, corrective and   supporting services required by exceptional children to the end that they   shall receive the benefits of a free public education appropriate to their   needs" & 1975 & 1973 & Earlier law from Hensley et al. used \\ \hline
Maryland & 1974 & 1973 & Laws of the State of Maryland 1973, Chapter 359; Senate Bill 649 & ``The state and its several counties shall make available free   educational programs for all handicapped children, including those children   who are severely handicapped" & 1979 & 1973 & Earlier law from Hensley et al. used \\ \hline
Massachusetts & 1974 & 1972 & Acts and Resolves of the General Court of Massachusetts in the Year 1972,   Chapter 766 & ``to provide a flexible and uniform system of special education   program opportunities for all children requiring special education" & 1974 & 1972 & Law passed in 1972, took effect in 1974 \\ \hline
Michigan & 1973 & 1971 & Public and Local Acts of the Legislature of the State of Michigan Passed   at the Regular Session of 1971, Public Act 198 & ``The intermediate board may, and for the 1973-1974 school year and   thereafter the intermediate board shall: (a) Develop and establish and   continually evaluate and modify in cooperation with its constituent school   districts, a plan for special education which shall provide for the delivery   of special eduation programs and services designed to develop the maximum   potential of every handicapped person" & 1973 & 1971 & Law passed in 1971, took effect in 1973 \\ \hline
Minnesota & 1975 & 1975 & Laws of Minnesota for 1976, Chapter 211; H.F. No. 1993 & ``Every district shall provide   special instruction and services, either within the district or in another   district, for handicapped children of school age who are residents of the   district and who are handicapped as set forth in section 120.03." & & 1957 & Law cited by Hensley et al. could not be found; earliest available law noted instead \\ \hline
Mississippi & 1978 & 1978 & Laws of the State of Mississippi, 1978, Chapter 461; Senate Bill No. 2620 & ``to provide competent educational services and equipment for exceptional children, for whom the   regular school programs are not adequate" & & 1973 & Law cited by Hensley et al. suggests that parents must first petition the school board for special education before services are mandated to be provided; removed in the 1978 law. \\ \hline
Missouri & 1974 & 1973 & Laws of Missouri Passed at the First Regular, First Extra, Second Regular   and Second Extra Sessions of the Seventy-Seventh General Assembly, HB 474 & ``to provide or to require public schools to provide to all   handicapped and severely handicapped children within the ages prescribed   herein, as an integral part of Missouri's system of gratuitous education,   special educational services sufficient to meet the needs and maximize the   capabilities of handicapped and severely handicapped children." & 1974 & 1973 & Law passed In 1973, took effect in 1974 \\ \hline
Montana & 1979 & 1974 & Laws and Resolutions of the State of Montana Passed by the Forty-Third   Legislature in Second Regular Session, Chapter 93 & ``After July 1, 1979, the board of trustees of every school district   must provide or establish and maintain a special education program for every   handicapped person as herein defined between the ages of six (6) and   twenty-one (21) in the district who cannot benefit sufficiently from the   regular programs of instruction by reason of his mental, physical, emotional   or learning problems" & 1979 & 1974 & Law passed in 1974, took effect in 1979 \\ \hline
Nebraska & 1976 & 1973 & Laws Passed by the Legislature of the State of Nebraska Eighty-Third   Legislature, First Session, 1973, Legislative Bill 403 & ``It shall be the duty of the board of education of every school   district to provide or contract for special education programs for all   resident children who would benefit from such programs" & 1976 & 1973 & Law passed in 1973, took effect in 1976 \\ \hline
Nevada & 1973 & 1973 & Statutes of the State of Nevada Passed at the Fifty-Seventh Session of   the Legislature, Chapter 806; Senate Bill 648; Assembly Bill 66 & ``The board of trustees of a school district shall make such special provisions as may be necessary for the education of handicapped minors" & 1973 & 1973 &  \\ \hline
New Hampshire & 1965 & 1965 & Laws of the State of New Hampshire Passed January Session, 1965, Chapter   378 & ``It is hereby declared to be the policy of the state to provide the   best and most effective education possible to all handicapped children in New   Hampshire" & 1965 & 1971 & Earlier law from NASDSE used \\ \hline
New Jersey & 1954 & 1954 & Additional Acts of the One Hundred and Seventy-seventh Legislature of the   State of New Jersey, Chapters 178-179 & ``It shall be the duty of each board of education to provide suitable   facilities and programs of education for all the children who are classified   as physically handicapped under this act"; ``It shall be the duty of   each board of education or training for all children who are classified as   educable or trainable under this act" & 1954 & 1954 &  \\ \hline
New Mexico & 1972 & 1972 & Laws of the State of New Mexico passed by the Second Regular Session of   the Thirtieth Legislature, Chapter 95 & ``The state shall require school districts over a five year period to   provide special education sufficient to meet the needs of all exceptional   children" & 1972 & 1972 &  \\ \hline
New York & 1967 & 1967 & Laws of the State of New York Passed at the One Hundred and Ninetieth   Session of the Legislature, Chapter 786 & ``The board of education of each city and of each union free school   district shall be required to furnish suitable education facilities for   handicapped children by means of home-teaching, transportation to school or   by special classes" & 1973 & 1956 & Law cited by Hensley et al. includes only physically disabled students;   1967 law includes a broader definition of ``handicapped" \\ \hline
North Carolina & 1974 & 1974 & State of North Carolina Session Laws and Resolutions Passed by the 1973   General Assembly at its Second Session 1974, Chapter 1238 & ``to ensure every child a fair and full opportunity to reach his full   potential and that no child as defined in this act shall be excluded from   service or education for any reason whatsoever" & 1973 & 1974 & No such law found in 1973 \\ \hline
North Dakota & 1980 & 1973 & Laws passed at the Forty-third Session of the Legislative Assembly of the   State of North Dakota, Chapter 171; House Bill 1090 & ``School districts shall provide special education to handicapped   children" & 1980 & 1973 & Law passed in 1973, took effect in 1980 \\ \hline
Ohio & 1976 & 1972 & General Laws of the One Hundred Eleventh General Assembly of Ohio,   Amended Substitute House Bill 455 & ``The state board of education shall authorize the establishment and   maintenance of programs for the education of all handicapped children of   compulsory school age" & 1976 & 1973 & The law highlighted by Hensley et al. is only permissive of special   education, not a mandate \\ \hline
Oklahoma & 1970 & 1970 & Oklahoma Session Laws 1970, Chapter 292; SB 403 & ``From and after September 1, 1970, it shall be the duty of each   school district to provide special education for all handicapped exceptional   children as herein defined" & 1970 & 1971 & Earlier law from NASDSE used \\ \hline
Oregon & 1974 & 1973 & Oregon Laws and Resolutions Enacted and Adopted by the Regular Session of   the Fifty-seventh Legislative Assembly, Chapter 510 & ``The Department of Education shall report the results of the surveys   to all agencies concerned with the needs of children and shall whenever   possible assist school districts to commence implementation of programs aimed   at the unmet needs revealed by the survey. If necessary the Department of   Education shall propose appropriate legislation to insure that the   educational needs of all children are met." & 1973 & 1973 & Law cited by Hensley et al. used, passed in 1973 and took effect in 1974 \\ \hline
Pennsylvania & 1956 & 1955 & Laws of the General Assembly of the Commonwealth of Pennsylvania Passed   at the Session of 1955; Act 429 & ``it shall be the duty of the board of directors of any district   having such children to provide and maintain, or to jointly provide and   maintain with neighboring districts, such special classes or schools." & 1976 & 1955 & Earlier law from Hensley et al. used \\ \hline
Rhode Island & 1952 & 1952 & Acts and Resolves Passed by the General Assembly of the State of Rhode   Island and Providence Plantations 1952, Chapter 2905 & ``In any city or town where there is an educable child of school age   resident therein who is physically, mentally, or emotionally handicapped to   such an extent that normal educational growth and development is prevented,   the school committee of such city or town shall provide such type of training   or instruction as recommended by the state department of education that will   best satisfy the needs of the handicapped child" & 1964 & 1952 & Earlier law from Hensley et al. used \\ \hline
South Carolina & 1977 & 1972 & Acts and Joint Resolutions of the General Assembly of the State of South   Carolina, Regular Session of 1972, Act 977 & ``The General Assembly declares that the public policy of this State   is to provide, when feasible, the resources, assistance, coordination, and   support necessary to enable the handicapped person to receive an education   within the context of his home and community" & 1977 & 1972 & Law passed in 1972, took effect in 1977 \\ \hline
South Dakota & 1972 & 1972 & Laws of South Dakota, 1972, Chapter 100, SB 108 & ``It shall be the responsibility of the school board to provide all of its resident exceptional children with   an appropriate educational program." & & 1972 &  \\ \hline
Tennessee & 1974 & 1972 & Public Acts of the State of Tennessee Passed by the Eigthy-Seventh   General Assembly, Chapter 839; House Bill 2053 & ``It is the policy of this state to provide, and to require school   districts to provide, as an integral part of free public education, special   education services sufficient to meet the needs and maximize the capabilities   of handicapped children." & 1974 & 1972 & Law passed in 1972, took effect in 1974 \\ \hline
Texas & 1969 & 1969 & General and Special Laws of the State of Texas Passed by the Regular   Session of the Sixty-First Legislature, Chapter 863; SB 230 & ``It is the intention of this Act to provide for a comprehensive   special education program for exceptional children in Texas" & 1976 & 1969 & Earlier law from Hensley et al. used \\ \hline
Utah & 1959 & 1959 & Laws of the State of Utah, 1959 Passed by the Regular Session of the   Thirty-Third Legislature, Chapter 83; House Bill 23 & ``it shall be the duty of the board of education of any school   district having such children, to provide and maintain from the funds of said   school district, or to provide jointly and maintain with neighboring   districts from the funds of each of the school districts so participating in   proportionate amounts, such special classes" & 1959 & 1969 & Earlier law from NASDSE used \\ \hline
Vermont & 1972 & 1972 & Acts and Resolves Passed by the General Assembly of the State of Vermont Fifty-First Biennial Session Adjourned Session, Act 207; S. 98 & ``Within the limits of funds made available for purposes of this   chapter and the availability of trained personnel, the commissioner shall   provide for the essential early education and for the special education of   handicapped children in such schools and public programs as he may   designate." & 1972 & 1972 &  \\ \hline
Virginia & 1973 & 1972 & Acts and Joint Resolutions of the General Assembly of the Commonwealth of   Virginia, Session 1972, Chapter 603; SB 143 & ``The Board of Education shall prepare and place in operation a   program of special education designed to educate and train handicapped   children between the ages of two and twenty-one years" & 1976 & 1972 & Earlier law from Hensley et al. used \\ \hline
Washington & 1973 & 1971 & 1971 Session Laws of the State of Washington, Chapter 66; Engrossed House   Bill 90 & ``It is the purpose of this 1971 amendatory act to ensure that all   handicapped children as defined in section 2 of this 1971 amendatory act   shall have the opportunity for an appropriate education at public expense as   guaranteed to them by the Constitution of this state" & 1973 & 1971 & Law passed in 1971, took effect in 1973 \\ \hline
West Virginia & 1974 & 1974 & Acts of the Legislature of West Virginia 1974, Chapter 123; House Bill   1271 & ``shall establish and maintain for all exceptional children between   five and twenty-three years of age special educational programs, including   but not limited to special schools, classes, regular classroom programs,   home-teaching or visiting-teacher services" & 1974 & 1974 &  \\ \hline
Wisconsin & 1973 & 1973 & Wisconsin Session Laws 1973, Chapter 89; Senate Bill 195 & ``It is the policy of this state to provide, as an integral part of   free public education, special education sufficient to meet the needs and   maximize the capabilities of all children with exceptional educational   needs." & 1976 & 1973 & Earlier law from Hensley et al. used \\ \hline
Wyoming & 1969 & 1969 & Session Laws of the State of Wyoming Passed by the Fortieth State   Legislature, Chapter 111 & ``Each and every child of school age in the State of Wyoming havng a   mental, physical or psychological handicap or social maladjustment which   impairs learning, shall be entitled to and shall receive a free and   appropriate education in accordance with his capabilities" & 1969 & 1969 & 

\end{longtable}
\end{landscape}
\normalsize

\end{document}